\documentclass[12pt,a4paper]{article}
\usepackage{jheppub}
\usepackage{graphicx} 
\usepackage{amsmath}
\usepackage{mathtools}
\usepackage{amssymb} 
\usepackage{dsfont}
\usepackage{xcolor}
\usepackage{cancel}
\usepackage{pgfplots}
\pgfplotsset{compat=1.18}
\usepackage{hyperref}
\usepackage[sep=5pt, offset=1.5em]{simpler-wick}
\usepackage[normalem]{ulem} 
\usepackage{amsthm}

\usepackage{braket}
\usepackage{circuitikz}
\usepackage{fixcmex}
\usepackage{tikz}
\usetikzlibrary{spy}

\usepackage{booktabs}
\usepackage{dsfont}
\usetikzlibrary{shapes.geometric,positioning}
\usepackage{cleveref}
\usepackage{verbatim}

\DeclareMathOperator{\Tr}{Tr}

\title{Cardy random matrix theory}

\author[a,b,c]{Marco Ambrosini,}

\author[d,e]{Alexandre Belin,} 

\author[f]{Jan de Boer}

\author[a,c]{and Julian Sonner}
\affiliation[a]{Department of Theoretical Physics, University of Geneva, 24 quai Ernest-Ansermet, 1214 Gen\`eve 4, Switzerland}
\affiliation[b]{Leinweber Institute for Theoretical Physics and Department of Physics, University of California, Berkeley, CA 94720, USA} 
\affiliation[c]{Geneva Quantum Center, University of Geneva, 24 quai Ernest-Ansermet, 1214 Gen\`eve 4, Switzerland}
\affiliation[d]{Dipartimento di Fisica, Universit\`a di Milano - Bicocca,
I-20126 Milano, Italy}
\affiliation[e]{INFN, sezione di Milano-Bicocca, I-20126 Milano, Italy} 
\affiliation[f]{Institute for Theoretical Physics,
University of Amsterdam, PO Box 94485,
1098 GL Amsterdam, The Netherlands}

\emailAdd{Marco.Ambrosini@unige.ch, alexandre.belin@unimib.it, J.deBoer@uva.nl, Julian.Sonner@unige.ch}

\abstract{We construct and study a ``Cardy random matrix model'': a random matrix theory whose only input, in a maximal-ignorance prescription, is the constraint of $\mathbb{S}$-modular invariance. We regularize this constraint by coarse-graining over energy windows of fixed width, and obtain a potential for a single random matrix whose eigenvalues play the role of the scaling dimensions of primaries in a 2d CFT. The potential drives the coarse-grained density of states to the Cardy profile above a gap at the black hole threshold. We perform Monte Carlo simulations to confirm this behavior and find good agreement for the coarse-grained density of states, while at the same time revealing GUE statistics for nearby eigenvalues. We identify the scaling limit in which these phenomena emerge. In this limit, the model realizes a hierarchy of scales: at the macroscopic scale given by the window size, the occupations are frozen and the bootstrap constraint rigidly fixes the spectrum to the Cardy profile, while at the mesoscopic scale (inside a window) the Vandermonde repulsion rearranges the eigenvalues as in a random matrix theory. It is only at the microscopic scale of the mean level spacing that the individual levels are resolved. Accordingly, our random matrix model displays an effective Thouless time, set by the inverse energy width of the coarse-graining window. Our model provides a first step towards solving the matrix/tensor model dual to 3D quantum gravity on a computer.}

\pgfplotsset{
  every axis plot/.append style={line width=1.2pt},
  every axis plot post/.append style={
    every mark/.append style={line width=2.0pt,draw=green,fill=red}
  }
}

\begin{document}

\maketitle

\section{Introduction}

The past few years have seen tremendous progress in formulating a CFT dual to pure gravity in AdS$_3$ \cite{Belin:2020hea,Chandra:2022bqq,Cotler:2020ugk,Mertens:2022ujr,DiUbaldo:2023qli,Haehl:2023tkr,Boruch:2025ilr,Belin:2023efa,Jafferis:2025vyp,deBoer:2023vsm,Collier:2023fwi, Collier:2024mgv,deBoer:2024mqg,deBoer:2025rct,Hartman:2025ula,Hartman:2025cyj,Chandra:2025fef,Yan:2023rjh,Chandra:2024vhm,Hung:2024gma,Belin:2026pko,Dymarsky:2026lnf,Jafferis:2026gzn}. The current understanding of the dual theory is not that of an individual compact CFT, but rather corresponds to an ensemble average over CFT data. This generalizes nicely the duality between JT gravity and a matrix integral to three dimensions, and explains why the spectrum of pure gravity obtained by Maloney, Witten and Keller (MWK) by summing over geometries with a torus boundary gives a continuous function of the energy \cite{Maloney:2007ud,Keller:2014xba}. It also explains the existence of wormholes solutions in 3D gravity that stretch between multiple asymptotically AdS Euclidean boundaries \cite{Maldacena:2004rf}.

The ensemble for 3D gravity is much more complicated than the matrix integral of JT gravity. This is a consequence of the fact that the CFT data contains both a spectrum and OPE coefficients, so there are more dynamical objects than the lone matrix in the JT ensemble. Perhaps even more importantly, the CFT data is subject to strong constraints coming from consistency: the CFT ensemble must be built such that crossing symmetry and modular invariance are obeyed. This has many important consequences,  notably the fact that the distribution of OPE coefficients contains arbitrarily high non-Gaussianities \cite{Belin:2021ryy,Anous:2021caj}, which must be correctly encoded in the probability distribution.

The state of the art for the dual to pure 3D gravity is a random tensor-matrix model introduced in \cite{Belin:2023efa} and developed further in \cite{Jafferis:2025vyp}. The model is built from a ``constraint$^2$" potential, and the probability distribution takes the form
\begin{equation}
\mathcal{Z}\sim \int dC dL_{0}d\bar{L}_{0}e^{-\frac{1}{a} V_{\textrm{crossing}}^2} \,.
\end{equation}
The potential depends both on the OPE coefficients $C_{ijk}$ and on the left and right moving scaling dimensions (the eigenvalues of $L_0$ and $\bar{L}_0$). It is built such that it vanishes on CFT data satisfying crossing symmetry and modular invariance. The potential is a square so it is purely positive, and in the limit $a\to0$ therefore enforces that the CFT data solves the bootstrap constraints which, among other things, should generate a modular invariant spectrum and the infinite tower of non-Gaussianities in the distribution of OPE coefficients. 

Unfortunately, many aspects of the tensor model still remain to be understood. If we start directly in the appropriate scaling limit as discussed in \cite{Jafferis:2025vyp}, then one can show that the triple-edge Feynman diagrams of the tensor model agree with Virasoro TQFT amplitudes \cite{Collier_2023}, which is the most complete formulation of bulk partition functions on manifolds of fixed topology.\footnote{see also \cite{Hartman:2025ula} for an alternative TQFT formulation when the bulk theory has different boundary conditions.} But this formulation of the tensor model requires one to work after having taken the correct scaling limit, and the way to take this limit is not understood. This should be contrasted with JT gravity, where extensive work on matrix models has enlightened the nature of the correct double-scaling limit, and where an explicit double-scaling procedure was described in \cite{Jafferis:2022wez}.

In this paper we will make progress on this front: we will study, in a simplified
model, how to take the scaling limits for these `constraint$^2$' potentials. This analysis is made possible by a careful regularization of the
modular constraints that the model imposes, and a subsequent scaling limit of the parameters involved. Two ingredients turn
out to be decisive:
first, the way in which the size of the matrix and the number of primaries retained below a certain UV cutoff
scale decides whether the ensemble produces a Cardy
density of states. Second, the resolution at which the modular constraint is
imposed, and which we will find to set the
scale below which the model behaves as a random matrix theory, must also be dealt with carefully. The next
subsection summarizes the construction and our analytical and numerical results.

\subsection{Overview}

We consider a simplified model where the constraint we wish to impose is that of modular invariance on the torus: our simplified model contains no tensors, i.e. no OPE coefficients, and is thus purely a matrix model. Moreover, we consider a CFT spectrum made out of a single quantum number: the scaling dimension.\footnote{Note that this is not a chiral 2d CFT, since in a chiral CFT the spin, i.e. the left moving dimensions, must be quantized. Here the conformal dimensions can take any real values.} Our model thus contains a single random matrix whose eigenvalues correspond to the scaling dimensions, and which are dynamically governed by a `constraint$^2$' potential enforcing
\begin{equation}
Z(\beta) = Z(4\pi^2/\beta) \,,
\end{equation}
for all inverse temperatures $\beta$.
To turn this condition into a matrix potential we use the `constraint$^2$'
prescription of \cite{Belin:2023efa}: instead of imposing the constraint exactly, one
softens the delta function that enforces it into a Gaussian of width $\sqrt{a}$,
so that individual members of the ensemble may violate it by a controlled amount.
What is left is a potential quadratic in the constraint, with $a$ measuring the
tolerance and $a\to 0$ recovering the strict case. In
\cref{sec:balian_review} we show that this prescription is closely related to the maximal
ignorance one in the sense of Balian \cite{Balian1968}, so that no information beyond
the constraint itself is injected into the ensemble.
 
The constraint needs some care before it can be used. Modular invariance is a property of the partition function as a function of temperature, but it can be
recast as a statement about Virasoro characters \cite{Collier:2019weq}, which can then further be projected onto a
single character. However, this projection identifies the spectral density of a discrete set of primaries, a sum of delta functions, with the smooth modular
$\mathbb{S}$-kernel, and is thus ill defined as it stands in order to use it in a potential. We resolve this in \cref{sec:cardy_matrix} by smearing the projection over a window of width
$\epsilon$ in the Liouville momentum $P=\sqrt{h-(c-1)/24}$. What survives is a
condition on the number $N_{P,\epsilon}$ of eigenvalues in each window, that is,
on the coarse-grained spectrum rather than on individual levels, and feeding it
into the `constraint$^2$' prescription gives the potential of the model we study:
\begin{equation}
\label{eq:cardypotential}
    \mathcal{Z} \; = \; \int \prod_j dh_j \; \Delta(\mathbf{h})^2 \,
    \exp\left[ -\frac{1}{4a\epsilon^2} \sum_P
    \Big( N_{P,\epsilon} - \epsilon \sum_i \mathbb{S}_{PP_i}(\mathds{1}) \Big)^2
    \right] ,
\end{equation}
with $\Delta(\mathbf{h})=\prod_{i<j}(h_i-h_j)$ the Vandermonde determinant of the
unitary class and $\mathbb{S}_{PP'}(\mathds{1})$ the modular $\mathbb{S}$-kernel
\cite{teschner2003liouvilletheoryquantumgeometry}. 
The potential is easy to read off. It sees the eigenvalues only through the bin
occupations, so it is blind to their arrangement inside a bin, and, since the
kernel couples every eigenvalue to every other, it is a double-sum object, which
is the main obstruction to solving the model directly. An under-occupied bin
attracts eigenvalues from its neighbors while an over-occupied one expels them,
so at equilibrium, neglecting the Vandermonde contribution, every occupation sits on the target set by the crossing kernel.
When the vacuum is occupied, that target is dominated at high energies by the
identity, $\epsilon\,\mathbb{S}_{P\mathds{1}}(\mathds{1})\equiv\epsilon\,\rho_{\rm Cardy}(P)\to
\sqrt{2}\,\epsilon\,e^{2\pi QP}$ with
$Q=b+b^{-1}$ and $c=1+6Q^2$. Living up to its name, the potential thus drives the
coarse-grained density of states to the Cardy profile \cite{cardyformula}, as an
\emph{output} of modular invariance and of maximal ignorance, not as an input.

We consider two variants of the model, depending on what we impose on the light states, i.e. those below the black hole threshold: 
\begin{itemize}

\item In the \emph{sourced} version, the light-state occupations are rigidly fixed and prescribed externally. They thus enter the constraints as sources. For example, we can declare that the only light state occupied is the identity. This is what a model for pure 3D gravity should have, and was the perspective taken in \cite{Belin:2023efa}. We can also consider a model where no light states are present at all: the ``source" for the identity operator is switched off. We use these versions of the model in the numerical analysis.

\item In the \emph{dynamical} version, the eigenvalues entering \eqref{eq:cardypotential} are allowed to explore energies below the black-hole threshold, so long as they stay above the unitarity bound $h\geq 0$. In this model, the light states are not rigidly fixed, but dynamically occupied. If it so happened that the identity was the only state occupied, this would come out as an output and would imply that pure gravity is favored by the maximum ignorance ensemble, similarly to what was studied in \cite{Belin:2025qjm}.\footnote{Given the results of \cite{Belin:2025qjm} for actual CFTs, we believe this to be unlikely, but it would be very interesting to understand better.} This version of the model allows us to determine the scaling limits under which the preferred configuration dynamically populates the vacuum, and the remaining heavy states arrange following the coarse-grained Cardy density. We can study these questions analytically, although only at the level of leading scaling limits, and the details of the light spectrum occupation are beyond the scope of this work.
\end{itemize}

For the bulk of this paper, we will study the sourced model, but we comment on our expectations for the dynamical model as well in \cref{sec:vac_occ_gap}.\footnote{We believe that at the level of accuracy at which only the leading exponential behavior of the density of states is considered, the two formulations display very similar high-energy physics if the number of states as function of the UV cutoff $N(\mathcal{E})$ is chosen to scale in the right way.} 
 As already mentioned, the sourced model is where the numerical analysis is framed. The expected behavior of the theory designed by the constraint$^2$ potential is confirmed numerically by the Markov chain Monte Carlo study in \cref{sec:numerics}. The measured occupations track the Cardy target at
high energies, and the density crosses over
sharply from a bounded profile of principal-series states to
exponential growth whose logarithmic slope reproduces $2\pi Q$ to within a few
percent; with the identity source switched off at the same
$N$, the spectrum stays bounded and never develops Cardy growth, confirming that
the exponential tower is sourced by the vacuum in the dual channel. 

Tracking the energy levels at the scale of mean level spacings, inside a window lying in the Cardy
regime, the model displays GUE statistics: Wigner-Dyson level
spacing and a linear ramp in the connected spectral form factor up to the
Heisenberg time. The model therefore reproduces
both the coarse-grained Cardy density (when the vacuum is occupied) and the fine-grained statistics one expects
of a chaotic CFT. Understanding the parameter regime where the model dynamically displays the physics we seek requires a closer look at the
parameters of the model, and an analytical understanding of the scaling-limit procedure.
We have:
\begin{itemize}
    \item $N$, the number of primary operators of the CFT, identified with the size of the
    matrix in standard matrix models, which we will eventually want to take to
    infinity.
 
    \item $a$, the coefficient that multiplies the `constraint$^2$' potential,
    which we will eventually send to zero in order to enforce the constraint
    strictly.
 
    \item $\epsilon$, the width of the coarse-graining window introduced above,
    which we may either send to zero or keep finite.
 
    \item $\mathcal{E}$, a UV cutoff on the Liouville momenta, without which the matrix integral 
    \eqref{eq:cardypotential} is ill-defined, as eigenvalues would run away to infinity; it is related to a
    cutoff $E$ on the energies by $\mathcal{E}=\sqrt{E-(c-1)/24}$, and eventually needs to be sent to infinity.
\end{itemize}

It was already  clear that the limits $N,\mathcal{E}\to\infty$, $a\to 0$ were needed,
but unfortunately unclear what type of multiple scaling was required to correctly
produce the appropriate model to describe a CFT. Part of the answer is that $N$
and $\mathcal{E}$ are not independent: in an ordinary matrix model the matrix
size is a free parameter, whereas here it counts the primaries below the cutoff,
and modular invariance fixes the asymptotic density of states to the Cardy form, so that the correct scaling $N(\mathcal{E})$ we need to chose is determined, up to sub-exponential
corrections, by the very constraint we are imposing. Prescribing how the matrix size scales
with the cutoff thus has a direct effect on the occupation numbers and the physics of the model. 
In \cref{sec:saddle_point}, for the heavy-states only configuration, we show that the double-scaling limit singled out by the saddle-point analysis, in which the occupation numbers localize to approximate modular-invariance, consists of:
\begin{equation}
\label{eq:doublescaling}
    N,\mathcal{E}\to\infty\, , \qquad a,\epsilon \to 0\, ,
    \qquad\text{with}\qquad
    n_b \equiv \epsilon N\, , \quad
    G \equiv \epsilon \mathcal{E}\, , \quad
    t \equiv a N
    \quad \text{fixed}\,.
\end{equation}
 When the vacuum is occupied, filling the Cardy well requires a stronger scaling discussed in \cref{sec:vac_occ_gap}, with \(N\) growing exponentially with the cutoff \(\mathcal E\).

\begin{figure}[t]
  \centering
  \definecolor{binyellow}{HTML}{B8860B}
  \definecolor{eigpurple}{HTML}{5B21B6}
 
  \begin{tikzpicture}[
    every axis/.append style={
      tick style={draw=none},
      xticklabel=\empty, yticklabel=\empty, xtick=\empty, ytick=\empty,
      label style={font=\small}},
    cardy/.style  ={black, thin},
    binned/.style ={binyellow, thick, const plot},
    binfill/.style={binyellow, thick, const plot, fill=binyellow!28},
    meso/.style   ={red!75!black, thin, fill=red!20},
    wall/.style   ={red!75!black, dashed, thin},
    ann/.style    ={red!70!black, font=\footnotesize},
    ]

  \def\CardyK{3.4}                
  \def\CardyHz{1}                 
  \def\CardyNrm{223.96}           
  \def\CardyRho{((exp(\CardyK*sqrt(x-\CardyHz))-exp(-\CardyK*sqrt(x-\CardyHz)))^2)/(4*sqrt(x-\CardyHz)*\CardyNrm)}

  \begin{axis}[name=main,
      width=5.9cm, height=4.9cm, scale only axis,
      axis lines=left, axis line style={-{Stealth[length=4pt]}},
      enlargelimits=false, clip=false,
      xmin=0.93, xmax=2.07, ymin=0, ymax=1.02,
      xlabel={$h$}, ylabel={$\rho(h)$},
      every axis x label/.style={at={(axis description cs:1,0)},anchor=north west},
      every axis y label/.style={at={(axis description cs:0,1)},anchor=south east},
      legend style={draw=none, fill=none, font=\footnotesize,
                    at={(0.03,0.97)}, anchor=north west, cells={anchor=west},
                    legend image post style={xscale=.7}},
    ]

    \addplot[binfill] coordinates {
      (1.00000,0.00291) (1.00694,0.00718) (1.02778,0.01299) (1.06250,0.02106)
      (1.11111,0.03292) (1.17361,0.05095) (1.25000,0.07901) (1.34028,0.12335)
      (1.44444,0.19418) (1.56250,0.30834) (1.69444,0.49359) (1.84028,0.79603)
      (2.00000,0.79603)}\closedcycle;
    \addlegendentry{$\bar\rho_\epsilon(h)$}

    \addplot[cardy, domain=1.0005:2, samples=300] {\CardyRho};
    \addlegendentry{$\rho_{\rm Cardy}(h)$}

    \draw[thin] (axis cs:1,0) -- (axis cs:1,-0.022);
    \node[font=\scriptsize, anchor=north] at (axis cs:1,-0.02) {$\frac{c-1}{24}$};

    \draw[red!75!black, thin] (axis cs:1.2335,0) rectangle (axis cs:1.5625,0.30);
    \coordinate (boxNE) at (axis cs:1.5625,0.30);
    \coordinate (boxSE) at (axis cs:1.5625,0);
  \end{axis}

  \begin{axis}[name=zoom,
      at={($(main.south east)+(1.7cm,0.55cm)$)}, anchor=south west,
      width=5.7cm, height=4.35cm, scale only axis,
      axis lines=box, axis line style={red!75!black, thin},
      enlargelimits=false, clip=true, clip mode=individual,
      xmin=1.2335, xmax=1.5625, ymin=0, ymax=0.30,
    ]

    \addplot[binfill] coordinates {(1.17361,0.05095) (1.25000,0.07901)
        (1.34028,0.12335) (1.44444,0.19418) (1.56250,0.19418)}\closedcycle;
 
    \addplot[meso, domain=1.17366:1.24083, samples=400]
       {0.05441*sqrt(x-1.16611)*sqrt(1.24083-x)/(sqrt(x-1.17361)*sqrt(1.25000-x))}\closedcycle;
    \addplot[meso, domain=1.25015:1.32944, samples=400]
       {0.08388*sqrt(x-1.24083)*sqrt(1.32944-x)/(sqrt(x-1.25000)*sqrt(1.34028-x))}\closedcycle;
    \addplot[meso, domain=1.34070:1.43194, samples=400]
       {0.13040*sqrt(x-1.32944)*sqrt(1.43194-x)/(sqrt(x-1.34028)*sqrt(1.44444-x))}\closedcycle;
    \addplot[meso, domain=1.44564:1.54833, samples=400]
       {0.20464*sqrt(x-1.43194)*sqrt(1.54833-x)/(sqrt(x-1.44444)*sqrt(1.56250-x))}\closedcycle;

    \addplot[cardy, domain=1.2335:1.5625, samples=100] {\CardyRho};
    \addplot[binned] coordinates {(1.17361,0.05095) (1.25000,0.07901)
        (1.34028,0.12335) (1.44444,0.19418) (1.56250,0.19418)};

    \draw[wall] (axis cs:1.25000,0) -- (axis cs:1.25000,0.30);
    \draw[wall] (axis cs:1.34028,0) -- (axis cs:1.34028,0.30);
    \draw[wall] (axis cs:1.44444,0) -- (axis cs:1.44444,0.30);

    \addplot[ycomb, no marks, eigpurple, line width=.3pt] coordinates {
      (1.23500,0.022)
      (1.25085,0.022) (1.25645,0.022) (1.26473,0.022) (1.27422,0.022) (1.28444,0.022)
      (1.29526,0.022) (1.30677,0.022) (1.31970,0.022)
      (1.34060,0.022) (1.34298,0.022) (1.34706,0.022) (1.35216,0.022) (1.35791,0.022)
      (1.36409,0.022) (1.37058,0.022) (1.37733,0.022) (1.38430,0.022) (1.39150,0.022)
      (1.39895,0.022) (1.40672,0.022) (1.41501,0.022) (1.42448,0.022)
      (1.44456,0.022) (1.44545,0.022) (1.44711,0.022) (1.44940,0.022) (1.45217,0.022)
      (1.45531,0.022) (1.45871,0.022) (1.46233,0.022) (1.46612,0.022) (1.47005,0.022)
      (1.47410,0.022) (1.47825,0.022) (1.48249,0.022) (1.48682,0.022) (1.49122,0.022)
      (1.49571,0.022) (1.50028,0.022) (1.50493,0.022) (1.50968,0.022) (1.51453,0.022)
      (1.51952,0.022) (1.52468,0.022) (1.53010,0.022) (1.53593,0.022) (1.54274,0.022)};
    \draw[eigpurple, font=\footnotesize]
         (axis cs:1.27422,0.014) -- (axis cs:1.2680,-0.016)
         node[anchor=north, inner sep=1pt]{$\sim e^{-S}$};

    \draw[{Stealth[length=4pt]}-{Stealth[length=4pt]}, red!75!black]
         (axis cs:1.34028,0.277) -- (axis cs:1.44444,0.277)
         node[midway, fill=white, inner sep=1pt, font=\footnotesize]{$\delta$};

    \node[font=\footnotesize, anchor=north] at (axis cs:1.34028,-0.002) {$b_k$};
    \node[font=\footnotesize, anchor=north west] at (axis cs:1.4475,-0.002) {$b_{k+1}$};
    \draw[red!75!black, thin] (axis cs:1.43194,0) -- (axis cs:1.43194,-0.013);
    \node[ann, anchor=north east, inner sep=1pt] at (axis cs:1.4335,-0.010) {$a_k$};
 
    \draw[ann] (axis cs:1.2632,0.118) -- (axis cs:1.2830,0.196)
         node[anchor=south west, inner sep=1pt]{$\rho(h)$};
  \end{axis}

  \draw[red!75!black, thin] (boxNE) -- (zoom.north west);
  \draw[red!75!black, thin] (boxSE) -- (zoom.south west);
 
  \end{tikzpicture}
  \caption{\label{fig:cardy-bins}%
    \emph{Left}: the smooth Cardy density $\rho_{\rm Cardy}(h)$ (black) and the
    coarse-grained density $\bar\rho_\epsilon(h)$ (yellow bins), obtained by
    averaging it over the microcanonical windows, of constant width $\epsilon$ in the Liouville
    momentum, used to regulate the modular
    constraints. Minimization of the `constraint$^2$' Cardy RMT potential
    fixes only the occupation numbers of the bins, that is the \textit{macroscopic physics} of the system, which provides a piecewise constant approximation to the continuous Cardy profile. \emph{Right}: if we zoom at scales comparable to the bins' width, $\delta$ in energy variables, we unveil the \textit{mesoscopic physics} of the model, governed by the Vandermonde eigenvalue repulsion at approximately frozen occupations number. The frozen occupations act as hard walls at the bin
    edges $b_k$, while the repulsion from the next
    wall opens a soft edge at $a_k$. Notice that this scale is still much larger than the \textit{microscopic} one, where we are able to distinguish the single eigenvalues (in purple), so that
    $e^{-S}\ll\epsilon\ll 1$.}
\end{figure}
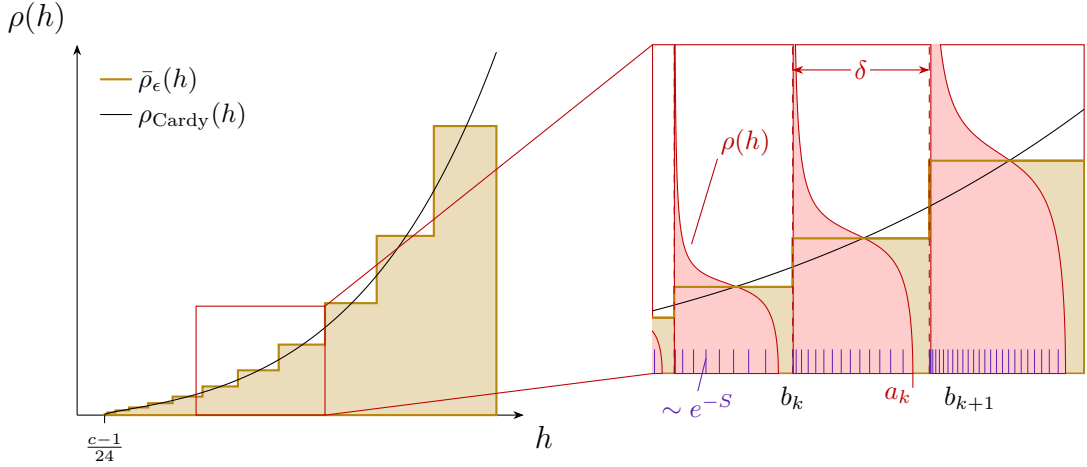

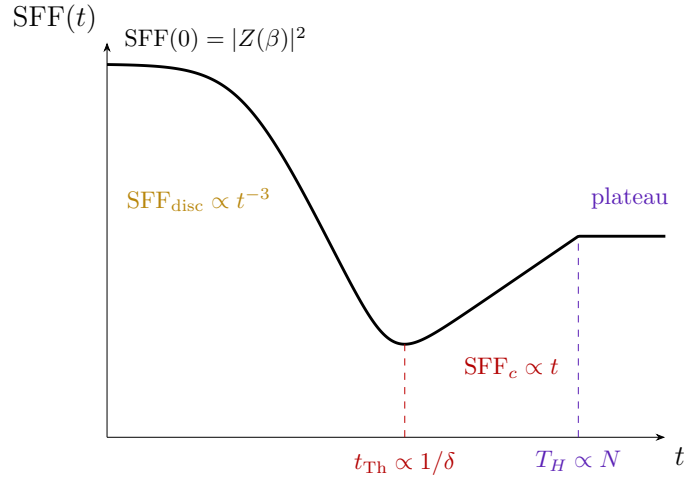
\begin{figure}[t]
  \centering
  \definecolor{binyellow}{HTML}{B8860B}
  \definecolor{eigpurple}{HTML}{5B21B6}

 \begin{tikzpicture}[scale=0.90]
  \def\Cramp{1.875e-5}          
  \def\Plat{9.375e-3}           
  \def\SFFdisc{(1+10^(2*\t))^(-1.5)}
  \def\SFFramp{\Cramp*sqrt(1+10^(2*\t))}

  \begin{loglogaxis}[
      width=8.2cm, height=5.8cm, scale only axis,
      xmin=0.08, xmax=2500, ymin=4e-5, ymax=1.8,
      axis lines=left, axis line style={-{Stealth[length=4pt]}},
      xlabel={$t$}, ylabel={$\mathrm{SFF}(t)$}, label style={font=\small},
      every axis x label/.style={at={(rel axis cs:1,0)}, anchor=north west},
      every axis y label/.style={at={(rel axis cs:0,1)}, anchor=south east},
      xtick=\empty, ytick=\empty, minor tick num=0, tick style={draw=none},
      clip=true, clip mode=individual
    ]

    \addplot[very thick, black, variable=\t, domain=-1.1:3.4, samples=320]
            ({10^\t},{\SFFdisc + min(\SFFramp,\Plat)});

    \addplot[red!75!black, dashed, thin] coordinates {(20,4e-5) (20,5.0e-4)};
    \addplot[eigpurple,   dashed, thin] coordinates {(500,4e-5) (500,9.375e-3)};
    \node[red!70!black, font=\footnotesize, anchor=north] at (axis cs:20,3.6e-5)
         {$t_{\rm Th}\propto 1/\delta$};
    \node[eigpurple, font=\footnotesize, anchor=north] at (axis cs:500,3.6e-5)
         {$T_H\propto N$};

    \node[font=\footnotesize, anchor=south] at (axis cs:0.6,1.06)
         {$\mathrm{SFF}(0)=|Z(\beta)|^{2}$};
    \node[binyellow, font=\footnotesize, anchor=north east]
         at (axis cs:2.0,0.045) {$\mathrm{SFF}_{\rm disc}\propto t^{-3}$};
    \node[red!70!black, font=\footnotesize, anchor=north west]
         at (axis cs:50,4.5e-4) {$\mathrm{SFF}_{c}\propto t$};
    \node[eigpurple, font=\footnotesize, anchor=south]
         at (axis cs:1300,1.5e-2) {plateau};

  \end{loglogaxis}
  \end{tikzpicture}
  \caption{\label{fig:sff_schematics}%
    Timescales of the spectral form factor in the Cardy matrix model at high energies
    (log--log, schematic, at fixed $\beta$), observed numerically in \cref{sec:numerics} and discussed analytically in \cref{sec:meso_phys}.  The three colors are those of
    figure~\ref{fig:cardy-bins} and label the same three scales: yellow for
    the macroscopic, coarse-grained Cardy physics, red for the mesoscopic physics
   inside a bin, and purple for the microscopic scale of the individual
    eigenvalues.
    At early times $t\lesssim\beta$ the form factor is flat,
    $\mathrm{SFF}(0)=|Z(\beta)|^2$.  It is then dominated by its disconnected
    component, which decays as a power law, and this is all the macroscopic (yellow) density of figure~\ref{fig:cardy-bins} can
    produce: no ramp and no plateau.  At $t\sim t_{\rm Th}\propto1/\delta$
    the connected in-bin contribution takes over and produces the
    linear ramp, so
    that the coarse-graining scale $\delta$ plays the role of an inverse Thouless
    time.  The ramp finally saturates at the Heisenberg time
    $T_H\propto N$, set by the mean level spacing, which at high-energies is $\sim e^{-S}$, of the purple
    eigenvalues of figure~\ref{fig:cardy-bins}. We remark that the connected contributions of the SFF still presents some regularization-dependent contribution on top of the universal linear behavior, which we can think of as analogous to the in-bin behavior of the mesoscopic density (red) in \cref{fig:cardy-bins}. In both cases, regularization-dependent effects can be removed by averaging over the bins' positions, before computing any observables.}
\end{figure}

The coexistence of a rigid coarse-grained density with GUE statistics at fine
resolution places the model in the hierarchy of \cref{fig:cardy-bins}, where
the potential fixes the macroscopic physics at scales coarser than $\epsilon$,
the Vandermonde governs the mesoscopic rearrangement of eigenvalues inside each
bin, as we study in \cref{sec:mesoscopic_inbin}, while individual levels are
resolved only at the microscopic scale $\sim e^{-S}$, with
$e^{-S}\ll\epsilon\ll 1$. In order to understand better the timescales associated with this hierarchy, in \cref{sec:top_recursion} we study the spectral form factor, whose expected behavior at the scales of interest is schematically reported in \cref{fig:sff_schematics}, as the
sharpest diagnostic of scale-separation. At the macroscopic scale, every occupation is frozen and, at any
resolution coarser than $\epsilon$, the density is essentially non-fluctuating; correlators
factorize and the disconnected piece $|\langle Z\rangle|^2$ dominates, showcasing the power-law decay characteristic of a continuous density with a square-root edge.

The correlations omitted by this coarse description become visible when the internal structure of the bins is resolved. Translating this problem in the language of topological recursion \cite{Eynard:2007kz}, brings us to consider a spectral curve with parametrically large genus in order to compute resolvent correlators. We are able to provide some analytical progress using a cut-pinching limit in a high-energy region of the spectrum, where the spectral curve degenerates to a genus-one problem and becomes 
tractable. In \cref{sec:meso_phys}, we compute the connected spectral form factor, which shows the expected linear ramp, valid for $t\gg 1/G$ with $\delta=2G$ the typical high-energy bin width in
energy variables. The onset of this ramp determines the Thouless time $t_{\rm Th}\sim
1/\delta\sim 1/G$: the onset of random matrix behavior is the inverse of the
width, in energy, of the window over which modular invariance was imposed. The intuition is that, via the constraint$^2$ potential, the
bootstrap constraint determines the spectrum at the macroscopic scale, no further than the resolution at which it
is imposed, after which it behaves like a random matrix theory. This also identifies the
corner of parameter space relevant for a genuinely chaotic CFT, where an
$\mathcal{O}(1)$ Thouless time is expected \cite{Chen:2024oqv}: it is reached
by holding $G$ fixed, independent of $N$, when performing the scaling limits.

\section{Defining the modular ``Cardy RMT"}

In this section we define the Cardy random matrix theory. We start by reviewing in \cref{sec:balian_review} how to build in general a matrix ensemble that knows about a prescribed set of constraints in the maximally ignorant sense
of \cite{Balian1968,jaynes1957information1,deBoer:2023vsm}, and how to soften them to obtain the `constraint$^2$'
potentials of \cite{Belin:2023efa}. Then, in \cref{sec:airy_quartic} we apply this machinery to a simple
case, a single constraint fixing the second moment of the spectrum, which returns the spectral properties of the Gaussian
ensemble, modulo some subtleties, which we discuss. Finally, in \cref{sec:cardy_matrix}, we turn to the constraint of interest: we recast
$\mathbb{S}$-modular invariance in terms of Virasoro characters and regulate it
(\cref{sec:S_modinv_reg}), discuss how the width of the regulating window should be chosen
(\ref{sec:bins_width}), and assemble the result into the Cardy random matrix theory potential
(\ref{sec:cardy_potential}).

\subsection{Maximal ignorance ensembles and the `constraint$^2$' potentials}\label{sec:balian_review}
The goal of this section is to construct a probabilistic ensemble of matrices, $M$, with probability measure $P[M]$, in such a way that it satisfies a given set of $K$ constraints $C_I$
\begin{equation}\label{eq:constraints_averaged}
    \left\langle C_I \right\rangle = 0,\qquad I = 1,2,\ldots K\,,
\end{equation}
where the angle brackets denotes an average with respect to the measure $P[M]$. 
We require that the distribution contains no further information than the one contained in the constraints, in the sense that the probability distribution $P[M]$ maximizes the Shannon information entropy
\begin{equation}    
S[P] = -\int [dM]P[M]\log P[M]
\end{equation}
subject to the satisfaction, upon ensemble averaging, of the above constraints, together with the normalization $\int[dM]P[M]=1$. It has been shown by Balian \cite{Balian1968} that the ensemble satisfying the above requirements, is given by setting to zero the constrained variation of the Shannon entropy, that is by demanding
\begin{equation}\label{eq:balian_variation}
\begin{aligned}
    \delta S[P]=-\int [dM]\delta P\biggr(\log P+1+f_0+&\sum_{I} f_IC_I\biggr) =0\,.
\end{aligned}
\end{equation}
Here $f_I$ are the Lagrange multipliers for each of the constraints, which have to be chosen such that \eqref{eq:constraints_averaged} are satisfied, while $f_0$ is set so that the distribution is correctly normalized. The resulting ensemble then has the measure
\begin{equation}\label{eq:balian_potential}
\begin{aligned}
 P[M]\propto \exp\left( -\sum_I f_I C_I[M] \right)\,.
\end{aligned}
\end{equation}
We could also consider a family of constraints that depend on a continuum parameter $\mathcal{I}$, in which case the sum in the exponent is replaced by an integral,
or indeed a case where some parameters are discrete and others continuous.\\

Next we explain how this construction is related to the so-called `constraint$^2$' potentials that were employed in the BDJNS-matrix/tensor model \cite{Belin:2023efa}\footnote{In this paper, we will be interested mostly to the matrix sector of the model in \cite{Belin:2023efa}. Thus, for sake of simplicity, we frame the discussion in terms of a distribution of matrices $M$, however a generalization to include other matrices and tensors poses no further conceptual complications.}. 
Our goal is to construct an ensemble of matrices that {\it softly} imposes a given set of constraints, in the sense of \cite{Belin:2023efa}, namely one which allows small violations of the constraints, uniformly controlled by a parameter, $a$. But before defining these {\it soft} ensembles, it is useful to first consider their {\it hard} cousins, in which the same constraints are imposed exactly at the level of expectation values over the ensemble. As we have seen above, the max-entropy construction is such a hard ensemble,
\begin{equation}\label{eq:Ztilt}
    Z_{\mathrm{tilt}}(f)=\int[dM]e^{-f\,C[M]},
\end{equation}
where the name `tilt' is common in probability theory\footnote{Note also that we have temporarily switched to just one single constraint, to keep notation lean. The multiple constraint case works exactly the same way, treating each constraint analogously.}. A more stringent way to ensure constraint satisfaction is given by the expression
\begin{equation}\label{eq:Zhard}
    Z_{\mathrm{exact}}=\int[dM]\delta\big(C[M]\big)\,,
\end{equation}
which imposes the constraint as a delta function inside the matrix integral, imposing the constraint in a microcanonical fashion, to draw a thermodynamic analogy. This achieves $\langle C\rangle=0$ by exactly constraining individual samples.
In order to see how the latter relates to the former, we start by writing the delta function in its Bromwich / Laplace form
\begin{equation}\label{eq:bromwich_delta}
    \delta(C)=\frac{1}{2\pi i}\int_{\gamma-i\infty}^{\gamma+i\infty}df\,e^{-fC},
\end{equation}
where as usual $\gamma \in \mathbb{R}$ should be set inside the region of convergence. Inserting this into the delta-constrained ensemble \eqref{eq:Zhard}, gives the exact identity
\begin{equation}\label{eq:exact_transform}
    Z_{\mathrm{exact}}=\frac{1}{2\pi i}\int_{\gamma-i\infty}^{\gamma+i\infty}df\;Z_{\mathrm{tilt}}(f).
\end{equation}
Thus, the delta-constrained partition function is the inverse Laplace transform of the unnormalized tilted partition function. If one evaluates this integral instead by steepest descent, then it localises on configurations satisfying
\begin{equation}\label{eq:saddle_cond}
    \frac{d}{df}\log Z_{\mathrm{tilt}}(f)=0
    \quad\Longleftrightarrow\quad
    \langle C\rangle_{f}\equiv\frac{\int[dM]\,C\,e^{-fC}}{\int[dM]\,e^{-fC}}=0\,,
\end{equation}
which is precisely the condition that the Lagrange multipliers ensure on average constraint satisfaction exactly. In other words, we have recovered the max-entropy ensemble.

\subsection*{Softening the model}
The above discussion results in matrix-valued measures subject to exactly satisfied constraints. Instead, as in \cite{Belin:2023efa}, we want to construct ensembles which impose these same constraints softly, allowing parametrically small violations. This is now easily done, by following the above route, but instead of the hard delta one uses a smeared version, or nascent delta function,
\begin{equation}\label{eq:nascent_bromwich}
    \delta_a(C)=\frac{1}{2\pi i}\int_{\gamma-i\infty}^{\gamma+i\infty}df\;
    e^{\,af^2-fC},\qquad a>0\,.
\end{equation}
The Gaussian smearing of the nascent delta commutes with the manipulations of the previous section, which therefore go through as before with the regulator going along for the ride. Instead of following these steps to obtain a smeared version of the max-entropy ensemble, we can set $f=\gamma+it$ and performing the resulting Gaussian integral over $t$,
\begin{equation}\label{eq:nascent_bromwich_eval}
\begin{aligned}
    \frac{e^{a\gamma^2-\gamma C}}{2\pi}\int_{-\infty}^{\infty}dt\;
    e^{-at^2+it(2a\gamma-C)} =\frac{1}{2\sqrt{\pi a}}\,e^{-C^2/4a}\,.
\end{aligned}
\end{equation}
If we perform this simple manipulation inside the matrix integral, we obtain our old friend, the constraint$^2$ potential
\begin{equation}\label{eq:constraint_squared_general_der}
   Z_{\rm soft} \propto \int  [dM]\exp \left(-\frac{1}{4a} \sum_I C_I^2  \right)\,.
\end{equation}
Let us conclude this section with a few comments.
\begin{itemize}
    \item \emph{The max-entropy philosophy.} We note that this is an application of the maximal-ignorance principle in the construction of our model, but distinct from Balian's argument above.  One may regard the softened condition as imposing $C_I[M]=\eta_I$, where the unknown constraint errors are unbiased and have a prescribed isotropic covariance, given by $a$. The distribution of $\eta_I$ that maximizes Shannon entropy subject only to these data is the Gaussian $\propto\exp[-\sum_I\eta_I^2/(4a)]$.  Averaging the hard constraint over this error distribution gives directly the Gaussian smearing of the Fourier-conjugate Lagrange multipliers obtained from the Bromwich contour above. The distributional covariance, $a$, then supplies the independent tolerance scale so that typical constraint violations are of order $\sqrt a$.
    \item \emph{Thermodynamic ensembles.} There is a thermodynamic analogy to the discussions and manipulations of this section. Under this analogy, the $\delta$-constrained ensemble becomes the microcanonical ensemble and the tilted ensemble at fixed multiplier $f$ is the canonical ensemble, with $f$ playing the role of the inverse temperature. The steepest-descent evaluation of \eqref{eq:exact_transform}, which replaces the multiplier integral by its saddle, is the same logic that establishes the equivalence of the two (exact in the thermodynamic limit).
    \item \emph{The Balian ensemble as a saddle.} The fixed Balian multiplier $f_*$, determined by $\langle C\rangle_{f_*}=0$, reproduces the full multiplier integral only when the latter localizes at its saddle, as in the usual canonical--microcanonical equivalence. The accuracy of this approximation is controlled by the fluctuations of the constraint, $\partial_f^2\log Z_{\rm tilt}=\operatorname{Var}_f(C)$, while higher derivatives give its higher cumulants. Hence equivalence requires an appropriate large-$N$ concentration property and does not follow from $\langle C\rangle_{f_*}=0$ alone. By contrast, the Gaussian multiplier integral defining the softened model is exactly evaluable at any $N$.
\item \emph{Annealed versus quenched multiplier averages.}
The identities above concern partition functions, and therefore un-normalized tilted measures. For a fixed multiplier, we can write the following expression for a normalized expectation value $\langle\mathcal{O}\rangle_f
= Z_{\rm tilt}(f)^{-1} \int[dM]\,\mathcal{O}[M]e^{-fC[M]}$.
The distinction is then a choice of model: in this work we integrate over the multiplier at the level of the unnormalized measure and normalize only afterward, which is the annealed prescription and produces precisely the constraint$^2$ ensemble. Integrating the normalized fixed-$f$ expectation values instead would define a quenched prescription and a different ensemble. At large $N$, the distinction may become immaterial if the two multiplier averages localize at the same saddle $f_*$ and normalized fixed-$f$ observables vary smoothly. 
\end{itemize}
\subsection{The constraint$^2$ Gaussian and its quartic potential}
\label{sec:airy_quartic}
In order to recover the content of the well-known and studied Gaussian matrix model with Airy edge, we introduce the single constraint fixing the second moment of the spectrum,
\begin{equation}\label{eq:GUE_constraint}
C[M]=\tfrac1N\Tr M^2 - g\,.
\end{equation}
From the expression \eqref{eq:GUE_constraint} above, it is immediate \cite{Balian1968} that this single constraint, together with the max-entropy principle give the Gaussian RMT, $P[M] \propto e^{- \Tr V}$, with potential $\Tr V[M]=\frac{N}{2g} \Tr M^2$ where the prefactor is fixed by the value $f = N^2/2g$ of the single Lagrange multiplier introduced in order to ensure the satisfaction of the constraint. The above manipulations demonstrate the well-known statement that Gaussian random matrix ensembles arise as the max-entropy distributions over normalizable matrices. This explains their ubiquitous appearance, for example to describe the level statistics of quantum chaotic Hamiltonians. In that context one would require the matrices to satisfy certain further symmetry properties, such as hermiticity etc.

With our later application to the Cardy matrix model in mind, we now want to analyse the constraint$^2$ version of this model. In this case, the potential becomes quartic, $P(M)\propto \exp\!\big[-\tfrac{1}{4a}(\tfrac1N\Tr M^2-g)^2\big]$, giving rise to the  partition function
\begin{equation}
    Z=\int \prod_i d\lambda_i\,\Delta(\lambda)^2
  \exp\left[-\frac{1}{4a}
    \left(\frac1N\sum_i\lambda_i^2-g\right)^2\right]\,,
\end{equation}
where we have changed to the eigenvalue basis, giving rise to the Vandermonde measure factor $\Delta(\lambda)=\prod_{i<j}(\lambda_i-\lambda_j)$. This is the soft-constraint version of the Gaussian random matrix model and we will now see that it gives rise to the same leading one-point density and universal local correlations. We define the collective variable $t = \frac{1}{N} \sum_i \lambda_i^2$, in terms of which we have
\begin{equation}
    Z = \int dt e^{-\frac{1}{4a}(t-g)^2} \Omega(t)\,, \qquad \textrm{where} \quad \Omega(t) = \int\prod_i d\lambda_i\,\Delta(\lambda)^2\,
    \delta\!\Big(t-\tfrac1N\textstyle\sum_i\lambda_i^2\Big)\,,
\end{equation}
which we recognise as the density of states for the collective variable $t$. In order to solve this integral, let us rescale $\lambda_i \rightarrow \sqrt{t} \mu_i$, so that
\begin{equation}
\Omega(t) = t^{N^2/2-1} \Omega(1)\,.
\end{equation}
If we define the 't Hooft coupling $\tilde a = a N^2$, we find
\begin{equation}
    Z = \int dt e^{-N^2 F(t)}\,,\qquad \textrm{with} \qquad F(t) = \frac{(t-g)^2}{4 \tilde a} - \frac{1}{2}\log t\,,
\end{equation}
so that there is a saddle point at $t_* = \frac{1}{2} \left( g + \sqrt{g^2 + 4 \tilde a} \right)$, the positive root of $t_*^2 - g\,t_* - \tilde a = 0$. Since $t$ localizes at $t_*$ with
fluctuations of order $N^{-1}$, the fixed-trace ensemble at $t_*$ is equivalent, for the leading density and local spectral correlations, to a canonical Gaussian ensemble with coupling $t_*$. Accordingly, these leading spectral observables can be computed
from
\begin{equation}
    Z_{\rm GUE}(t_*) =
    \int \prod_i d\lambda_i\,\Delta^2(\lambda)
    e^{-\frac{N}{2t_*}\sum_i\lambda_i^2}\,.
\end{equation}
This has the  semicircular density of eigenvalues 
\begin{equation}
    \rho_{\rm Wigner}(\lambda) = \frac{2}{\pi R^2} \sqrt{R^2 - \lambda^2}
\end{equation}
with $R = 2\sqrt{t_*}$. In other words, the level density agrees with the standard GUE at effective coupling $t_* = \tfrac12\!\left(g+\sqrt{g^2+4\tilde a}\right)$, which reduces to the bare coupling $g$ as $\tilde a\to0$. Nevertheless, the softness of the constraint leaves a trace in the fluctuations of the collective variable, $t$,
\begin{equation}\label{eq:t_variance_soft_gaussian}
  N^{2}\,\mathrm{Var}(t)=\frac{2\,\tilde a\,t_*^{2}}{t_*^{2}+\tilde a}\,
\end{equation}
differing from the usual GUE case at the same effective coupling, which gives instead $N^{2}\,\mathrm{Var}_{\mathrm{GUE}}(t)=2\,t_*^{2}$. This is consistent, since the limit $\tilde{a}=0$ gives back the exact (i.e. microcanonical) ensemble where $t$ does not fluctuate.

Since the $t$-integral localizes at $t_*$, the fixed-trace ensemble is locally equivalent, at large-$N$, to the canonical GUE with coupling $t_*$. After unfolding by the semicircle density, these are governed at leading order by the usual GUE sine kernel. Thus the soft constraint modifies the fluctuations of the collective mode $t$, as seen in \eqref{eq:t_variance_soft_gaussian}, but without changing the leading local random-matrix statistics\footnote{The same equivalence gives the usual Airy physics at the soft edge. Indeed, the $\mathcal{O}(N^{-1})$ displacement of the edge induced by fluctuations of $t$ is parametrically smaller than the $\mathcal{O}(N^{-2/3})$ Airy scale.}. This separation is relevant for the Cardy RMT studied in the  remainder of this paper. There the regulated modular constraints control a set of coarse spectral observables, while the Vandermonde governs the spectral degrees of freedom left unresolved by those constraints and produces the local random-matrix statistics studied below.
\subsection{Cardy RMT}\label{sec:cardy_matrix}

Our goal is to construct a matrix model which takes into account the constraints of modular invariance, so that asymptotically its density of states behaves like the Cardy density. In this section we will  simplify our model and consider a theory of a single random matrix, $M$.  Using the constraint$^2$ construction, reviewed in \cref{sec:balian_review}, we will build a matrix potential accounting for the invariance under modular $\mathbb{S}$-transform for a purely holomorphic approximate CFT\footnote{We stress again that this should \textit{not} be viewed as a model for  a chiral CFT, because in a chiral CFT spin is quantized, while for us the weights $h$ will be able to take any real value. Our model is much closer to a purely scalar CFT.}. Note that we use the word approximate in the sense of \cite{belin2024approximatecftsrandomtensor}, which is reflected in the constraint$^2$ model as the softening factor $a$. This toy example, which we call the ``Cardy matrix model", or simply ``Cardy RMT", encoding only the spectral information contained in $\mathbb{S}$-modular invariance, can be viewed as a first step to understanding ensembles describing chaotic CFTs. Indeed, in order to describe a complete CFT we would need also to consider both left and right moving weights, implement $\mathbb{T}$-modular invariance (i.e. spin quantization), and incorporate OPE coefficients to impose the sphere four-point function and torus one-point function crossing constraints. We will comment on the definition of the complete ensemble, as well as how considerations made for the Cardy toy model apply to this case, in \cref{sec:discussion}.

\subsubsection{Regularizing the $\mathbb{S}$-modular invariance constraint}\label{sec:S_modinv_reg}
In this section, we state the set of constraints used to build the Cardy matrix model. We start by reviewing the (holomorphic) $S$-modular invariance constraint, and how to recast it in terms of Virasoro characters (following \cite{ DiFrancesco:1997nk, Collier:2019weq}). We then devote the majority of this section to define a regularization procedure so that we can use this set of conditions to build the `constraint$^2$' potential defining the Cardy RMT (\cref{sec:cardy_potential}). 

We start by considering the holomorphic sector of a 2D CFT, where $L_0$ is the generator of the Virasoro algebra associated to dilatations. The partition function of this theory is
\begin{equation}\label{eq:z_def_m}
    Z(\tau)=\mathrm{tr}_\mathcal{H}\left(q^{L_0-c/24}\right),\quad\mathrm{where}\quad q=e^{2\pi i \tau}.
\end{equation}
We can decompose the Hilbert space $\mathcal{H}$ into irreducible chiral modules as $\mathcal{H}=\bigoplus_h \mathcal{V}_h$, where $\mathcal{V}_h$ is the Virasoro module generated from the primary state $\ket{h}$ and $\bigoplus_h$ runs over all primaries with their appropriate multiplicities. Define, as usual, the character of $\mathcal{V}_h$ as $\chi_h(\tau)=\mathrm{tr}_{\mathcal{V}_h}\left(q^{L_0-c/24}\right)$, implies the well-known decomposition
\begin{equation}\label{eq:char_def_Z_diag}
 Z(\tau)=\sum_h \chi_h(\tau)\,.
\end{equation}
Expressed directly on the partition function, $\mathbb{S}$-modular invariance is the  condition
\begin{equation}\label{eq:s_mod_inv_def}
    Z(\tau)-Z(-1/\tau)=0\,,
\end{equation}
which we will momentarily proceed to massage and use to build a constraint$^2$ potential, treating it as a generalization of the normalization condition of \eqref{eq:GUE_constraint}. Before carrying out these steps, we comment on the random matrix interpretation we wish to instate. We want to build a theory of a single random matrix, which we identify as the RMT avatar of $(L_0-c/24)$ by the discussion above. Then, the condition \eqref{eq:s_mod_inv_def}, together with \eqref{eq:z_def_m}, by the procedure discussed in \cref{sec:balian_review} will return a potential $V[M]$. As we have seen, the characters repackage states by irreducible representations, so if we use the constraints \eqref{eq:s_mod_inv_def} with \eqref{eq:char_def_Z_diag} instead, we will obtain an equivalent potential $V(h_1,\dots h_N)$, where $h_1,\dots h_N$ are the weights of the $N$ primaries in the theory. In the RMT interpretation, this is the matrix potential after the diagonalization of $M$, and the energies of the $N$ eigenvalues are identified with the conformal weights of the $N$ primaries. We assume $M$ in the unitary symmetry class, so that the measure change $[dM]\to\prod dh$ in the matrix integral comes with the Vandermonde determinant $\prod_{i<j}(h_i-h_j)^2$.

With the bare measure in place, let us go back to the discussion of the $\mathbb{S}$-modular constraint. We start by considering the Virasoro characters written in terms of the Liouville momenta $P$,
\begin{equation}\label{eq:P_to_h}
    P=\sqrt{h-\frac{c-1}{24}} \quad\mathrm{with}\quad h>0.
\end{equation}
The character for a non-vacuum, non-degenerate Virasoro primary of momentum $P$ is 
\begin{equation}\label{eq:character_P_eta_expr}
   \chi_P(\tau)=\frac{q^{P^2}}{\eta(\tau)},\qquad \eta(\tau)=q^{1/24}\prod_{n=1}^{\infty}(1-q^n),  
\end{equation}
where we recall that $q=e^{2\pi i \tau}$ and $\eta(\tau)$ is the Dedekind $\eta$-function.
For the identity, which is a degenerate expression, we instead have the expression
\begin{equation}\label{eq:chi_identity}
    \chi_{\mathds{1}}(\tau)=\frac{q^{-Q^2/4}(1-q)}{\eta(\tau)}=\chi_{h=0}(\tau)-\chi_{h=1}(\tau)\quad\mathrm{where}\quad Q^2=\frac{c-1}{6}\,.
\end{equation}
 So, the partition function \eqref{eq:char_def_Z_diag} is written in terms of the Liouville momenta as
\begin{equation}\label{eq:Z_characters}
    Z(\tau)=\sum_i \chi_{P_i}(\tau)=\int dP\,\Tilde{\rho}(P)\chi_P(\tau),
\end{equation}
where, in the first equality, the sum runs over all primaries labeled, with multiplicity, by $i$ and with momentum $P_i$, and in the second we introduced their (un-normalized) density $\Tilde{\rho}(P)=\sum_i \delta(P-P_i)$. 
 Our notation is that, whenever we write the integral over Liouville momenta for a partition function like \eqref{eq:Z_characters}, its domain is the full support of $\Tilde{\rho}$. This includes also the light spectrum, i.e. imaginary values of $P$.
 
Now, we introduce the $\mathbb{S}$-kernel of \cite{teschner2003liouvilletheoryquantumgeometry}, which relates modular-transformed characters as
\begin{equation}\label{eq:chi_mod_transf}
    \chi_P(-1/\tau)=\int_{\mathbb{R}^+} dP'\,\mathbb{S}_{P'P}(\mathds{1})\chi_{P'}(\tau)
\end{equation}
so that the modular-transformed partition function can be written as 
\begin{equation}
 Z(-1/\tau)=\int dP\int_{\mathbb{R}^+}dP'\,\mathbb{S}_{P'P}(\mathds{1})\Tilde{\rho}(P)\chi_{P'}(\tau)\,.
\end{equation}
Notice that $h(P)=h(-P)$, so that characters of opposite Liouville momenta refer to the same primary operator. In the above expressions we already factored this out, and importantly restricted the $\mathbb{S}$-transform integration to its support: positive real momenta.\footnote{Note that an alternative notation is frequently used in the literature, for example in \cite{Collier:2019weq}, which integrates over the full real line with the measure \(dP/2\) accounting for the reflection symmetry \(P\sim -P\).}

We have the following explicit expression for the $\mathbb{S}$-kernel as a function of momenta \cite{Collier:2019weq}:
\begin{equation}\label{eq:S_kernel_expression}
    \begin{gathered}
\mathbb{S}_{P P^{\prime}}(\mathds{1})=2 \sqrt{2} \cos \left(4 \pi P P^{\prime}\right) \\
\mathbb{S}_{P \mathds{1}}(\mathds{1})=\mathbb{S}_{P, \frac{i}{2}\left(b+\frac{1}{b}\right)}(\mathds{1})-\mathbb{S}_{P,-\frac{i}{2}\left(b-\frac{1}{b}\right)}(\mathds{1})=4 \sqrt{2} \sinh (2 \pi b P) \sinh \left(2 \pi \frac{P}{b}\right) ,
\end{gathered}
\end{equation}
where $c=1+6(b+b^{-1})^2=1+6Q^2$, and the separate expression for $\mathbb{S}_{P \mathds{1}}(\mathds{1})$ originates again from the degeneracy of the vacuum representation, as in \eqref{eq:chi_identity}. Notice  the following limiting behavior, ensuring that at high energies non-vacuum contributions are exponentially suppressed with respect to the identity:
\begin{equation}\label{eq:S_kernel_highenergy}
\frac{\mathbb{S}_{PP'}}{\mathbb{S}_{P\mathds{1}}}\underset{P\to\infty}{\to}\begin{dcases*}
e^{-4\pi \alpha' P} & $\alpha'=Q/2+i P'\in (0,Q/2)$\\
2\cos{4\pi PP'}e^{-2\pi QP} & if $P'\in\mathbb{R}$
\end{dcases*}
\end{equation}
This is the fundamental reason behind the emergence of the Cardy density in the high-energy region of the spectrum \cite{Collier:2019weq}.
Using the $\mathbb{S}$-kernel, we can write the modular $\mathbb{S}$-transform invariance as:
\begin{equation}\label{eq:S_invariance}
\begin{aligned}
  {\cal C}_{\mathbb{S}}[\tau] =  \int dP\,\Tilde{\rho}(P)\chi_P(\tau)-\int dP\ \int_{\mathbb{R}^+} dP'\,\mathbb{S}_{P'P}(\mathds{1})\Tilde{\rho}(P)\chi_{P'}(\tau)=0\,,\qquad \forall \tau\,.
\end{aligned}
\end{equation}
So far all manipulations were completely standard, see e.g. \cite{DiFrancesco:1997nk} or the more recent incarnations \cite{Collier:2019weq,Belin:2023efa}.
To move towards a constraint$^2$ implementation of these conditions, we want to project onto a given Virasoro character, so that we can eventually impose modular invariance character-by-character. For this purpose, we recall the orthogonality relations between characters \cite{Verlinde:1989ua, Collier_2023}:
\begin{equation}\label{eq:char_proj_exact}
    \int \frac{d^2\tau}{\sqrt{\Im(\tau)}} f(\tau) \chi_P(\tau) \chi_{P'}(\tau)^* = \delta(P-P')
\end{equation}
for a suitably defined function $f(\tau)$. This orthogonality relation is the only property of the function $f(\tau)$ that we will use in this section, the reader may refer to \cref{app:f_details} for a more detailed characterization. Let us consider the projection of the $\mathbb{S}$-modular transform on the character labeled by real Liouville momentum $P$: 
\begin{equation}\label{eq:mod_inv_projected}
\begin{aligned}
   0 &=  \int\frac{d^2\tau}{\sqrt{\Im(\tau)}}{\cal C}_{\mathbb{S}}[\tau]f(\tau)\chi_{P}(\tau)^*\\
   &=\int dP'\Tilde{\rho}(P')\delta(P'-P)-\int dP'\;\mathbb{S}_{PP'}(\mathds{1})\Tilde{\rho}(P')\\
   &\overset{?}{=}\Tilde{\rho}(P)-\sum_i\mathbb{S}_{PP_i}(\mathds{1})\,,
\end{aligned}
\end{equation}
where, in the last equality, we used that $\Tilde{\rho}(P)=\sum_i \delta(P-P_i)$. We note that \eqref{eq:mod_inv_projected} holds only for the projection onto heavy blocks, i.e. for real $P$. This is sufficient to describe the sourced model, and we will address the generalization to light states when we discuss the dynamical model.

Let us now discuss the meaning of \eqref{eq:mod_inv_projected} further. The density $\Tilde{\rho}$ is defined as an atomic sum of delta functions, while the sum of functions $\mathbb{S}_{PP'}(\mathds{1})$ \eqref{eq:S_kernel_expression} is analytical for a finite number $N$ of primaries, so the condition above is ill-defined, equating a function with a distribution. Similarly, if the number of primaries is infinite, because of the erratic oscillations appearing in the sum of \eqref{eq:S_kernel_expression}, we need to interpret \eqref{eq:mod_inv_projected} as a statement of zero distance between distributions. However, equality between distributions is only understood upon integrating against test functions, so, at the present stage, using the condition \eqref{eq:mod_inv_projected} to define a `constraint$^2$' potential remains an ill defined procedure. We can resolve this issue by fixing a set of test functions and then build the matrix potential from the constraints obtained by integrating \eqref{eq:mod_inv_projected} against all chosen test functions. Before doing so, let us pause and reflect on the implications of a given specific choice of these test functions.

\subsubsection*{Remark: test functions and metric on constraint space}
Let us pause briefly to record some observations on the general form of the modular constraint \eqref{eq:mod_inv_projected}. Fixing test functions $\{\phi_I\}$ turns this constraint, written succinctly $\mathcal{C}\equiv(\mathds{1}-\mathbb{S})\Tilde{\rho}$, into the numbers $C_I=\langle\phi_I,\mathcal{C}\rangle\equiv\int dP\,\phi_I(P)\,\mathcal{C}(P)$. The constraint$^2$ potential is their sum of squares, that is to say the squared norm of $\mathcal{C}$ in the metric, $K=\sum_I\phi_I\otimes\phi_I$:
\begin{equation}\label{eq:constraint_metric}
\begin{aligned}
 \tfrac{1}{4a}\sum_I C_I^2
  &=\tfrac{1}{4a}\,\|\mathcal{C}\|_K^2
  =\tfrac{1}{4a}\,\langle\mathcal{C},\,K\,\mathcal{C}\rangle\\
  &=\tfrac{1}{4a}\,\big\langle\Tilde{\rho},\,W_K\,\Tilde{\rho}\big\rangle\,,
\end{aligned}
\end{equation}
where $W_K=(\mathds{1}-\mathbb{S})^{\dagger}\,K\,(\mathds{1}-\mathbb{S})$.
The second line uses that $\mathcal{C}=(\mathds{1}-\mathbb{S})\Tilde{\rho}$ is linear in the density, so that the constraint$^2$ becomes a two-body confining interaction $W_K$ among the eigenvalues. The regularization enters only through the metric $K$ on the space of constraint distributions. Different families probe the constraint at different, possibly sliding, scales. We can think of at least three natural choices:

\begin{itemize}
\item[(i) ]characteristic functions of disjoint $\epsilon$-intervals. This is the simplest choice, but gives a single-scale metric blind to structure below $\epsilon$
\item[(ii)] a wavelet basis, giving rise to a covariant multi-scale metric resolving all scales
\item[(iii)] the automorphic Maass--Eisenstein basis on $\mathrm{PSL}(2,\mathbb{Z})\backslash\mathbb{H}$, covariant under the full modular group. 
\end{itemize}
In practice, in this paper we adopt the single-scale binning (i), which lends itself to direct treatment, both analytically and numerically. It is interesting to note that (iii) makes an interesting contact with the approach of \cite{DiUbaldo:2023qli, Boruch:2025ilr}, although the constraint relaxation parameter $a$ does not appear to have a place in their formalism.

We now proceed with option (i). Notice that the distributions we are equating in \eqref{eq:mod_inv_projected} define a (possibly signed) measure on $\mathbb{R}$, and in this case characteristic functions of intervals are admissible test functions, as they generate the Borel $\sigma$-algebra. If we chose the set of such test functions associated with disjoint intervals with a certain fixed width $\epsilon$, we would obtain only a pseudo-metric on the space of distributions. Indeed, two distributions giving the same mass to each interval but differing at lower scales will be at zero distance from each other under this set of test functions. From a more physically motivated perspective, we can interpret this finite $\epsilon$ prescription as a coarse graining, which becomes finer as we send $\epsilon\to0$.

We will proceed momentarily to describe a regularization procedure that similarly resolves the issues of the condition \eqref{eq:mod_inv_projected} and leaves us with a set of well-defined constraints for the Cardy matrix model. Then we will discuss the integration procedure, as explained above, and show that we obtain, modulo overall multiplicative factors, the same set of constraints. Because of this, in the rest of the paper, we will use both perspectives somewhat interchangeably. However, the intended takeaway of this comment is that we do not have to send $\epsilon\to0$; rather, we can also make sense of the finite $\epsilon$ case as a coarse-graining procedure.
\subsubsection*{Regularization of the (constraint)$^2$ potential}
Let us now consider a regularized version of the orthogonality relation \eqref{eq:char_proj_exact}, with a kernel $f_\epsilon(\tau)$ that we think of as a convolution of $f(\tau)$ with a smearing function of characteristic width $\epsilon$:
\begin{equation}\label{eq:char_project_regularized}
    \int \frac{d^2\tau}{\sqrt{\Im(\tau)}} f_\epsilon(\tau) \chi_P(\tau) \chi_{P'}(\tau)^* = \delta_\epsilon(P-P') \,.
\end{equation}
Here $\delta_\epsilon$ is a nascent delta function, so that when we send $\epsilon\to0$ we have that $\delta_\epsilon$ distributionally converges to the standard Dirac delta. For example, we can imagine this to be a step function equal to $1/\epsilon$ in a certain window of width $\epsilon$ around $P=P'$ and 0 outside\footnote{considering these characteristic functions of intervals is going to be sufficient for our purposes, particularly in the limits considered in \cref{sec:saddle_point}. We note that in applications where smoothness is required one can equivalently choose gaussian functions with height controlled by $\frac{1}{\epsilon}$ and width $\epsilon$.}.
When integrated against a test function with argument $P$, this regularized $\delta_\epsilon$ effectively acts as a projector of $P'$ onto a window of width $\epsilon$ centered around $P$. With this regularization, \eqref{eq:mod_inv_projected} becomes
\begin{equation}\label{eq:mod_inv_reg_noconst}
    \begin{aligned}
        \int dP'\,\Tilde{\rho}(P')\delta_\epsilon(P'-P)-&\int dP'\;\int_{\mathbb{R}^+}dP''\,\mathbb{S}_{P''P'}(\mathds{1})\Tilde{\rho}(P')\delta_\epsilon(P-P'')=\\&=\frac{1}{\epsilon}\int_{P,\epsilon} dx\Tilde{\rho}(x)-\frac{1}{\epsilon}\int dP'\,\int_{P,\epsilon}dx\mathbb{S}_{xP'}(\mathds{1})\Tilde{\rho}(P')\,,
    \end{aligned}
\end{equation}
where by $\int_{P,\epsilon}$ we denoted the integral on the interval of width $\epsilon$ centered around $P$. If we use the explicit expression of the $\mathbb{S}$-kernel \eqref{eq:S_kernel_expression}, we obtain the following result from the integral
\begin{equation}\label{eq:mod_invariance_integrated}
    \frac{1}{\epsilon }\biggr(N_{P,\epsilon}- \sum_{i\neq\mathds{1}} \frac{\sqrt{2}\cos{4\pi PP_i}\sin{2\pi \epsilon P_i}}{\pi P_i}-N_{\rm vac}\mathcal{S}_\mathds{1}(P,\epsilon)\biggr)=0\,.
\end{equation}
Here $N_{P,\epsilon}$ denotes the number of primary states in the window of width $\epsilon$ around the momentum $P$, $N_{\rm vac}$  is the occupation number of the vacuum (which we will take to be $0$ or $1$ since we will work with the sourced model), and we have defined \\
\begin{equation}\label{eq:Sp1_integrated}
    \begin{aligned}
     \mathcal{S}_\mathds{1}(P,\epsilon)&\equiv\int_{P,\epsilon}dx \mathbb{S}_{x\mathds{1}}(\mathds{1} )\\
     &=\frac{2\sqrt{2}}{\pi(b+b^{-1})} \cosh\left(2\pi(b+b^{-1}) P\right)\sinh \left(\pi  \left(b+b^{-1}\right)\epsilon \right)
     \\&-\frac{2\sqrt{2}}{\pi(b-b^{-1})} \cosh\left(2\pi(b-b^{-1}) P\right)\sinh \left(\pi  \left(b-b^{-1}\right)\epsilon \right).
\end{aligned}
\end{equation}
We notice that the constraint condition \eqref{eq:mod_invariance_integrated}, originating from smearing the delta function in \eqref{eq:char_project_regularized}, can also be equivalently obtained by another regularization procedure. Indeed, consider the coarse graining of \eqref{eq:mod_inv_projected}, implemented by integrating $P$ in an interval of width $\epsilon$ and centered at $P$, that resolves the distribution-theoretic issue by recasting it into a distance of real numbers condition. As mentioned in passing above, this procedure, modulo a $1/\epsilon$ prefactor, is equivalent to integrating \eqref{eq:mod_inv_projected} against the set of characteristic functions of disjoint intervals of width $\epsilon$. Indeed, also in this case we obtain the set of constraints \eqref{eq:mod_invariance_integrated}. As we will comment in \cref{sec:saddle_point}, the different prefactor between these regularization procedures is reflected in a rescaling of the regimes where we state our results. The more important point is that, if we adopt this coarse-graining perspective $\epsilon$ is not just a regulator to be removed, but can be kept finite.\\

Before concluding this subsection, we discuss how the projected constraints \eqref{eq:mod_inv_projected} are modified when taking into account the degenerate vacuum module, as \eqref{eq:mod_inv_reg_noconst} is rigorously derived for non-degenerate characters. Indeed, if we consider for example the projected constraint on the $h=1$ character, because of the expression of the vacuum character \eqref{eq:chi_identity}, we would obtain:
\begin{equation}\label{eq:constr_block_h=1}
    \frac{1}{\epsilon}\biggr(N_{P_{h=1},\epsilon}-N_{\rm vac}-\sum_{i\neq\mathds{1}} \frac{\sqrt{2}\cos{4\pi P_{h=1}P_i}\sin{2\pi \epsilon P_i}}{\pi P_i}-N_{\rm vac}\mathcal{S}_\mathds{1}(P_{h=1},\epsilon)\biggr)=0.
\end{equation}
So the intuition is that, in the $P_{h=1}=\sqrt{1-(c-1)/24}$ block, the $h=1$ component present in the vacuum character \eqref{eq:chi_identity} manifests as an addition of $N_{\rm vac}$ eigenvalues to the occupation number. In this paper,  we will be mostly interested in the case where there is only a single state in the vacuum $N_{\rm vac}=1$ and this is the only light state. Moreover, considering the aforementioned correction slightly modifies the low-energy behavior, while most of the features we discuss are determined by the high-energy region of the spectrum where the majority of states live. More rigorously, we can state that \eqref{eq:mod_invariance_integrated} holds for every occupied block if we consider the large $c\geq 25$  case and there are no light states except the vacuum, and so in particular no states with dimension $h=1$. For these reasons, in the next sections, we will directly use the set of constraints \eqref{eq:mod_invariance_integrated} and neglect the possible modification of \eqref{eq:constr_block_h=1}.\\

\subsubsection{Comments on the scale of coarse-graining}
\label{sec:bins_width}
We proceed to discuss how to choose $\epsilon$ in our prescription, in order to simplify the expression of the modular constraints \eqref{eq:mod_invariance_integrated}. Mirroring previous discussions of coarse-graining in CFTs \cite{ Das_2017,Collier:2019weq, Mukhametzhanov:2019pzy, Ganguly_2020, Mukhametzhanov_2020}, the idea is to choose the width of the microcanonical bins so that they are small enough for the $\mathbb{S}$-kernel to be approximately constant in each interval, but large enough to contain sufficiently many states. Here we discuss the upper bound on $\epsilon$ given by the former condition, considering both the cases when the vacuum is occupied and when it is not. A more thorough analysis would derive both lower and upper bounds on the coarse-graining window, for example using Beurling-Selberg estimates, as was done in \cite{Mukhametzhanov_2020} in order to obtain universal bounds on corrections to the Cardy formula.

Let us introduce a UV cutoff at momentum $\mathcal{E}$, whose role we will return to later. We start by discussing the case where the vacuum is not occupied $N_{\rm vac}=0$, and the theory has $N$ primaries above the threshold at $\frac{c-1}{24}$. Then, in order to ensure that the $\mathbb{S}$-kernel is approximately constant in each bin, let us consider the second term of the integrated constraint, \eqref{eq:mod_invariance_integrated}, compared to the $\mathbb{S}-$kernel evaluated at the bin center multiplied by the bin width. The strictest condition comes from the UV end of the sum, $P_i = {\cal E}$, so that we need to impose
\begin{equation}\label{eq:eps_consistency_novac}
    \frac{\sin{\pi\epsilon\mathcal{E}}}{\pi\mathcal{E}}\approx\epsilon\qquad\implies\qquad \epsilon\mathcal{E}\ll 1\,.
\end{equation}
Using \eqref{eq:P_to_h}, we can understand this relation in terms of the energy variables instead of momenta. We find that the width of the bin in units of energy $h$, for a bin centered at momentum $P$, is to leading order in $\epsilon$
\begin{equation}\label{eq:delta_eps}
    \delta h(P,\epsilon) = 2P\epsilon\,.
\end{equation}
Notice that if we fix $\epsilon$, then $\delta h (P,\epsilon)$ increases for increasing momenta $P$, and vice versa. We define the following parameter $\delta\equiv 2 \mathcal{E}\epsilon$, which will be useful in the double-scaling procedures to be discussed in \cref{sec:vac_occ_gap}. Then the condition \eqref{eq:eps_consistency_novac}, in terms of $\delta$, holds if $\delta\ll 1$.

On the other hand, if the vacuum is occupied, we expect that the requirement of the exponentially large $\mathbb{S}_{P\mathds{1}}(\mathds{1})$ being approximately constant in the $P$-bin will require a parametrically smaller $\epsilon$ with respect to \eqref{eq:eps_consistency_novac}. Indeed, we find the series expansion of \eqref{eq:Sp1_integrated} around $\epsilon= 0$:
\begin{equation}\label{eq:Sp1_integrated_expansion}
\begin{aligned}
    \mathcal{S}_\mathds{1}(P,\epsilon)=& \;\epsilon\mathbb{S}_{P\mathds{1}}(\mathds{1})+\frac{\sqrt{2}\epsilon^3\pi ^2}{3 b^2}\left(\left(b^2+1\right)^2 \cosh \left(\frac{2 \pi  \left(b^2+1\right) P}{b}\right)\right)+\\&-\frac{\sqrt{2}\epsilon^3\pi ^2}{3 b^2} \left(\left(b^2-1\right)^2 \cosh \left(\frac{2 \pi  \left(b^2-1\right) P}{b}\right)\right)+\mathcal{O}(\epsilon^5)
\end{aligned}
\end{equation}
The requirement of $\mathbb{S}_{P\mathds{1}}(\mathds{1})$ being approximately constant in the $P$-bin follows from $\mathcal{S}_\mathds{1}(P,\epsilon)/\epsilon\approx\mathds{S}_{P\mathds{1}}(\mathds{1})$. In particular, this condition is equivalent to the contribution from the second term in \eqref{eq:Sp1_integrated_expansion} being suppressed, which in the high momentum region of the spectrum translates to
\begin{equation}\label{eq:eps_consistency_vac}
    \frac{\sqrt{2}\pi ^2 \left(b^2+1\right)^2}{6 b^2} \epsilon^3 e^{\frac{2 \pi  \left(b^2+1\right) \mathcal{E}}{b}}=\frac{\sqrt{2}\pi ^2 \left(c-1\right)}{36} \epsilon^3 e^{2\pi \sqrt{\frac{c-1}{6}} \mathcal{E}}\ll1\,.
\end{equation}
Notice that the equation above is obtained by imposing that the bin occupation number changes by much less than a unit, if we consider the correction to the constant $\mathbb{S}$-kernel approximation. The most stringent constraint has again come from the UV part of the spectrum. 

It is also possible to impose a more lenient condition,  by demanding that the constant term in \eqref{eq:Sp1_integrated_expansion} is much greater than the first order correction:
\begin{equation}\label{eq:eps_consistency_subexp_acc}
    \epsilon\ll\frac{\sqrt{6}}{\pi Q} \,,
\end{equation}
which will be sufficient for us. 

We note that it could also be useful to consider bins with non-constant widths $\epsilon(P)$ dependent on the momenta of the bin center, for example choosing $\epsilon(P)\propto1/P$ results in constant $\delta$ energy bins, but we leave a detailed discussion of this for future work.

To sum up, in this section we have reformulated the idea that the modular constraint holds in an integrated sense \cite{Collier:2019weq, belin2024approximatecftsrandomtensor}, by introducing a suitable regulating procedure, parametrized by the bin width $\epsilon$. We have discussed different upper bounds on $\epsilon$, their regimes of validity and their consequences on the $\mathbb{S}$-kernel approximation. Before moving on, we wish to clearly state which of these assumptions we will adopt in the rest of the paper, unless otherwise stated.
\begin{itemize}
    \item In the numerics of \cref{sec:numerics}, we study the model in the small $\epsilon$ case and stop at $\mathcal{E}\sim\mathcal{O}(1)$, because  cranking up the value of the UV cutoff is, computationally,  exponentially complex. In this case, we are in the limit \eqref{eq:eps_consistency_novac}, where, due to the $\mathbb{S}$-kernel being approximately constant in each bin, we can rewrite the constraints \eqref{eq:mod_invariance_integrated} in the following simplified form:
\begin{equation}\label{eq:mod_constrant_blocks}
    \frac{1}{\epsilon }\biggr(N_{P,\epsilon}-\epsilon \sum_i \mathbb{S}_{PP_i}(\mathds{1})\biggr)=0.
\end{equation}
This is now the set of modular constraints, regulated into functional form, that can be implemented as a constraint$^2$ matrix model and studied numerically via Monte Carlo.
    \item In the rest of the paper, when we assume that the vacuum is occupied and we restrict to the high-energy region of the spectrum, we can adopt the less stringent condition \eqref{eq:eps_consistency_subexp_acc}, so that the vacuum $\mathbb{S}$-kernel is approximately constant, but only up to sub-leading exponential correction. However, to recover leading exponential Cardy growth at high energies, we must also impose that the bins are wide enough to control the oscillatory sum over heavy $\mathbb{S}$-kernels. More precisely, we are examining the case \eqref{eq:delta_eps} of energy windows growing $\propto\sqrt{\Delta}$ up to $\delta=2G$. Then, from \cite{Mukhametzhanov:2019pzy}, corrections to Cardy are at most $\mathcal{O}(1)$ if we choose $G>\sqrt{3}/\pi$, and thus subleading in the high-energy region of the spectrum where $P/\mathcal E\to1$. 
\end{itemize}
We will discuss further how $\epsilon$ needs to be scaled with the other parameters of the theory in \cref{sec:vac_occ_gap}, and we will explicitly check the consistency of the proposed limiting procedure with the considerations made here to obtain \eqref{eq:mod_constrant_blocks}.\\

\subsubsection{The Cardy RMT potential}\label{sec:cardy_potential}

We want to build a matrix model that approximately imposes the coarse-grained $\mathbb{S}$-transform invariance \eqref{eq:mod_constrant_blocks} for all bins. To this end, we can use the  `constraint$^2$' construction of \cref{sec:balian_review}, applied to the set of constraints \eqref{eq:mod_constrant_blocks}, for all the momenta $P$ associated to the center of the bins. We obtain the following probability distribution for the Cardy matrix model:

\begin{equation}\label{eq:part_funct_cardy}
\begin{aligned}
    {\cal Z}=\int\prod_j dh_j\,\Delta(\mathbf{h})^2 \exp\left[{-\frac{1}{4a\epsilon^2}\sum_P \bigr(N_{P,\epsilon}-\epsilon\sum_i\mathbb{S}_{PP_i}(\mathds{1})\bigr)^2}\right],
\end{aligned}
\end{equation}
where we recall that $\sum_P$ runs over the center momenta of all principal-series bins we used to coarse-grain our spectrum, and the eigenvalues $h_i$, to be identified with the conformal dimensions of the primaries, are related to the $P_i$ appearing in the potential via \eqref{eq:P_to_h}.\footnote{Implicitly, in everything that follows, we should always think of $P$ as a function of $h$. The Vandermonde is however written directly in terms of scaling dimensions, which are the physical eigenvalues of the Hamiltonian we are considering.} 

Let us also note that the sum over $P$ implicitly comes with a UV cutoff. The role of the UV cutoff is also to regularize the model: modular invariance is imposed at all temperatures, which in particular means it is imposed also at very high temperatures. To satisfy it at high temperatures, there must be many states at very high energies. Combined with the Vandermonde that pushes eigenvalues away from one another, this will cause a runaway effect for the eigenvalues which needs to be regulated.  

We also emphasize that we are considering only real momenta bins because the dynamical states are only those above the black-hole threshold. This does not remove the information
contained in the light spectrum (in particular the identity), but light states enter the potential only as specific fixed terms in the potential which can be viewed as sources. In the dynamical version of the model, the integral over $P$ would be replaced by an integral over $h$ which runs over the entire positive real line. Unitarity enforces the lower bound on this integral $h\geq0$. In the sourced model, unitarity is obvious since all states with real $P$ have a unitary weight.

 We assumed for simplicity the unitary universality class, but we envision no further difficulties in generalizing the discussion here to the orthogonal or other classes.

 Consistently imposing a UV cutoff $\mathcal{E}$ on \eqref{eq:part_funct_cardy}, requires that we specify what enters the modular constraint from momenta regions above $\mathcal{E}$. Because the $\mathbb{S}$-kernel
is nonlocal in momentum, allowing some state population above $\mathcal E$ still contributes to the constraints below it. So populating the region beyond $\mathcal{E}$ is equivalent to specifying a boundary condition at the UV cutoff. Consider a light spectrum with occupations $N_\ell$, imaginary momenta
$P_\ell$, and modular image $\rho_L(P)
    \equiv \sum_{\ell\in\mathrm{light}}
        N_\ell\,\mathbb{S}_{P P_\ell}(\mathbf{1})$ (in particular $\rho_L(P)=N_{\mathrm{vac}}\rho_{Cardy}(P)$ when the vacuum is the only occupied light-state). We will consider two natural boundary conditions at the UV cutoff:
\begin{itemize}
    \item Prescribe that the spectrum stops at the cutoff $\mathcal{E}$. For this reason we will refer to this UV boundary condition as `\emph{clipped}'.
    \item Alternatively, we can impose that an occupied light state is inserted together
with its full modular tail, so that above $\mathcal E$ the density is
frozen to the light-induced profile, obtained from modular transforming $\rho_L$. Then the UV cutoff separates a
fluctuating window $[0,\mathcal{E}]$, described by \eqref{eq:part_funct_cardy}, from the fixed modular background that the light
spectrum itself generates.  We refer to this prescription as `\emph{completed}' boundary conditions. 
\end{itemize}
We will study numerically both of these boundary conditions in \cref{sec:numerics}, and comment on analytical properties of the completed prescription in \cref{sec:vac_occ_gap}. Before concluding the section, we write the potentials associated to the two cutoff conditions considered, and recognize that their difference boils down to the presence of an edge term.

\paragraph{Potential for clipped b.c.}
In this case, we can just consider \eqref{eq:part_funct_cardy} without adding any contributions to the Cardy potential due to states beyond $\mathcal{E}$:
\begin{equation}\label{eq:cardy_potential}
    V(h_1,\dots, h_N)\equiv\frac{1}{\epsilon^2}\sum_P \bigr(N_{P,\epsilon}-\epsilon\sum_i\mathbb{S}_{P\,P_i(h_i)}(\mathds{1})\bigr)^2,
\end{equation}
which depends on the eigenvalues $\mathbf{h}=(h_1,\dots,h_N)$, as well as on the centers and width $\epsilon$ of the chosen binning prescription.
We want to write our potential in \cref{eq:cardy_potential} explicitly as a sum over eigenvalues. We consider the characteristic functions for the interval associated to the $P$-bin:
\begin{equation}
    \chi_{P,\epsilon}(x)\equiv\Theta(x-(P-\epsilon/2))-\Theta(x-(P+\epsilon/2)).
\end{equation}
Using these bins' characteristic functions, we can rewrite \eqref{eq:cardy_potential} as
\begin{equation}
\begin{aligned}
    V(\mathbf{h})&=\frac{1}{\epsilon^2}\sum_P\sum_{i,j}\biggr(\chi_{P,\epsilon}(P_i)-\epsilon\mathbb{S}_{P P_i}(\mathds{1}) \biggr)\biggr(\chi_{P,\epsilon}(P_j)-\epsilon\mathbb{S}_{ PP_j}(\mathds{1}) \biggr)\\
    &=V_{\rm vac}+\frac{2}{\epsilon^2}\sum_{P}\sum_{i,j\neq\mathds{1}}\biggr(\chi_{P,\epsilon}(P_i)\chi_{P,\epsilon}(P_j)-\epsilon\chi_{P,\epsilon}(P_i)\mathbb{S}_{PP_j}(\mathds{1})\biggr)\,.
\end{aligned}
\end{equation}
In the final expression, the first term is the contribution from the identity module, to which we return shortly. The second term is the sum over above-threshold states on the principal series, which have real $P$. Summing only over real momenta $P\in \mathbb{R}_{+}$, we have the regulated $\mathbb{S}-$kernel orthogonality
\begin{equation}
\begin{aligned}
   \epsilon \sum_{P\in \mathbb{R}}\mathbb{S}_{PP_i}(\mathds{1}) \mathbb{S}_{PP_j}(\mathds{1}) &= \delta_\epsilon (P_i - P_j)\,,
   \end{aligned}
\end{equation}
where, when $\epsilon\to0$, we can equivalently use the following $\epsilon-$regularized delta function
\begin{equation}
    \delta_\epsilon (P_i - P_j) = \frac{1}{\epsilon}\sum_{P\in\mathbb{R}}\chi_{P,\epsilon}(P_i)\,\chi_{P,\epsilon}(P_j)\,,
\end{equation}
which again holds when the sum extends over real momenta along the principal series. The remaining term in the potential comes from the identity, and reads
\begin{equation}
    V_{\rm vac}=N_{\rm vac}\sum_{P\in\mathbb{R}}\mathds S_{P\mathds1}(\mathds1)\left[N_{\rm vac}\mathds S_{P\mathds1}(\mathds1)-\frac{2}{\epsilon}\left(N_{P,\epsilon}-\epsilon\sum_{i\neq\mathds1}\mathds S_{PP_i}(\mathds1)\right)\right]\,.
\end{equation}
Notice that, when summed over the eigenvalues, the two theta-functions appearing in $\chi_{P,\epsilon}$ precisely give $N_{P,\epsilon}$, the occupation number of the bin centered at $P$ of width $\epsilon$. As a consequence, we can write the Cardy potential as a double sum over eigenvalues,
\begin{equation}
V(\mathbf{h})=V_{\rm vac} + \sum_{i,j}V_{(dt)}(h_i,h_j)\,,
\end{equation}
where we defined
\begin{equation}\label{eq:cardy_potential_doubletrace}
\begin{aligned}
     V_{(dt)}(h_i,h_j)=\frac{2}{\epsilon^2}\sum_{P}\biggr(\chi_{P,\epsilon}(P_i)\chi_{P,\epsilon}(P_j)-\epsilon\chi_{P,\epsilon}(P_i)\mathbb{S}_{PP_j}(\mathds{1})\biggr),
\end{aligned}
\end{equation}
and $P_{i,j}$ are again given as functions of the eigenvalues $h_{i,j}$ from \eqref{eq:P_to_h}. In particular, we notice that
the `constraint$^2$' potential cannot be written down as a single sum over eigenvalues. Indeed, it is expressed as a double sum, morally originating from a double-trace matrix potential. The double-trace nature of the potential \eqref{eq:cardy_potential} is one of the aspects contributing to the difficulty of directly solving the Cardy matrix model with standard techniques. 

\paragraph{Potential for completed b.c.} For completed boundary conditions, we must additionally include the contribution
of the fixed density above $\mathcal{E}$. We define the finite-edge term:
\begin{equation}
    \mathcal{B}_{L,\mathcal{E}}(P)
    \equiv\int_0^{\mathcal{E}}dP'\,
        \mathbb{S}_{P P'}(\mathds{1})\rho_L(P').
    \label{eq:cardy-boundary-term}
\end{equation}
As an example, in the case where the vacuum is the only light state $\rho_L(P)=\rho_0(P)\equiv \mathbb{S}_{P\mathds 1}(\mathds 1)$, one
finds at leading exponential order:
\begin{equation}\label{eq:vacuum-edge-explicit}
 \mathcal B_{0,\mathcal E}(P)
 \simeq
 \frac{2}{\pi}\,
 e^{2\pi Q\mathcal E}\,
 \frac{Q\cos(4\pi P\mathcal E)+2P\sin(4\pi P\mathcal E)}
      {Q^{2}+4P^{2}}\,,
\end{equation}
an oscillation of period $1/2\mathcal E$ in momentum whose amplitude is
of the same exponential order as the total number of states sourced by
the vacuum.

Applying a second full $\mathbb{S}$-transform to $\rho_L$ returns the light spectrum, whose principal-series projection vanishes. Consequently, the prescribed contribution of the tail is\footnote{The light-induced profile grows exponentially, so its transform over the infinite principal series is not an ordinary oscillatory integral. In order to make sense of the modular tail integral, and its connection to \eqref{eq:cardy-boundary-term}, we need the complex-deltas prescription discussed in \cref{app:f_details}.}
\begin{equation}\label{eq:edge-identity}
 \int_0^{\infty}\mathrm dP'\,\mathbb{S}_{PP'}(\mathds{1})\rho_L(P')
 =
 0\,,
 \quad\Longrightarrow\quad
 \mathcal B_{L,\mathcal E}(P)
 =
 -\int_{\mathcal E}^{\infty}\mathrm dP'\,\mathbb{S}_{PP'}(\mathds{1})\rho_L(P')\,.
\end{equation}
 The edge term is thus the below-cutoff half of a cancellation whose other half lives above the cutoff: it is a UV boundary effect of truncating an oscillatory transform\footnote{In particular it is not a property of the light spectrum, but rather a Gibbs ringing effect.}. Then, since the modular image of the tail is subtracted in the constraint, we can obtain the potential in the completed b.c. by simply adding $+\mathcal{B}_{L,\mathcal{E}}(P)$ to the clipped constraint:
\begin{equation}
    V(\mathbf{h})
    \equiv\frac{1}{\epsilon^2}
    \sum_{0<P<\mathcal{E}}
    \left[
        N_{P,\epsilon}
        -\epsilon\sum_{i}
            \mathbb{S}_{P P_i(h_i)}(\mathbf{1})
        +\epsilon\mathcal{B}_{L,\mathcal{E}}(P)
    \right]^2,
    \label{eq:cardy-potential-completed}
\end{equation}
and from this point we can repeat the same manipulations performed for the `clipped' case.
The `completed' ensemble is therefore obtained by replacing the potential in \eqref{eq:cardy_potential} with
\eqref{eq:cardy-potential-completed} upon implementing the UV cutoff in \eqref{eq:part_funct_cardy}: the frozen tail enters as an external source for the modular constraints, not as additional dynamical eigenvalues.

\subsubsection*{The high-energy behavior}
Let us now look at the potential for bins in the high energy region of the spectrum whose momenta are of order $\mathcal{E}$, assuming that the vacuum is occupied and the bins' width is chosen according to \cref{sec:bins_width}. From \eqref{eq:S_kernel_highenergy}, we find that the contribution from the identity dominates the sums over eigenvalues above the black hole gap in each bin\footnote{With possible sub-leading exponentials contributions coming from the lightest non-vacuum primary.}, so that the $P$-bin contribution to the potential asymptotically becomes
\begin{equation}\label{eq:high_energy_potential_term_approx}
      \bigr(N_{P,\;\epsilon}-\epsilon\sum_i \mathbb{S}_{ PP_i}(\mathds{1})\bigr)^2\to_{P\to\infty}(N_{P,\;\epsilon}-\epsilon \mathbb{S}_{P\mathds{1}}(\mathds{1}))^2.
\end{equation}
In the equation above we can already recognize
\begin{equation}\label{eq:cardy_formula}
    \rho_0(P)\equiv\mathbb{S}_{P\mathds{1}} (\mathds{1})\approx\sqrt{2}e^{2\pi \sqrt{\frac{c-1}{6}}P}=\rho_{\rm Cardy}(P),
\end{equation}
where $\rho_{\rm Cardy}$ is the (holomorphic) asymptotic Cardy density for the primaries of large conformal dimension. Neglecting the repulsive contribution of the Vandermonde, $\rho_{\rm Cardy}(P)$ thus appears as the leading spectral density of the Cardy RMT. We will give a more complete analysis of the saddle point equations of the full potential, including the Vandermonde contribution in \cref{sec:saddle_point} below, where we will establish this fact more carefully.

Notice that we can find the more familiar Cardy behavior as a function of energy $\propto \exp{\bigr(2\pi \sqrt{(c-1)/6}\sqrt{h-(c-1)/24}\bigr)}$, by using \eqref{eq:P_to_h}.
It is easy to understand the behavior of this single-bin contribution to the potential: when the occupation number $N_{P,\epsilon}<\epsilon \mathbb{S}_{P\mathds{1}}(\mathds{1})$ it attracts eigenvalues from neighboring bins, while when $N_{P,\epsilon}>\epsilon \mathbb{S}_{P\mathds{1}}(\mathds{1})$ it repels them. The equilibrium configuration is achieved when $N_{P,\epsilon}\approx\epsilon \mathbb{S}_{P\mathds{1}}(\mathds{1})$. Thus in the $a\to0$ limit, the Cardy potential, living up to its name, imposes that the occupation number of each bin is given by the appropriate coarse graining of the Cardy density. At high energies, this occupation number is the only eigenvalue dependence in the potential, because the dominant $\mathbb{S}_{P\mathds{1}}(\mathds{1})$ just depends on the vacuum being occupied and the bin position, $P$. Notice, however, that at low energies it is still possible, without extra assumptions, that eigenvalue dependent contributions from $\mathbb{S}_{PP_i}(\mathds{1})$ are giving non-negligible or even leading contributions. With the vacuum occupied, we expect both clipped and completed boundary conditions to reproduce the same leading Cardy growth at high energies, with the distinguishing edge-term \eqref{eq:vacuum-edge-explicit} shifting the characteristic momentum at which this regime sets in. We will study numerically both boundary conditions, and confirm these heuristic considerations, in \cref{sec:numerics}.

\section{Numerical and analytical study of the model}\label{sec:results_tot}

In this section we begin our analysis of the Cardy RMT with a numerical Monte-Carlo study. The main take-away results from our numerical and subsequent analytical study are
\begin{itemize}
\item We study two variants of the model, with and without an identity source. The former results in a high-energy Cardy density of states, while the latter gives rise to a bounded spectrum.
    \item The density of states deep in the Cardy phase is exponentially large and changes exponentially fast. This means that the wave-length of level-spacing oscillations ($\sim$ level repulsion) is observed predominantly on intra-bin scales of the constraint$^2$ model.
\end{itemize}
We will establish the general phenomenology of the model numerically, and confirm these features analytically in later sections. As we will see, the two analyses probe largely complementary regimes: because the model’s complexity grows exponentially, computational constraints restrict numerical studies to a UV cutoff $\mathcal{E}\sim\mathcal{O}(1)$ and small $N$, whereas analytic methods exploit double-scaling limits that are naturally formulated in the regime $\mathcal{E},N\to\infty$.

\subsection{Monte Carlo numerical implementation}\label{sec:numerics}
We now test the heuristic picture of \cref{sec:cardy_potential} directly, by sampling the Cardy matrix model \eqref{eq:part_funct_cardy} with a Markov chain Monte Carlo (MCMC) algorithm. Throughout, we work with the sourced model, i.e. the light spectrum is prescribed: we will consider the vacuum as the only light state, and take $N_{\text{vac}}=0,1$. The dynamical variables are the $N$ conformal dimensions $h_i$, or equivalently the momenta $P_i=\sqrt{h_i-(c-1)/24}$, subject to the Vandermonde repulsion and the $\mathbb{S}$-kernel constraint$^2$ potential. We compare the coarse grained density and its crossover to Cardy for both clipped and completed boundary conditions at the UV cutoff, and finally, we examine the fine-grained
spectral statistics.\\

For convenience, we work at the self-dual point $b=1$ ($c=25$, $Q\equiv b+b^{-1}=2$), where the mathematical expressions of the relevant kernels simplify. The vacuum element becomes $\mathbb{S}_{P\mathds{1}}(\mathds{1})=4\sqrt{2}\,\sinh^2(2\pi P)$ \eqref{eq:S_kernel_expression}, the Cardy exponent $2\pi Q=4\pi$ is an integer multiple of $\pi$, while the principal-series kernel $\mathbb{S}_{P'P}(\mathds{1})=2\sqrt{2}\cos(4\pi P'P)$ is independent of $b$. To avoid artifacts due to working at the self-dual point $b=1$, we have also verified that non-self-dual values ($c=26$ and $c=38.5$) reproduce the same qualitative behavior, including the crossover to Cardy growth.

A central feature of the simulation is that the identity is \emph{not} a dynamical eigenvalue: it sits at the fixed imaginary momentum $P_{\mathds{1}}=iQ/2$ and enters the constraint as an external source $\epsilon\,\mathbb{S}_{P\mathds{1}}(\mathds{1})$. It is this source that supplies the exponential growth $\sim e^{2\pi Q P}$ of the target bin occupations \eqref{eq:cardy_formula} and makes the Cardy density a self-consistent saddle. Switching the source off (formally $\mathbb{S}_{P\mathds{1}}\to0$) defines the complementary `no-vacuum' problem.

We recall that with clipped b.c. the spectrum stops at the UV cutoff, while with completed b.c. the density above the cutoff is instead frozen to the light-induced profile $N_{\mathrm{vac}}\rho_0(P)$ and contributes to the
constraints through the prescribed boundary term. This background does not introduce additional sampled eigenvalues: the Vandermonde measure over the fluctuating spectrum is the same in both cases. When the identity source is switched off, the prescribed background also vanishes, so the two boundary prescriptions reduce to the same no-vacuum ensemble.

We evaluate the eigenvalue integral \eqref{eq:part_funct_cardy} by Monte Carlo. Two features of the model make this delicate. First, at small $a$ the constraint$^2$ potential is very stiff: it fixes the number of eigenvalues in each bin almost rigidly, so an efficient exploration of configurations must redistribute eigenvalues both within individual bins and between them, on all scales from well below the bin width $\epsilon$ up to the full spectral width $\mathcal{E}$. Second, configurations with different bin occupations are separated by large potential barriers. Rather than work at the target coupling directly, we therefore lower $a$ gradually from a weakly coupled, nearly free matrix down to the desired value, which can be seen as the numerical counterpart of the analytic $a\to0$ limit of \cref{sec:saddle_point}. The regulators are fixed throughout by the double-scaling variables $n_b=\epsilon N$ and $G=\epsilon\mathcal{E}$ of \eqref{eq:limits}. We have collected details of the sampling algorithm, including proposal kernel, annealing schedule, initialization, and the parameters of every run, in \cref{app:numerics}.

\begin{figure}[h]
    \centering
    {\small\textbf{(a) Clipped boundary conditions}\par}
    \includegraphics[width=0.84\linewidth]{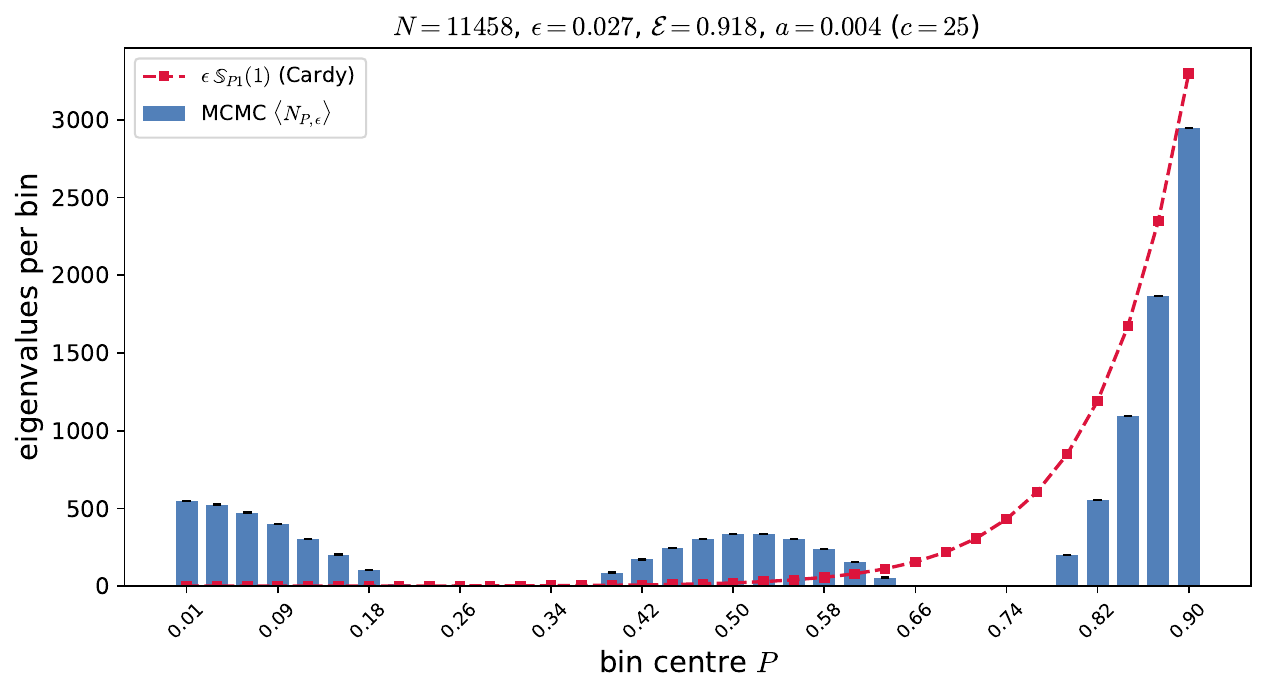}
    \par\medskip
    {\small\textbf{(b) Completed boundary conditions}\par}
    \includegraphics[width=0.84\linewidth]{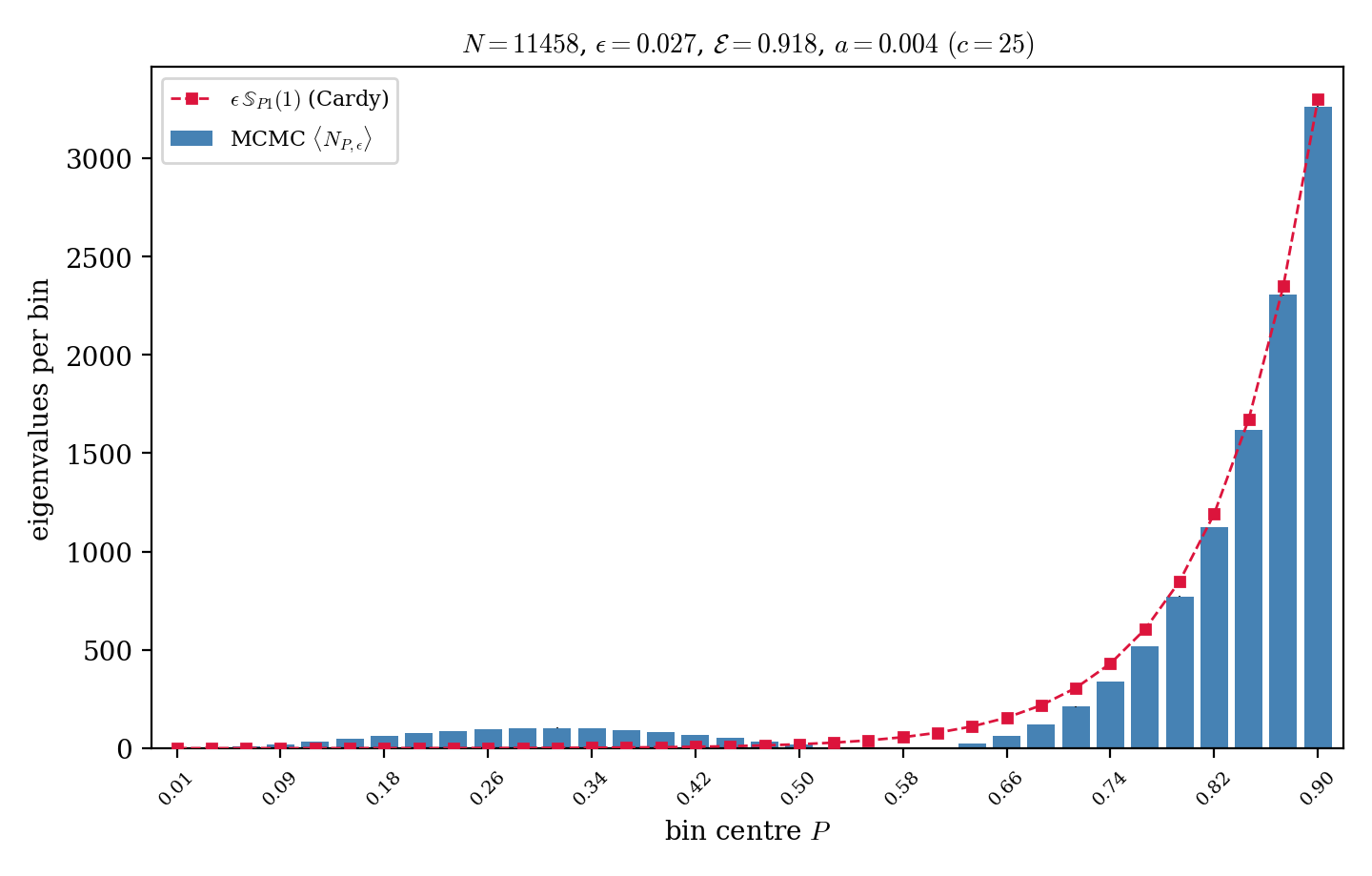}
    \caption{Equilibrium bin occupations at $c=25$, $N=11458$,
    $\epsilon=0.027$, $\mathcal E=0.918$ and $a=0.004$, with the identity
    source switched on. Blue bars denote the MCMC averages
    $\langle N_{P,\epsilon}\rangle$; the red dashed curve is the
    bin-centre Cardy prediction $\epsilon \mathbb{S}_{P\mathds{1}}(\mathds{1})$. The two runs use the same numerical parameters and sampling
    procedure; only the cutoff prescription is changed.
    \textbf{(a)} The clipped model displays pronounced low- and
    intermediate-momentum excesses, predominantly due to the edge-term \eqref{eq:vacuum-edge-explicit}. We can interpret the oscillations as a \emph{Gibbs ringing} artifact due to the $\mathbb{S}$-transform truncation.
    \textbf{(b)} With completed conditions, the bins' occupations are closer to Cardy. The remaining excess at low-momenta comes from eigenvalue repulsion and the dynamical principal-series states (the
eigenvalue-dependent $\mathbb{S}_{PP_i}(\mathds{1})$ terms), which are subleading to Cardy growth at high-energies.}
    \label{fig:mcmc_occupations}
\end{figure}
We illustrate the overall behavior of the model in
\cref{fig:mcmc_occupations}, comparing clipped and completed boundary
conditions at the same numerical parameters. In the high-energy region,
the measured bin occupations $\langle N_{P,\epsilon}\rangle$ reproduce
the coarse-grained Cardy density
$\epsilon\,\mathbb{S}_{P\mathds{1}}(\mathds{1})$ in both cases, so that
the potential drives the occupation numbers to
the Cardy profile. For clipped conditions, the low-$P$ mismatch with Cardy is predominantly driven by the edge-term \eqref{eq:vacuum-edge-explicit}, as anticipated in
\cref{sec:cardy_potential}.
  With completed conditions, this contribution is subtracted, and the occupations track the Cardy
profile down to lower momenta. In this case, the considerably smaller excess at low P, where the
bounded principal-series kernel $\mathbb{S}_{PP_i}(\mathds{1})$ contributes to the constraint target at leading
order, is precisely the eigenvalue-dependent contribution anticipated at the end of \cref{sec:cardy_potential}.

\begin{figure}[h]
\centering
{\tiny\textbf{(a) Clipped boundary conditions}\par}
    \includegraphics[width=0.74\textwidth]{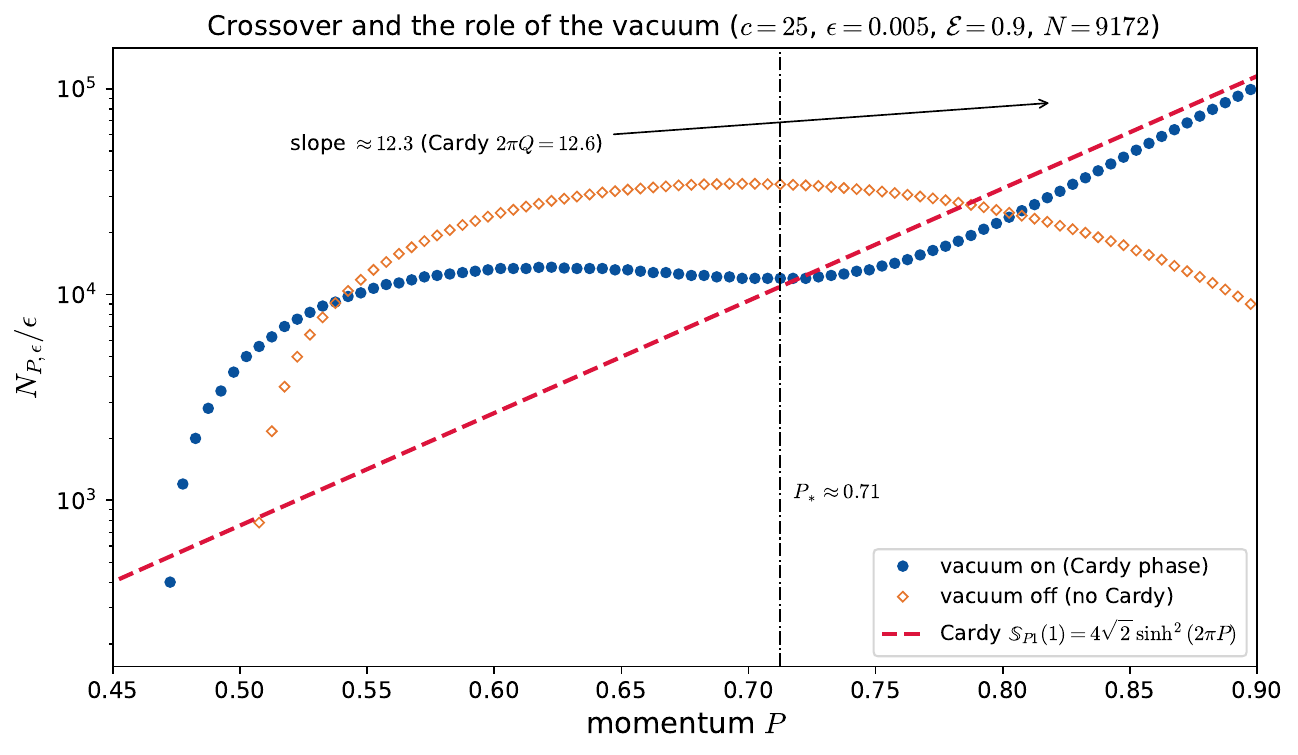}
    \par\medskip
    {\tiny\textbf{(b) Completed boundary conditions}\par}
    \includegraphics[width=0.74\linewidth]{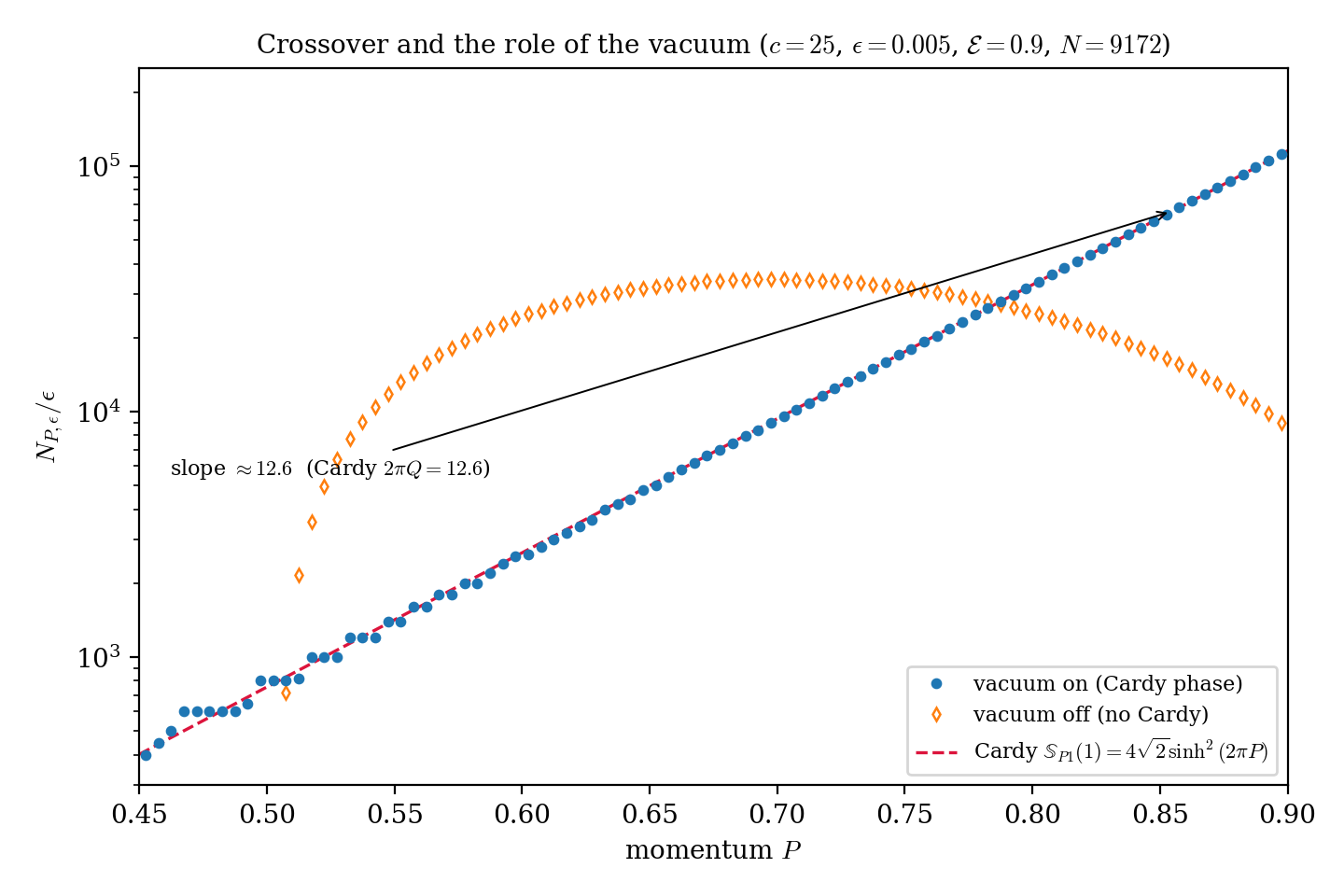}
    \caption{Crossover to Cardy growth and the role of the vacuum ($c=25$, $\epsilon=0.005$, $\mathcal{E}=0.9$, $N=9172$, eigenvalues initialized at the Cardy density). \emph{Filled circles} (blue): the equilibrium spectral density $N_{P,\epsilon}/\epsilon$ with the identity (vacuum) source on (Cardy phase). \textbf{(a)} With clipped conditions, a broad plateau precedes the
    crossover to Cardy growth near the indicated $P_*\simeq0.71$, where the Cardy curve $\mathbb{S}_{P\mathds{1}}(\mathds{1})$ (red dashed) overtakes the plateau. Above $P_\ast$ the fitted log-slope matches the Cardy exponent $2\pi Q$ (here $4\pi$) to within a few percent. \textbf{(b)} With completed conditions, the sourced density tracks
    Cardy over a wider range, due to the absence of the edge-term \eqref{eq:vacuum-edge-explicit}.  \emph{Open diamonds} (orange): the same ensemble at the same $N$ with the vacuum source switched off: lacking the identity drive, the density stays bounded and never develops Cardy growth (it turns over at high $P$). Because $N$ is fixed, occupying the vacuum redistributes eigenvalues from the bulk into the high-energy tower, so the two data sets cross near $P_\ast$. The exponential black-hole tower is thus sourced by the identity in the dual channel.}
    \label{fig:mcmc_crossover}
\end{figure}

To isolate the Cardy regime cleanly, we restrict in
\cref{fig:mcmc_crossover} to a high-energy window and resolve it at the
finer binning scale $\epsilon=0.005$. Both boundary prescriptions
showcase a crossover to exponential Cardy growth, but its onset is
different. For clipped conditions, the spectral density displays a
sharp crossover from an approximately flat plateau to Cardy growth at the momentum $P_\ast$, where
$\mathbb{S}_{P\mathds{1}}(\mathds{1})$ overtakes the low-momenta state
density. The restriction of the Cardy regime to the upper part of the spectrum is
consistent with the edge term $\mathcal{B}_{0,\mathcal{E}}$ \eqref{eq:vacuum-edge-explicit}: at fixed $Q$ and
large cutoff, its envelope relative to
$\rho_0(P)=\mathbb{S}_{P\mathds{1}}(\mathds{1})$ is
$O(e^{2\pi Q(\mathcal E-P)}/\mathcal E)$ for $P\sim\mathcal E$.
Requiring $\mathcal{B}_{0,\mathcal{E}}$ to be subleading with respect to the vacuum $\mathbb{S}$-kernel,
$e^{2\pi Q(\mathcal E-P)}\ll\mathcal E$, restricts the uniformly
controlled vacuum-dominance approximation to a momentum window of
width $O(\log\mathcal E)$ below the cutoff. With completed conditions,
the prescribed modular tail cancels this edge term, removing this
UV-window restriction: at the matched Cardy count $N=M_{\mathds{1}}\equiv\int_{0}^\mathcal{E}dP\rho_{Cardy}(P)$,\footnote{$N$ is an integer, so we really mean the floor of $M_{\mathds{1}}$.}  the continuum
profile $\rho=\rho_0$ solves the constraint throughout the fluctuating
interval. Correspondingly, the Cardy match extends across the high-energy
region rather than emerging only near the cutoff.

Above the respective crossover scales, the numerically measured
log-slope matches the Cardy exponent $2\pi Q$ to within a few percent.
Varying the central charge confirms that this slope tracks the
$b$-dependent exponent rather than a fixed value. For the clipped
runs at $b=1$ and $b=2$ (i.e.\ $c=25$ and $c=38.5$), we measure slopes
$12.3$ and $15.5$, against the predicted $2\pi Q=12.6$ and $15.7$,
respectively, while the completed runs give $12.7$ and $15.8$.
This crossover is the spectral-density manifestation of the
low-momenta state/Cardy competition discussed in
\cref{sec:vac_occ_gap}, with its onset sensitive to the boundary
prescription.\footnote{At coarser $\epsilon$, the clipped simulations
also display a spurious density-wave, with alternating over- and
under-occupied bins below $P_\ast$. This finite-binning artifact is
suppressed as $\epsilon$ is reduced, and we have checked that it does
not affect the Cardy regime. It should be distinguished
from the continuum edge term, which persists at fixed $\mathcal E$
as $\epsilon\to0$ and is canceled by the completed boundary
conditions.}
Our analysis also isolates the role of the vacuum: repeating the run
at fixed $N$ with the identity source switched off (open diamonds in
\cref{fig:mcmc_crossover}), and hence with no light-induced completion
tail, the spectrum never develops Cardy growth, confirming that the
exponential black-hole tower is sourced by the identity in the dual
channel.
The complementary in-bin distribution, reflecting the Vandermonde repulsion via the hard-wall ansatz \eqref{eq:rho_ansatz}, is analyzed separately in \cref{sec:mesoscopic_inbin} (\cref{fig:inbin_density_ansatz}).\\
 
\begin{figure}
\centering
    \includegraphics[width=0.98\textwidth]{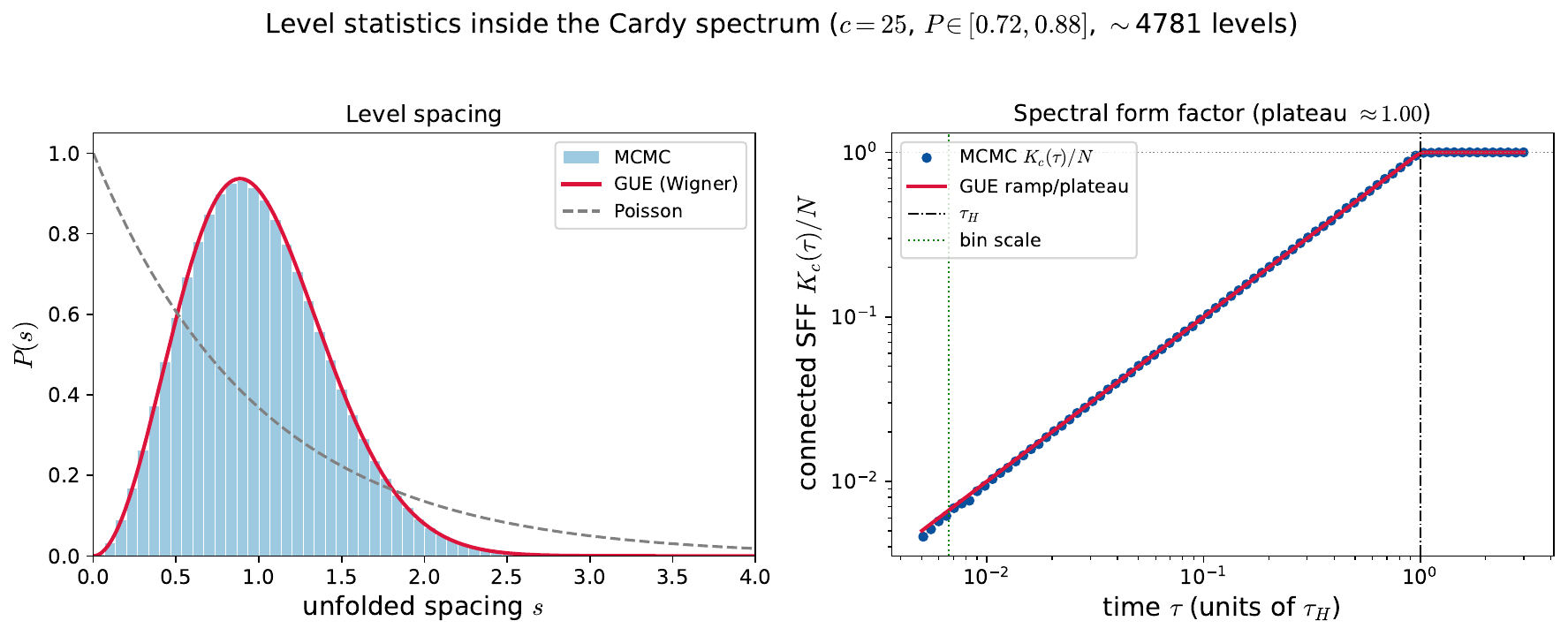}
    \caption{Spectral statistics inside the Cardy part of the spectrum ($c=25$, window $P\in[0.72,0.88]$, $\sim$4800 levels at $\epsilon = 0.005$). \emph{Left:} the unfolded nearest-neighbour level spacing $P(s)$ follows the GUE Wigner--Dyson distribution (red), with repulsion $P(s\to0)\to0$, far from Poisson (indicated as dashed line). \emph{Right:} the connected spectral form factor $K_c(\tau)/N$ exhibits the GUE linear ramp (slope $\approx1$) up to the Heisenberg time $\tau_H$, marked in black dots. The late-time form factor saturates to the GUE plateau (\cref{sec:meso_phys}). For this figure, we use simulations at fixed coupling $a=0.04$, without annealing.}
    \label{fig:mcmc_sff}
\end{figure}

Finally, we further probe the random-matrix content of the model, by examining fine-grained spectral statistics, which the coarse modular constraint does not fix. For this, we choose the momentum window $P\in[0.72,0.88]$, so that we lie squarely within the Cardy regime and have a large number of eigenvalues (one sees this in \cref{fig:mcmc_crossover}). Resolving the spectrum at the level-spacing scale, \cref{fig:mcmc_sff} shows that the eigenvalues exhibit GUE statistics: the nearest-neighbour spacing distribution follows the GUE Wigner--Dyson law, and the connected spectral form factor displays the linear ramp up to the Heisenberg time.  Despite the modular constraint rigidifying the spectrum at macroscopic scales, the form factor at nearby scales shows clean RMT behavior. We establish the universal ramp of the form factor analytically in \cref{sec:meso_phys}, while the plateau is established numerically in \ref{sec:numerics}.

Having established some of the most interesting phenomenological manifestations of the Cardy-RMT numerically, we turn our focus to an analytical study for the same.

\subsection{Saddle point equations}\label{sec:saddle_point}
We want to derive the saddle point equations for the double trace potential \eqref{eq:cardy_potential_doubletrace}. To this end, we define an effective action, such that \eqref{eq:part_funct_cardy} is written as
\begin{equation}\label{eq:part_funct_seff}
    \mathcal{Z}=\int \prod dh_i e^{-N^2S_{\rm eff}},
\end{equation}
with the effective action receiving a contribution from the exponentiated Vandermonde determinant
\begin{equation}
    S_{\rm eff}=\frac{1}{4a(\epsilon N)^2}\sum_P\biggr(N_{P,\epsilon}-\epsilon \sum_i \mathbb{S}_{PP_i}(\mathds{1})\biggr)^2-\frac{2}{N^2}\sum_{i<j}\log|h_i-h_j|\,.
\end{equation}
 We see that the effective potential consists of the $\mathbb{S}-$kernel contribution balanced by the inter-eigenvalue logarithmic repulsion term from the Vandermonde.
As we will elaborate more on in the next section, our prescription when taking the $N\to\infty$ limit, is to double-scale the regulators so that $a,\,\epsilon\to0$ with $\epsilon N$ and the `t Hooft coupling $t=aN$ kept fixed. In these limits, \eqref{eq:part_funct_seff} localizes on its saddle point configurations, determined by the equations $\delta S_{\rm eff} / \delta h_i =0$, obtained by imposing that the following formal expression is equal to zero:
{\small\begin{equation}\label{eq:saddle_point_eq}
\begin{aligned}
     \frac{2}{N^2}\sum_{j\neq i}\frac{1}{h_i-h_j}-\frac{1}{2t\epsilon^2 N}\sum_P\biggr(N_{P,\epsilon}-\epsilon \sum_j \mathbb{S}_{PP_j}(\mathds{1})\biggr)\biggr(\frac{\delta(P_i-(P\pm\epsilon/2))}{2P_i}-\epsilon \mathbb{S}'_{PP_i(h_i)}(\mathds{1})\biggr)\,.
\end{aligned}
\end{equation}}
Here we adopted the notation $\delta(P_i-(P\pm\epsilon/2))\equiv\delta(P_i-(P-\epsilon/2))-\delta(P_i-(P+\epsilon/2))$.
Notice that the second term is enhanced at least by a $1/\epsilon$ factor after performing the double-scaling procedure, so we can solve \eqref{eq:saddle_point_eq} order by order in $\epsilon$. In the standard matrix model literature, for example in the case of the Gaussian RMT, this usually happens when $t\to0$, which localizes the potential in its minimum. Here, the more general case suffices $a\to0$, because, due to the $1/\epsilon$ prefactor appearing in the potential \eqref{eq:cardy_potential}, we have an `effective `t Hooft coupling' $\mathrm{a}N\equiv a\epsilon^2 N$. Of course, one could consider the `constraint$^2$' potential for the equivalent set of \eqref{eq:mod_constrant_blocks} without the $1/\epsilon$ factor, and this would bring the argument made here to the more standard $t\to0$ case. Analogously, we notice that, in the case where we keep $\epsilon$ finite, the arguments made here similarly need to be translated to the $t\to 0$ limit.

In the limits discussed above, the leading term in \eqref{eq:saddle_point_eq} wants to localize the eigenvalues to the modular invariant configuration with zero potential\footnote{notice that the LHS below is an integer, so an exact zero-potential configuration does not exist generically. However, we imagine that rounding errors are negligible when the occupation numbers are large, as it happens for the Cardy case.}
\begin{equation}
    N_{P,\epsilon}\approx \epsilon \sum_i \mathbb{S}_{PP_i}(\mathds{1})\mathrel{\underset{P\to\infty}{\approx}} \epsilon\mathbb{S}_{P\mathds{1}}(\mathds{1}),
\end{equation}
where the last approximation holds in the high-energy region when the vacuum is occupied. After the bin occupations numbers have been fixed to the Cardy expectation, solving the sub-leading order of \eqref{eq:saddle_point_eq} gives us that the eigenvalue configuration inside each bin is determined by the Vandermonde repulsion. So intuitively, in the parameter limit we consider, the Cardy potential determines the `global' structure of the coarse-grained spectrum, by fixing the bin-occupation numbers, while the random matrix nature of the model is manifest in the regularization-dependent `mesoscopic' eigenvalue configuration inside the bins. We expand on this in \cref{sec:mesoscopic_inbin}.

One interesting output of this analysis is that the Cardy potential \eqref{eq:cardy_potential} indicates a preferred number of eigenvalues to insert in the system:
\begin{equation}\label{eq:ntildeprime_def}
   M_{\mathds{1}}\approx\epsilon\sum_P\mathbb{S}_{P\mathds{1}}(\mathds{1}),
\end{equation}
where again for simplicity we assume the vacuum is occupied and focus on the high-energy region. In particular, $M_{\mathds{1}}$ is the number of eigenvalues prescribed to achieve regularized modular invariance for the exponentially growing Cardy density. We emphasize that $N$ is in principle a free parameter that we can pick in any way we want. But what we see is that if we choose the number of eigenvalues given by \eqref{eq:ntildeprime_def}, we have that the Cardy configuration is a natural saddle of the
matrix model.

For a parametrically bigger/smaller $N$, the natural expectation is that the mismatch arranges itself along deformations of the density that are as close as possible to modular invariance, as we discuss at length in following section. This is already an indication, that the scaling of $N$ with the cutoff $\mathcal{E}$ has a dramatic effect on the physics of the model, and indeed we will show in \cref{sec:vac_occ_gap} that it determines in which phase the RMT lies.

\subsection{Modular invariance and prolate spheroidal modes}
\label{sec:vac_occ_gap}
As discussed in the previous section, the eigenvalues of the Cardy
matrix model first arrange themselves so as to minimize the potential
\eqref{eq:cardy_potential}. In this section we compare the minima of the continuum Cardy potential for different prescribed light spectra, using completed boundary conditions. The light spectrum is held fixed in each minimization. We first compare the cases with no light states and with a single vacuum, identifying the crossing of their minimized potentials as a “Cardy transition.” We then extend this comparison to generic light sources. These results identify which source configurations have the lowest constraint cost and suggest a mechanism for light-state selection in a possible dynamical extension of our model.

We show that the way in
which the number of states $N(\mathcal E)$ scales with the cutoff
determines which minimal configurations survive in the limit
where the regulators are removed, and hence in which ``phase'' the
model lies. The feature we are after is the presence, and the
onset, of the black-hole gap at energy $(c-1)/24$. The basic intuition
comes from the crossing kernel \eqref{eq:S_kernel_expression}: when $P$ and
$P'$ are both real, or both imaginary, $\mathbb S_{PP'}(\mathds{1})$ is
proportional to a cosine, whose modulus is bounded by unity; when
exactly one of them is imaginary it becomes a hyperbolic function,
growing exponentially at high energies. A modular-invariant exponentially-growing spectrum must therefore be sourced by a
state below the black-hole threshold, and we will argue that this is the
mechanism which decides whether light states are occupied and which
configurations are exponentially suppressed. In particular, our goal will be to compare the `constraint$^2$' potential for
vacuum occupation $N_{\rm vac}=0,\;1$, and to show that beyond
a critical number of eigenvalues $N_c$ the configuration with the occupied vacuum
has the lower potential.

\subsubsection*{Potential in the continuum approximation}
Since the occupation numbers in the regime of interest are large, and we are interested only in the macroscopic Vandermonde-independent features of the system, a convenient way to estimate the minima of the potential
\eqref{eq:cardy_potential} will be to take a continuum limit by replacing the bin
occupations with a smooth density $\rho(P)$ on the window
$[0,\mathcal E]$. Moreover in this section, we will limit our study to the model in the `\emph{completed}' boundary conditions, in order to focus on properties descending from modular invariance and avoid dealing with the edge-term truncation artifact.

Consider a prescribed light spectrum with
occupations $N_\ell$ at imaginary momenta $P_\ell$, with the following associated $\mathbb{S}$-transform profile on the real momenta $\rho_L(P)
 \equiv
 \sum_{\ell\in{\rm light}}
 N_\ell\, \mathbb{S}_{P P_\ell}(\mathds{1})$.
The continuum limit of the constraint \eqref{eq:mod_constrant_blocks} in the $P$-bin reads as
\begin{equation}\label{eq:hard-truncation-residual}
 f^{\rm clip}_L(P)
 =
 \rho(P)
 -
 \int_0^{\mathcal E}\mathrm dP'\,\mathbb{S}_{PP'}(\mathds{1})\rho(P')
 -
 \rho_L(P)\,,
\end{equation}
where $f^{\rm clip}_L(P)$ is defined to be the continuum constraint residual, and the light states contribute only via their $\mathbb S$-kernel images. We can rewrite the above constraint in terms of the boundary term $\mathcal B_{L,\mathcal E}(P)\equiv
 \int_0^{\mathcal E}\mathrm dP'\,\mathbb{S}_{PP'}(\mathds{1})\rho_L(P')$  as:
\begin{equation}\label{eq:edge-term-def}
\begin{aligned}
   & f^{\rm clip}_L(P)
 =
 \bigl(\rho-\rho_L\bigr)(P)
 -
 \int_0^{\mathcal E}\mathrm dP'\,\mathbb{S}_{PP'}(\mathds{1})
 \bigl(\rho-\rho_L\bigr)(P')
 \;-\;
 \mathcal B_{L,\mathcal E}(P)\,.
\end{aligned}
\end{equation}
Squaring and integrating over real momenta the expression above, would give the Cardy potential under \emph{clipped} UV cutoff conditions. As discussed in \cref{sec:cardy_potential}, passing to the \emph{completed} prescription requires adding the edge-term to \eqref{eq:edge-term-def}. So the continuum limit of the \emph{completed} Cardy potential is  given by:
\begin{equation}\label{eq:completed-residual}
\begin{aligned}
    f_L(P)
 &=
 \rho(P)
 -\int_0^{\mathcal E}\!\mathrm dP'\,\mathbb{S}_{PP'}(\mathds{1})\rho(P')
 -\rho_L(P)
 -\underbrace{\int_{\mathcal E}^{\infty}\!\mathrm dP'\,
   \mathbb{S}_{PP'}(\mathds{1})\rho_L(P')}_{=-\,\mathcal B_{L,\mathcal E}(P)}
 =\\&=
 \bigl(\rho-\rho_L\bigr)(P)
 -\int_0^{\mathcal E}\!\mathrm dP'\,\mathbb{S}_{PP'}(\mathds{1})
 \bigl(\rho-\rho_L\bigr)(P')\,.
\end{aligned}
\end{equation}
In particular, the edge term cancels identically and $\rho=\rho_L$ is an exact zero of
the constraint at every finite cutoff. This is the reason why the numerics in \cref{sec:numerics}, performed at the Cardy eigenvalue count $N=M_{\mathds{1}}$, match the Cardy profile in the completed boundary conditions. In the next sections, we will compare the macroscopic minima associated with different prescribed light spectra, where the modular invariance information embedded in the constraint  \eqref{eq:completed-residual} resides. As prescribed by the `completed' boundary
condition, the spectrum above the cutoff is not dynamical but frozen
to the $\rho_L$ background: for $P>\mathcal E$ the density is fixed to
$\rho_L(P)$, and only the window $[0,\mathcal E]$ fluctuates.

\subsubsection{The `Cardy transition'}
If we take a continuum limit and ignore the effect of the Vandermonde
level repulsion, we can compute the value $I$ of the Cardy potential for
the macroscopic saddle point configuration of \cref{sec:saddle_point}
as:
\begin{equation}\label{eq:continuum-light-potential}
	I_L[\rho]
 =
 \frac{1}{4a\epsilon}
 \int_0^{\mathcal E}\mathrm dP\,
 \left[
  \rho(P)-\rho_L(P)
  -
  \int_0^{\mathcal E}\mathrm dP'\,
  \mathbb{S}_{P P'}(\mathds{1})
  \bigl(\rho(P')-\rho_L(P')\bigr)
 \right]^2,
\end{equation}
where the density is subject to the following positivity and
normalization constraints
\begin{equation}\label{eq:fP_cont_mod_inv}
	 \rho(P)\geq0,
 \qquad
 \int_0^{\mathcal E}\rho(P)\,\mathrm dP
 =
 N-\sum_{\ell\in{\rm light}}N_\ell.
\end{equation}
In this section we will limit our analysis to a finite number of light
states, so in particular in the constraints above we can neglect the
difference between the number of heavy states and $N$ for our scaling
arguments.
In equation \eqref{eq:continuum-light-potential}, we included the UV
cutoff with completed conditions, and,
for example, if only the identity were occupied then $\rho_L(P)$ would
be the usual Cardy density of states,
$\rho_L(P)=\rho_0(P)$. In the following part of this section, we will be interested
in studying the value of the minimized potential for $N_{vac}=0,1$
defined as follows:
\begin{align}
 \mathcal R_{N_{vac}}(N,\mathcal E)
 \equiv
 \inf_{\substack{
  \rho(P)\geq0\\[1mm]
  \int_0^{\mathcal E}\rho(P)\,\mathrm dP=N-N_{vac}
 }}
 \int_0^{\mathcal E}\mathrm dP\,
 \Bigg[
  &\rho(P)-N_{vac}\rho_0(P)
 \nonumber\\
  &-
  \int_0^{\mathcal E}\mathrm dP'\,
  \mathbb{S}_{P P'}(\mathds{1})
  \bigl(
   \rho(P')-N_{vac}\rho_0(P')
  \bigr)
 \Bigg]^2 .
 \label{eq:exact-sector-residual}
\end{align}
We start by considering the case where only heavy-states are occupied,
that is $\rho_L=0$, together with the following limit procedure, as in \cref{sec:saddle_point}, to
remove the regulators:
\begin{equation}\label{eq:limits}
    N, \mathcal{E}\to\infty;\; \epsilon,a\to 0\quad\mathrm{with}\quad \epsilon N \equiv n_b,\;\epsilon\mathcal{E}\equiv G,\;t\equiv aN\quad\mathrm{fixed}.
\end{equation}
At this point, we notice that the $\mathbb{S}$-kernel integration acts like a Fourier transform on the finite window $[0,\mathcal{E}]$. The Fourier transform with bounded support is a self-adjoint operator and its eigenfunctions and eigenvalues have been studied extensively, see \cite{SlepianPollak1961,Slepian1965} for the results relevant to our discussion. These eigenfunctions are known
as \emph{prolate spheroidal} wave functions, and their eigenvalues are all
strictly smaller than one in modulus at finite cutoff. In particular,
the eigenfunction with the eigenvalue closest to one is non-negative and,
up to normalization, uniquely maximizes the normalized overlap with its bounded Fourier transform. This eigenfunction, denoted as
$s_{\cal E}(P)$, is normalized by $s_{\cal E}(0)=1$ and has eigenvalue
$\lambda_0(\mathcal{E})$ so that:
\begin{equation}
s_{\cal E}(P) \equiv S_{00}\left(4\pi {\cal E}^2, \frac{P}{\cal E} \right),\qquad \int_0^{\mathcal E}\mathrm dP'\,
 S_{P P'}(1)s_{\mathcal E}(P')
 =
 \lambda_0(\mathcal E)s_{\mathcal E}(P).
\end{equation}
In particular, we can find the following large-cutoff $\mathcal{E}$
expressions \cite{fuchs1964eigenvalues}:
\begin{equation}\label{eq:prolate-gaussian-limit}
    s_{\mathcal E}(P)
 \longrightarrow
 e^{-2\pi P^2},\quad \lambda_0({\cal E}) \sim 1- 4\pi {\cal E} e^{-8\pi {\cal E}^2},
\end{equation}
and we denote the spectral gap of the transform by
$\varsigma_{\mathcal E}\equiv1-\lambda_0(\mathcal E)$. It will also be
useful to define the following quantity:
\begin{equation}
 r_{\mathcal E}
 \equiv
 \frac{
  \int_0^{\mathcal E}
  s_{\mathcal E}(P)^2\,\mathrm dP
 }{
  \left(
   \int_0^{\mathcal E}
   s_{\mathcal E}(P)\,\mathrm dP
  \right)^2
}\longrightarrow_{\mathcal{E}\to\infty} 2.
 \label{eq:prolate-ratio}
\end{equation}
Another general result we can obtain from the lowest prolate eigenvalue
is the following inequality for every real
$g(P)\in L^2([0,\mathcal E])$:
\begin{equation}\label{eq:prolate-spectral-bound}
    \int_0^{\mathcal E}\mathrm dP\,
 \left[
  g(P)
  -
  \int_0^{\mathcal E}\mathrm dP'\,
  \mathbb{S}_{P P'}(\mathds{1})g(P')
 \right]^2\geq
 \varsigma_{\mathcal E}^2
 \int_0^{\mathcal E}g(P)^2\,\mathrm dP.
\end{equation}
Now consider the (correctly normalized) positive trial density given by
the lowest prolate eigenfunction
\begin{equation}
 \rho_{\rm p}(P)
 =
 \frac{
  N\,s_{\mathcal E}(P)
 }{
  \int_0^{\mathcal E}
  s_{\mathcal E}(P')\,\mathrm dP'
 }.
 \label{eq:positive-prolate-trial}
\end{equation}
If we insert this expression, and in particular using the large-cutoff
asymptotics, in the potential \eqref{eq:continuum-light-potential}, we
can obtain an upper estimate of $\mathcal R_0(N,\mathcal E)$
\eqref{eq:exact-sector-residual}:
\begin{equation}
 \mathcal R_0(N,\mathcal E)
 \leq
 \varsigma_\mathcal{E}^2
 r_{\mathcal E}N^2\simeq 32 \pi^2 {\cal E}^2  N^2 e^{-16 \pi {\cal E}^2}.
 \label{eq:no-vacuum-upper-bound}
\end{equation}
Conversely, a lower bound is obtained from
\eqref{eq:prolate-spectral-bound} together with Cauchy-Schwarz
\begin{equation}
 \int_0^{\mathcal E}\rho(P)^2\,\mathrm dP
 \geq
 \frac{1}{\mathcal E}
 \left(
  \int_0^{\mathcal E}\rho(P)\,\mathrm dP
 \right)^2
 =
 \frac{N^2}{\mathcal E}\implies \mathcal R_0(N,\mathcal E)
 \geq
 \frac{
  \varsigma_\mathcal{E}^2N^2
 }{
  \mathcal E
 }.
 \label{eq:positive-density-CS}
\end{equation}
The exact no-vacuum residual therefore satisfies:
\begin{equation}
 \frac{
  \varsigma_\mathcal{E}^2N^2
 }{
  \mathcal E
 }
 \leq
 \mathcal R_0(N,\mathcal E)
 \leq
 \varsigma_\mathcal{E}^2
 r_{\mathcal E}N^2.
 \label{eq:no-vacuum-rigorous-bounds}
\end{equation}
The potential is thus lower bounded, and there is no configuration that leads to a vanishing potential at finite UV cutoff. Moreover, from the linearity of the constraint and the invariance of
positivity under rescalings of $\rho$, we obtain that $\mathcal R_0$ respects the following homogeneity property
\begin{equation}
 \mathcal R_0(N,\mathcal E)
 =
 N^2\,\mathcal R_0(1,\mathcal E)\,,
 \label{eq:no-vacuum-homogeneity}
\end{equation}
and in particular it is
strictly increasing in $N$ at every finite cutoff. 
 
Now we move to consider the situation where the vacuum is occupied
$N_{vac}=1$. In this case, we find that there exist a preferred number
$M_{\mathds{1}}$ of eigenvalues that can be accommodated in the new
zero-potential configuration, as in \eqref{eq:ntildeprime_def}, is:
\begin{equation}\label{eq:ntildeprime_expression}
    M_{\mathds{1}} \;=\; \int^{\mathcal{E}} \! dP\,\mathbb{S}_{P\mathds{1}}(\mathds{1})
    \;\sim\; \frac{\sqrt{2}}{2\pi Q}\,e^{2\pi Q\mathcal{E}}\,.
\end{equation}
where we recall that $Q=\sqrt{(c-1)/6}$. Notice that in the integral we
used the high-energy expression \eqref{eq:cardy_formula} of the vacuum
S-kernel, that is the Cardy density, because most of the $M_{\mathds{1}}$
eigenvalues lie in the high-energy region of the spectrum. On this line,
we didn't specify the lower extremum of integration, which depends on
the requested sub-exponential accuracy of this approximation. More
broadly, for a generic light-state spectrum the most obvious solution
with zero potential is to take $\rho(P) = \rho_L(P)$. However, this analogously
requires that $N=\int_0^{\cal E} \rho_L(P)$, which in general will not
be the case, unless we allow ourselves to optimize over $\rho_L(P)$ as
well. Then we would simply find a light spectrum such that
$N=\int_0^{\cal E} \rho_L(P)$. This is a single equation for a large
number of unknowns, so it is not very restrictive.

If instead $N \neq \int_0^{\cal E} \rho_L(P)$, the constraint cannot be
satisfied exactly, and the question becomes how the density minimizes
the residual. The difference $\rho-\rho_L$ must then be as close as
possible to an invariant function of the finite-window
$\mathbb S$-transform, which is precisely the problem analyzed above in
the vacuum-less case: the cheapest available direction is the lowest
prolate mode. It is therefore natural to postulate that the minimum is
attained on
\begin{equation}\label{eq:prolate-ansatz}
 \rho(P)
 =
 \rho_L(P)+c\,s_{\mathcal E}(P)\,,
 \qquad
 c
 =
 \frac{N-\int_0^{\mathcal E}\rho_L}
      {\int_0^{\mathcal E}s_{\mathcal E}}\,,
\end{equation}
with $c$ fixed by the normalization. When $c\geq0$, that is when the
light spectrum accommodates fewer states than are present, this ansatz
is manifestly non-negative and hence admissible: the excess states are
stored in the almost-self-dual prolate mode at exponentially small cost.
When $c<0$, which happens for $N<\int_0^{\cal E}\rho_L(P)$, the
candidate density fails positivity. This is particularly evident in the region near the black hole threshold: here the vacuum
profile vanishes quadratically, $\rho_0(P)=16\sqrt2\,\pi^2P^2+O(P^4)$,
while $s_{\mathcal E}(0)=1$, so any negative prolate correction drives
$\rho(P)$ negative in such a neighborhood of $P=0$. The true minimizer must
then be corrected by higher prolate modes, which is hard to control
analytically. We will ignore this point for now and come back to it
below; until then, in order to state our arguments regarding the `Cardy transition', we will make use of less optimal bounds that avoid this positivity issue.

Independently of any ansatz, a rigorous lower bound is available: if we
apply \eqref{eq:prolate-spectral-bound} to $g(P)=\rho(P)-\rho_0(P)$,
whose integral is fixed to $N-1-M_{\mathds{1}}$ by the normalization
\eqref{eq:fP_cont_mod_inv}, together with Cauchy-Schwarz as we did
above, we obtain:
\begin{equation}
 \mathcal R_1(N,\mathcal E)
 \geq
 \frac{
  \varsigma_{\mathcal E}^2
 }{
  \mathcal E
 }
 \left(
  N-1-M_\mathds{1}(\mathcal E)
 \right)^2.
 \label{eq:vacuum-residual-lower-bound}
\end{equation}
For $1\leq N\leq M_\mathds{1}+1$, combining
\eqref{eq:no-vacuum-upper-bound} and
\eqref{eq:vacuum-residual-lower-bound} proves that $ \mathcal R_0(N,\mathcal E)
 <
 \mathcal R_1(N,\mathcal E)$
whenever
\begin{equation}
 N
 <
 \frac{
  M_\mathds{1}(\mathcal E)+1
 }{
  1+\sqrt{\mathcal E r_{\mathcal E}}
 },
 \label{eq:rigorous-no-vacuum-region}
\end{equation}
therefore the no-vacuum sector is rigorously selected throughout this
region. On the other hand, at the Cardy matching point $N
 =
 M_\mathds{1}(\mathcal E)+1$
the density $\rho(P)
 =
 \rho_0(P)$
is admissible in the $N_{vac}=1$ sector and has exactly zero residual:
\begin{equation}
 \mathcal R_1
 \bigl(
  M_\mathds{1}+1,\mathcal E
 \bigr)
 =
 0.
 \label{eq:zero-vacuum-residual}
\end{equation}
By contrast, because the finite-window cosine transform has no non-zero fixed
function, we have:
\begin{equation}
 \mathcal R_0
 \bigl(
  M_\mathds{1}+1,\mathcal E
 \bigr)
 >
 0,
 \label{eq:positive-no-vacuum-residual}
\end{equation}
The minimized vacuum residual is a convex function of the total number of heavy eigenvalues $N-1$, which we will refer to as `mass' in this section. Indeed, if $\rho_1$ and $\rho_2$ are admissible densities of
masses $M_1$ and $M_2$, then
$\theta\rho_1+(1-\theta)\rho_2$ is an admissible density of mass
$\theta M_1+(1-\theta)M_2$, and the square appearing in
\eqref{eq:exact-sector-residual} is convex. Moreover,
$\mathcal R_1$ has a unique zero at
$N=M_\mathds{1}+1$: a vanishing residual would otherwise give a
non-zero fixed function of the finite-window cosine transform.
Consequently, $\mathcal R_1$ is non-increasing on
$1\leq N\leq M_\mathds{1}+1$, while
\eqref{eq:no-vacuum-homogeneity} shows that $\mathcal R_0$ is strictly
increasing.
There is therefore a unique crossing $N_c(\mathcal E)$ between $\mathcal R_0$ and $\mathcal R_1$ in this interval, satisfying the following bounds:
\begin{equation}
 \frac{
  M_\mathds{1}(\mathcal E)+1
 }{
  1+\sqrt{\mathcal E r_{\mathcal E}}
 }
 \leq
 N_c(\mathcal E)
 \leq
 M_\mathds{1}(\mathcal E)+1.
 \label{eq:rigorous-vacuum-transition-bounds}
\end{equation}

This bound is not special to the
vacuum: as we show below for a generic light state, the spectral lower
bound and the prolate trial always confine the sector crossing between a
fixed fraction of the appropriate $M_{\mathds{1}}$ and $M_{\mathds{1}}$ itself. Now we come back to the positivity constraint issue:
within the window \eqref{eq:rigorous-vacuum-transition-bounds}, positivity acts asymmetrically and moves the lower
bound toward higher $N$. Intuitively, without the positivity constraint
a deficit $D=M_\mathds{1}+1-N$ of states could be removed along the
negative lowest-prolate direction, carrying $O(D)$ mass at the
exponentially soft cost $O(\varsigma_{\mathcal E}^{2}D^{2})$. However, this mode
is largest near $P=0$, where the vacuum profile vanishes quadratically,
so $\rho\geq0$ forbids precisely this cheap cancellation and forces a
parametrically larger residual.
The physical crossing therefore sits closer to $N=M_\mathds{1}+1$ than
the lower bound in \eqref{eq:rigorous-vacuum-transition-bounds}
suggests, and a more precise bound requires a more careful treatment of higher prolate functions.

\subsubsection*{Scaling and stiffness in the continuum potential}
 The preceding comparison concerns minima of the constraint potential. Let us now examine the parameter scaling associated with their crossing and the quadratic cost of deformations of the density of states. First, from \eqref{eq:rigorous-vacuum-transition-bounds}, the Cardy regime requires exponentially many states,
$N=e^{\lambda\mathcal E+o(\mathcal E)}$: at fixed $G$ this forces
$n_b=\epsilon N=GN/\mathcal E\to\infty$, exponentially in the cutoff. Second, we examine the quadratic cost of density deformations. Writing $g=\rho-\rho_L=\sum_n c_n\hat s_n$
in orthonormal prolate modes, the continuum potential takes
the diagonal form
\begin{equation}\label{eq:prolate-fluctuation-action}
 I_L[g]
 =
 \frac{1}{4a\epsilon}
 \sum_n
 \bigl(1-\lambda_n\bigr)^{2}c_n^{2}\,.
\end{equation}
The quadratic coefficient therefore defines a characteristic
scale for variations in each mode:
\begin{equation}\label{eq:prolate-gaussian-spread}
 \Delta c_n
 \sim
 \frac{\sqrt{2a\epsilon}}{1-\lambda_n}\,.
\end{equation}
Freezing the softest mode, that is requiring its spread \eqref{eq:prolate-gaussian-spread} to vanish along the nearly self-dual direction $\hat s_0\propto s_{\mathcal E}$, therefore requires:
\begin{equation}\label{eq:kappa-stiffness}
 \kappa_{\mathcal E}
 \equiv
 \frac{\varsigma_{\mathcal E}^{2}}{a\epsilon}
 =
 \frac{N\,\varsigma_{\mathcal E}^{2}}{t\,\epsilon}
 \longrightarrow
 \infty\,.
\end{equation}
In particular, since $\varsigma_{\mathcal E}^2\sim16\pi^2\mathcal E^2
e^{-16\pi\mathcal E^2}$, while $N$ is only exponential in $\mathcal E$,
this cannot hold at fixed 't~Hooft coupling $t=aN$: $t$ must be sent to
zero, faster than $e^{-16\pi\mathcal E^{2}}$ up to exponentially large
corrections. Therefore we adopt the limit
\begin{equation}\label{eq:hard-limit}
 N, \mathcal{E}\to\infty;\; \epsilon,a\to 0\quad\mathrm{st.}\quad k_\mathcal{E}\to\infty\quad\mathrm{and}\quad\lambda=\frac{\log N}{\mathcal{E}},\;G=\epsilon\mathcal{E}\;\;\mathrm{held\;fixed}.
\end{equation}
In this limit, the two configurations whose exchange we want to probe become
separated by a parametrically large potential barrier, and the full
partition function is dominated by the sector with the smaller residual:
\begin{equation}\label{eq:sector-laplace}
 \lim_{t\to0}
 \bigl[-4a\epsilon\log Z_{N_{vac}}\bigr]
 =
 \mathcal R_{N_{vac}}(N,\mathcal E)\,,
 \qquad
 \frac{Z_1}{Z_0}
 \sim
 \exp\!\left(
 -\frac{\mathcal R_1-\mathcal R_0}{4a\epsilon}
 \right)\,,
\end{equation}
where we denoted by $Z_{N_{vac}}$ the partition function in the occupation sector with $N_{vac}$ eigenvalues in the vacuum. In particular, for the cases $N_{vac}=0,1$ considered here, \eqref{eq:hard-limit} guarantees
\begin{equation}
 N_{vac}^{\rm dom}
 =
 \begin{cases}
  0,
  & 1\leq N<N_c(\mathcal E),
  \\[2mm]
  1,
  & N_c(\mathcal E)<N.
 \end{cases}
 \label{eq:dominant-vacuum-sector}
\end{equation}
Since $r_{\mathcal E}\to2$ when $\mathcal{E}\to\infty$, the polynomial difference between the two
bounds in
\eqref{eq:rigorous-vacuum-transition-bounds} does not affect their
exponential scaling, and we conclude that:
\begin{equation}
 \lim_{\mathcal E\to\infty}
 \frac{
  \log N_c(\mathcal E)
 }{
  \mathcal E
 }
 =
 2\pi Q.
 \label{eq:vacuum-transition-exponent}
\end{equation}
In particular, every polynomial scaling of $N(\mathcal E)$ lies in the
no-vacuum region, whereas the
vacuum is necessarily occupied by the time the Cardy matching value
$N=M_\mathds{1}+1$ is reached. Now, if we consider a dynamical model, in which one eigenvalue can flow below the
black-hole threshold, both sectors naturally appear in the partition function. In
the scaling limit~\eqref{eq:hard-limit} the sector with the smaller residual
dominates, which provides the mechanism for the onset of vacuum occupation in this
single-light-state model.

 \subsubsection{The picture for a generic light-state occupation}
We now generalize the analysis we performed above for the vacuum occupation to a single non-degenerate light-state with $P_\Delta=i\alpha_\Delta$,
$\alpha_\Delta^2+\Delta=(c-1)/24$ and $0<\alpha_\Delta<Q/2$. Its modular
image on the principal series and the associated matching count are
\begin{equation}\label{eq:generic-light-profile}
\begin{aligned}
     \rho_\Delta(P)
 =
 S_{P P_\Delta}(1)
 =
 2\sqrt2\cosh(4\pi\alpha_\Delta P)\,,
 \qquad
 M_\Delta(\mathcal E)
 \equiv
 \int_0^{\mathcal E}\rho_\Delta(P)\,\mathrm dP
 \sim
 \frac{\sqrt2}{4\pi\alpha_\Delta}
 e^{4\pi\alpha_\Delta\mathcal E}\,.
\end{aligned}
\end{equation}
Let
$\mathcal R_\Delta(N,\mathcal E)$ be the residual obtained from
\eqref{eq:exact-sector-residual} by the replacement
$N_{vac}\rho_0\to\rho_\Delta$. Repeating the arguments used for
\eqref{eq:no-vacuum-upper-bound} and \eqref{eq:positive-density-CS} on
$g_\Delta=\rho-\rho_\Delta$, we obtain

\begin{equation}\label{eq:generic-light-bounds}
 \frac{\varsigma_{\mathcal E}^{2}}{\mathcal E}
 \bigl(N-1-\ M_\Delta\bigr)^{2}\leq \mathcal R_\Delta(N,\mathcal E)
 \leq
 \varsigma_{\mathcal E}^{2}r_{\mathcal E}
 \bigl(N-1- M_\Delta\bigr)^{2}\,,
\end{equation}
 The important
common feature of the limits in \eqref{eq:generic-light-bounds} is that every one-light-state
sector is governed, up to the common polynomial uncertainty between $1/\mathcal E$ and
$r_{\mathcal E}$, by the same squared mass mismatch:
\begin{equation}\label{eq:generic-light-common-mismatch}
 \mathcal R_\Delta(N,\mathcal E)
 \sim
 \varsigma_{\mathcal E}^{2}
 \bigl(N- M_\Delta(\mathcal E)\bigr)^{2}\,.
\end{equation}
So, the distinctiveness of the light state enters, at leading
exponential order, only through $ M_\Delta(\mathcal E)$.
To make this concrete, consider two allowed light states with
\begin{equation}\label{eq:ordered-light-states}
 0<\alpha_1<\alpha_2\leq\frac{Q}{2}\,,
 \qquad
  M_{\alpha_1}
 <
  M_{\alpha_2}\,.
\end{equation}
For $ M_{\alpha_1}\leq N\leq M_{\alpha_2}$, the prolate
upper bound in the $\alpha_1$ sector and the spectral lower bound in the
$\alpha_2$ sector imply at least one pairwise crossing satisfying
\begin{equation}
 \frac{
   M_{\alpha_2}
  +
  \sqrt{\mathcal E r_{\mathcal E}}\,
   M_{\alpha_1}
 }{
  1+\sqrt{\mathcal E r_{\mathcal E}}
 }
 \leq
 N_c^{(1\to2)}(\mathcal E)
 \leq
  M_{\alpha_2}(\mathcal E)\,\implies\lim_{\mathcal E\to\infty}
 \frac{\log N_c^{(1\to2)}(\mathcal E)}{\mathcal E}
 =
 4\pi\alpha_2\,.
 \label{eq:generic-light-crossing-window}
\end{equation}
If, neglecting positivity, one assumes that the lowest prolate mode
gives the complete minimizer in both sectors, we report that the nominal crossing is
\begin{equation}
 N_{c,{\rm p}}^{(1\to2)}
 \equiv
 \frac{
   M_{\alpha_1}
  +
   M_{\alpha_2}
 }{2}.
 \label{eq:generic-light-relaxed-crossing}
\end{equation}
However, as already commented for the vacuum transition, below $ M_{\alpha_2}$, the $\alpha_2$ sector requires a
negative prolate correction, and the positivity constraint requires considering higher prolate corrections. So, analogously to \eqref{eq:rigorous-vacuum-transition-bounds}, the exact crossing point depends on the
low-momentum profile and is not fixed by the weaker
bounds above, however this does not affect the existence of the transition and the exponential result
\eqref{eq:generic-light-crossing-window}.

\begin{figure}[h]
    \centering
    \includegraphics[width=1.0\linewidth]{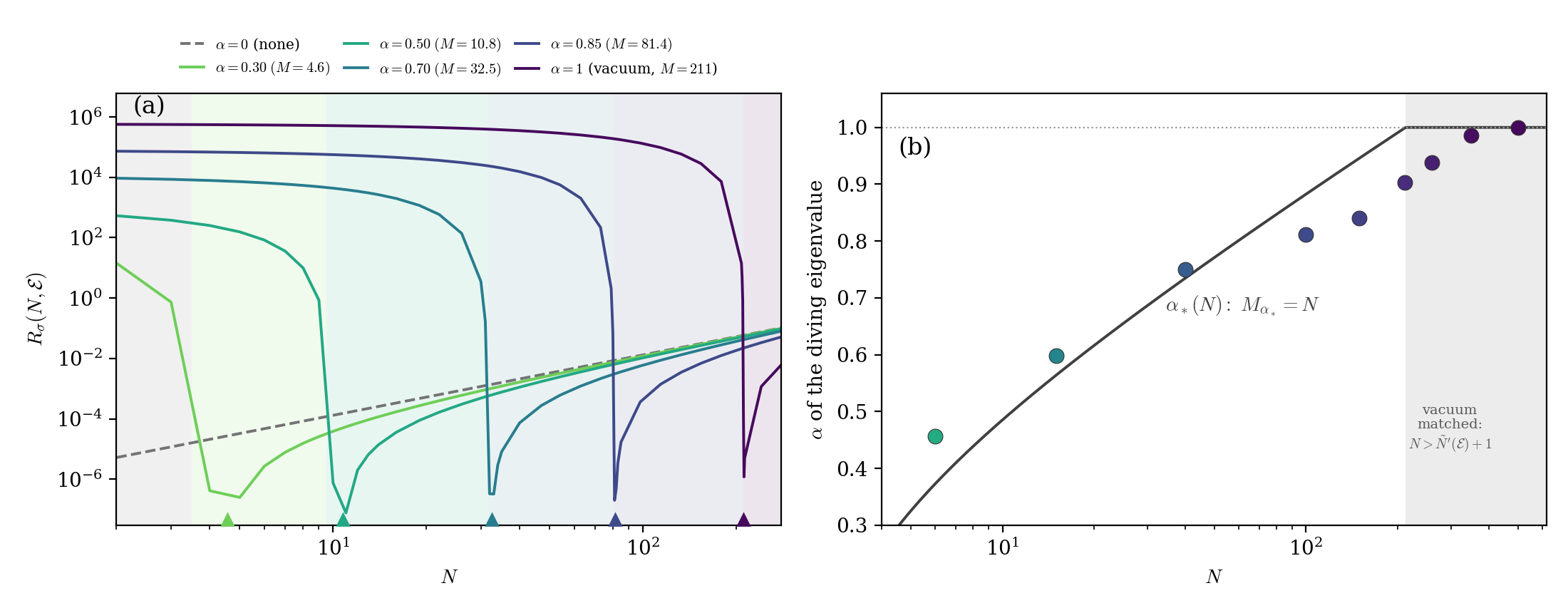}
    \caption{Competition of single-light--state configurations at $c=25$, $\mathcal{E}=0.6$, $\epsilon=0.01$ and $a=0.05$.
\textbf{(a)} Sector residuals $R_\sigma(N)$, defined analogously to \eqref{eq:exact-sector-residual} for a generic light state $\sigma\in\{\mathrm{none}, \alpha=\{0.30,0.50,0.70,0.85\}, \mathrm{vacuum}\}$, computed by convex quadratic programming (QP) optimization. The
least-residual sector (shaded) fixes the light content, the colored triangles give its
matching mass $ M_\sigma$. Notice that a purely heavy spectrum is preferred as long as its residual,
$R_0\sim(1-\lambda_0)^2N^2$, set by the leading near-null prolate mode
($1-\lambda_0\simeq8\times10^{-4}$), stays below that of lightest considered sector at $\alpha=0.30$. Notice that for each \emph{surplus} side $N>M_\sigma$ the prolate ansatz is approximately correct, while for a deficit $N<M_\sigma$ the positivity constraint forces the perturbation to $\rho_L$ onto steep directions of $(1-\mathbb{S}_\mathcal{E})$.
\textbf{(b)} Momentum $\alpha$ of the single diving eigenvalue, obtained from Monte Carlo initialized uniformly on the heavy-states. Cranking up $N$ makes the single eigenstate descend towards the vacuum as expected.
The measured $\alpha$ undershoot near
saturation the continuum matching profile: this freezing is a finite--parameter artifact which could be ameliorated by parallel tempering in
$a$.}
    \label{fig:light-state-transition}
\end{figure}

The picture emerging is therefore the following: entering an exponentially
populated regime first favors the occupation of a state below the
black-hole threshold, while increasing $\lambda$, the exponential growth rate of
$N$, favors light states of progressively larger imaginary momentum.
Indeed, since $M_\alpha(\mathcal E)$ is strictly increasing in
$\alpha$, if the light momentum is treated as a continuous parameter and
\begin{equation}
 N(\mathcal E) = e^{\lambda\mathcal E+o(\mathcal E)},
 \qquad 0<\lambda<2\pi Q,
 \label{eq:generic-light-exponential-N}
\end{equation}
there is a unique matching momentum $\alpha_{\Delta_*}$ solving
$M_{\Delta_*}=N-1$.
Cranking up $N$ therefore increases $\alpha_{\Delta_*}$ and lowers the
dimension of the preferred light state, moving it continuously away
from the black-hole threshold and toward the vacuum\footnote{for a discrete
light spectrum the continuous motion is replaced by a staircase of
pairwise transitions \eqref{eq:generic-light-crossing-window} through
states of decreasing dimension.}. This holds until we reach the maximal unitary exponent $\lambda=2\pi Q$, corresponding to the vacuum (see \cref{fig:light-state-transition} for a numerical confirmation). We recall that the vacuum character is degenerate, so its exact kernel differs from
the non-degenerate expression \eqref{eq:generic-light-profile} by the
familiar $h=1$ subtraction. However, the difference is exponentially subleading at
high momentum, and the leading occupation scale approaches the vacuum
result continuously.

We can perform a slight generalization of our analysis, still allowing a single occupied light state, but with arbitrary multiplicity. In this case, we can nicely summarize the result of this `flow to the vacuum' by treating its occupation 
$N_{vac}$ as a continuous order parameter. The minimization of the mass mismatch
$(N-N_{vac}M_{\mathds{1}})^2$ fixes
\begin{equation}\label{eq:continuous-N0}
 N_{vac}
 =
 \frac{N}{M_{\mathds{1}}}
 \simeq
 \sqrt{2}\,\pi Q\, N\, e^{-2 \pi Q {\cal E}}\,,
\end{equation}
which vanishes, tends to a constant, or diverges as
${\cal E}\to\infty$ according to whether the leading parametric scaling of $N$ grows slower than,
precisely as, or faster than $e^{2\pi Q\mathcal E}$: three ``phases''
distinguished by the vacuum occupation. In the last case no single
unitary light state can match the growth, and several light states must
be occupied, their modular images adding up, with each additional state
accounting for the next subleading exponential in $N(\mathcal E)$;
vacuum dominance of the leading Cardy behavior then requires an
HKS-type sparseness condition on the light spectrum
\cite{Hartman:2014oaa}. 

Allowing generic prescribed light states illustrates an
important limitation of the mass-matching criterion: say that $N=M_{\mathds{1}}$, then the mass mismatch is minimized by $N_{vac}=1$. However, another minimal configuration is reached by putting a large number of eigenvalues $N_\lambda=\lambda e^{2\pi \mathcal{E}(Q-\lambda/2\pi)}/(2\pi Q)$ in the light-state with exponent $\lambda$. This means that we have many degenerate configurations in the light-spectrum, and generically by multiplicity the one selected will have an exponential number of eigenvalues in a state close to the threshold. So this multi-light-state comparison seems to signal that the spectrum that we would like to associate to an holographic CFT is atypical, similarly to the discussion in \cite{Belin:2025qjm}. Whether such configurations are statistically favored depends on the measure on the light spectrum, which we do not specify here. A possibility is that introducing the other bootstrap constraints may degenerate some of these configurations or we may need to insert by hand a term akin to a chemical potential on light-states to typically localize to a Cardy phase. A systematic treatment of a multi-light-state
dynamical model, including the positivity-constrained locations of its
transitions, is left for future study.

\paragraph{Infinite-window self-dual densities.}
As a side remark, notice that as ${\cal E}\rightarrow \infty$ there are a large number of non-negative functions which are its own Fourier transform. Examples of such functions are $$\rho(P) = e^{-2\pi a P^2} + \frac{1}{\sqrt{a}} e^{-2\pi P^2/a}.$$ In that sense a chiral theory which is S-duality invariant is not particularly unique. However, if we start with a regulator ${\cal E}$ then there are no functions that transform into themselves, as we saw above that the relevant integral transform has operator norm less than one. The numerics with finite cutoff are therefore dominated by the eigenvalue closest to one, and the corresponding function is the $s_{00}$ we encountered before. In the large ${\cal E}$ limit, this function approaches $\rho(P)=e^{-2 \pi P^2}$. Hence, this is the distinguished density from a numerical point of view. It is amusing to compute the partition function of this "chiral Liouville" theory and one finds
\begin{equation}
	Z=Z_0 \frac{1}{\sqrt{1-i\tau}\, \eta(\tau)}
\end{equation} 
which is invariant under $\tau\rightarrow -1/\tau$ but obviously not under $\tau\rightarrow \tau+1$.

As another side remark, one can repeat the above considerations in the absence of Virasoro descendants. Now the starting point is 
\begin{equation}
	e^{-x/\beta} = \int_0^{\infty} \left( \delta(x') - \sqrt{\frac{x}{x'}} J_1(2\sqrt{ x x'}) \right) e^{-\beta x'}
\end{equation}
and there is also an expression for "light states"
\begin{equation}
	e^{+x/\beta} = \int_0^{\infty} \left( \delta(x') + \sqrt{\frac{x}{x'}} I_1(2\sqrt{ x x'}) \right) e^{-\beta x'}
\end{equation}
Most of the above observations carry through. There is a large number of self-dual non-negative functions in case the integral runs from $0$ to $\infty$, but once we introduce a cutoff ${\cal E}$ in $x'$ then the relevant kernel has operator norm less than $1$. Numerics would be dominated by the light spectrum plus the function with eigenvalue closest to one. That function appears to not be a known function and it appears to approach $e^{-x}$ as ${\cal E}\rightarrow \infty$. So this would be the spectrum obtained in the large ${\cal E}$ limit in the absence of a light spectrum.\\

\subsection{The in-bin mesoscopic density of states}\label{sec:mesoscopic_inbin}
Up to this point, we based our considerations mainly on the occupation numbers
$N_{P,\epsilon}$, without regard for their internal structure. This internal structure is the
content of the present section, which is organized in three main steps:
\begin{itemize}
    \item We first observe that freezing the occupations, as the limit \eqref{eq:limits} does,
    is equivalent to placing hard walls\footnote{In RMT a hard wall is an impenetrable boundary of the
eigenvalue domain: the Vandermonde repulsion pushes eigenvalues against it, so
they accumulate at the wall instead of thinning out gradually, as they do at a
soft edge where a confining potential balances the repulsion.} at the edges of every bin: the integration domain
    factorises into disjoint intervals, inside which the eigenvalues move in a constant
    potential, under the sole effect of the Vandermonde repulsion.
 
    \item We then posit an ansatz for the density inside a bin \eqref{eq:rho_ansatz}, guided by the hard-wall matrix
    models of \cite{Chekhov_2006}, in which each bin carries a hard wall at its lower edge and a soft
    edge in its interior, the latter generated by the repulsion of the more densely filled bin
    above it.
 
    \item Finally we fix the soft-edge positions from the occupation numbers themselves, through
    the filling-fraction condition \eqref{eq:epsk_integralCk}, and simplify the resulting elliptic
    integrals in a limit in which neighbouring cuts nearly pinch off. We verify in \cref{app:cut_pinching_consistency} that this limit is self-consistent within a high-energy window
    of parametrically large width in the UV cutoff $E$.
\end{itemize}

Let us consider the configurations discussed in \cref{sec:saddle_point}, where the bin occupation numbers are fixed by minimizing the potential. In this limit\footnote{Note that the relative scaling of level repulsion and coarse-graining scale is a technical assumption that allows us to make significant analytic progress. The Cardy matrix model, more generally, is well defined also for parameter regions where level repulsion and coarse scales are comparable.}, the Vandermonde repulsion manifests at smaller scales, and in this section we will discuss how it determines the eigenvalue distribution inside each bin.
Let us consider the partition function \eqref{eq:part_funct_cardy}, for simplicity, in the case where the only occupied light state is the vacuum and the majority of the contribution comes from the high-energy region.
In this section, we choose $N\sim M_{\mathds{1}}$ so that, in the limits of \cref{sec:vac_occ_gap}, $N_{P,\epsilon}$ is fixed $\forall P$ to $N_{P,\epsilon}=\epsilon \mathbb{S}_{P\mathds{1}}(\mathds{1})$ up to subleading corrections. 

Once the potential is minimized, the condition of fixing the occupation numbers, can be equivalently understood as imposing hard-wall constraints at the edges of each bin. This follows from \eqref{eq:part_funct_cardy}: in the strict-constraint limit $a\to0$ the Gaussian weight collapses onto the domain $N_{P,\epsilon}=\epsilon\,\mathbb{S}_{P\mathds{1}}(\mathds{1})\,,\;\forall P$, and, since $N_{P,\epsilon}$ merely counts the eigenvalues contained in the $P$-th bin, the limit confines each (ordered) eigenvalue to its own bin. This factorizes the integration domain into disjoint intervals separated by impenetrable walls. Inside these constrained bins, the eigenvalues move in a constant potential, under the sole effect of the Vandermonde repulsion. Random matrix theory with possible inclusion of hard walls has been studied in \cite{Chekhov_2006, chekhov2018topologicalrecursionhardedges}, in particular in the neighborhood of these hard-edges one expects an eigenvalue density behavior $\rho(x)\sim x^{-1/2}$, in contrast with the soft-edge case $\rho(x)\sim x^{1/2}$. Moreover, \cite{Chekhov_2006} contains an ansatz for the resolvent $y(x)$ of a matrix model with hard walls in positions $b_j$ and soft edges in $a_i$, which in the case of constant potential at hand reduces to
\begin{equation}\label{eq:wall_resolvent_ansatz}
    y(x)=\frac{-M(x)}{2}\Tilde{y}(x), \quad\mathrm{where}\quad\Tilde{y}(x)=\prod_{i,\,j}^{}\sqrt{\frac{x-a_i}{x-b_j}}\,,
\end{equation}
and where $M(x)$ is determined by the potential as follows.
If we call respectively $r$ and $s$ the number of walls and soft edges, and we have a potential $V(x)$, then $M(x)$ is given by
\begin{equation}\label{eq:wall_M_def}
    M(x)=\oint_{\mathcal{C}_\infty}\frac{d\xi}{2\pi i}\frac{V'(\xi)+P_{\frac{r-s}{2}}(\xi)-P_{\frac{r-s}{2}}(x)}{(\xi-x)\Tilde{y}(\xi)}\,,
\end{equation}
where $\mathcal{C}_{\infty}$ is a complex integration contour around infinity, and $P_{\frac{r-s}{2}}(x)$ is a polynomial of degree $\frac{r-s}{2}$. Now, we specify to the configuration considered in this section where we have a constant $V(x)$ and, as we will momentarily check for consistency, $r=s+2$. In this case, notice that we have the asymptotic behavior $\Tilde{y}(x)\sim \frac{1}{x}$ and we can set $P_1(x)=\alpha x+\gamma$, so that we obtain a constant $M(x)=\mathcal{M}_0$. The leading coefficient $\alpha$ is determined by ensuring the asymptotic behavior of the resolvent $y(x)\sim \frac{1}{x}$ \cite{Chekhov_2006}, but in the case where $M(x)$ is constant we can use this condition to find directly $\mathcal{M}_0=-2$. At this point, we have $y(x)=\Tilde{y}(x)$, and we can obtain the normalized density of eigenvalues as
\begin{equation}\label{eq:resolvent_to_eigenv}
    \rho(x)=-\frac{1}{\pi}\Im\Tilde{y}(x)\,.
\end{equation}
Even assuming a constant potential and $r=s+2$, it is still cumbersome to make progress without a further simplifying ansatz. In order to justify such an ansatz, we want to understand further how the soft edges arise from the Vandermonde repulsion.
Let us assume that we have a single small bin $[-a_0,a_0]$. Then the Vandermonde-driven eigenvalue density inside will be $\rho(x)\sim (\sqrt{a_0-x}\sqrt{x+a_0})^{-1}$, consistent with the hard walls condition in $\pm a_0$. However, if we add another bin, say $[a_0,2a_0]$, with the same conditions, it is possible that the high eigenvalue density near $a_0$, by Vandermonde repulsion, creates a gap in the spectrum, and hence a soft edge in whose neighborhood we have a square root behavior. In full generality, we should determine the eigenvalue density by finding the positions $x_{edge}$ of two edges in each bin (see, for example, \cite{eynard2018randommatrices, Livan_2018}): if $|x_{\rm edge}|<a_0$ then we have the soft-edge eigenvalue density feature $\sim\sqrt{x-x_{edge}}$ near $x_{edge}$, while if $|x_{\rm edge}|>a_0$ the hard wall characteristic behavior $\sim (x-a_0)^{-1/2}$ near $a_0$ is imposed instead. This procedure in principle works even if we allow for a residual potential inside the bins, but in practice becomes cumbersome to implement. For the case at hand, we will instead adopt a simplifying ansatz.

\begin{figure}
\centering
        \includegraphics[width=\textwidth]{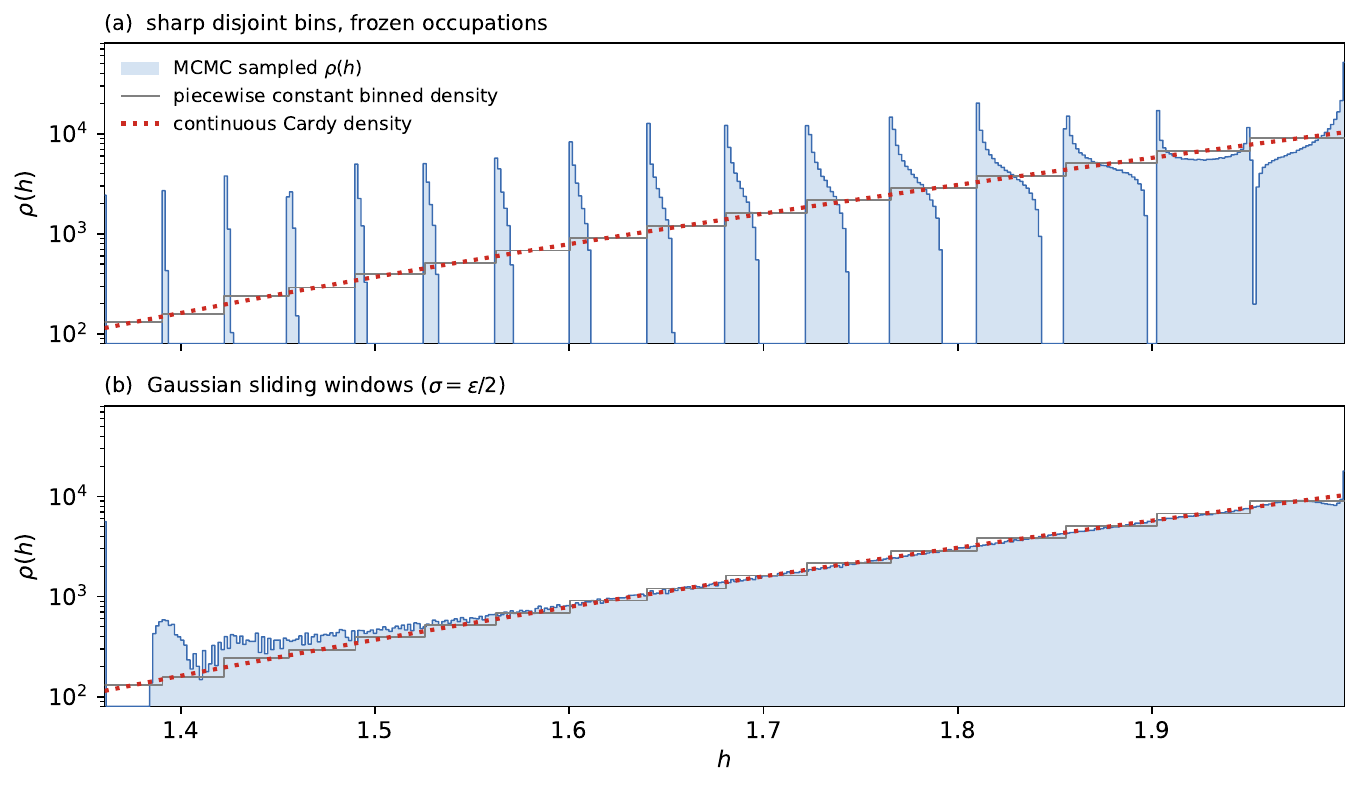}
    \caption{The in-bin mesoscopic density $\rho(h)$ (blue; MCMC with eigenvalues interacting only through the Vandermonde repulsion at fixed Cardy occupations. Simulation run at $c=25$ in the momentum window $P\in[0.6,1.0]$ and at bin/window width $\epsilon=0.025$ and $N\simeq1643$ eigenvalues. Note the log scale). The red dotted curve is the continuous Cardy density $\rho(h)=A\,\sinh^2(2\pi P)/(2P)$ and the gray staircase the $\epsilon$-binned density. \emph{Top (a):} sharp disjoint bins with frozen occupations, showing the imprint of the hard-wall regularization of \cref{sec:mesoscopic_inbin}, which prevents eigenvalues from crossing the bin edges. The Vandermonde repulsion then piles them against the walls, producing the comb of hard-wall ($\rho\sim x^{-1/2}$) spikes and intervening soft edges of the ansatz \eqref{eq:rho_ansatz}. \emph{Bottom (b):} the same occupations with the sharp bins replaced by Gaussian sliding windows ($\sigma=\epsilon/2$) averaged over their centres. This is the bin-centre averaging discussed in \cref{sec:mesoscopic_inbin}, softening the constraint into an analytic pair potential. The spikes vanish and $\rho(h)$ follows the smooth Cardy profile at the same macroscopic occupations. This illustrates our statement in the main text that spikes are a scheme artifact of the sharp binning, not a physical feature of the mesoscopic spectrum.}
    \label{fig:inbin_density_ansatz}
\end{figure}
 Notice that the occupation numbers in the high-momentum region are $N_{P,\epsilon}=\epsilon\mathbb{S}_{P\mathds{1}}(\mathds{1})>N_{P',\epsilon}=\epsilon\mathbb{S}_{P'\mathds{1}}(\mathds{1})$ when $P>P'$. So, we assume, for each bin $[i_k,i_{k+1}]$, that its left boundary $b_k=i_k$ is a hard wall, while a soft edge is created in an unknown position $a_k<i_{k+1}$ by the eigenvalue repulsion coming from the hard wall region of the next bin in $i_{k+1}$. We expect that this assumption of fixing the left edges to the hard walls of the bins holds if the occupation number of the bins are growing sufficiently fast with the energies, as it happens in the high-momentum region, where the density is exponentially increasing, and the number of eigenvalues in each bin is parametrically scaled to infinity (see \cref{fig:inbin_density_ansatz}). This brings us to an eigenvalue density ansatz:
\begin{equation}\label{eq:rho_ansatz}
\rho(x)=\frac{1}{\pi}\frac{\sqrt{x-a_0}\dots\sqrt{x-a_{n-1}}}{\sqrt{x-b_0}\dots\sqrt{x-b_n}\sqrt{b_{n+1}-x}},
\end{equation}
 Notice that, for the last bin, we have no repulsion coming from the right, so we have two hard walls instead of a wall and a soft edge, and consistently $r=s+2$. The center of the $k$-th bin in energy variables $h_k=(i_k+i_{k+1})/2$ is related to the center $P_k$, appearing for example in the S-kernels, via \eqref{eq:P_to_h}. Similarly, the two extremes are $b_0=(c-1)/24$ and $b_{n+1}=E$, and the latter is related to $\mathcal{E}$ using again \eqref{eq:P_to_h}. This is considerably simpler than the general case, because, we are halving the incognita associated to the cuts determining the support of the eigenvalue density. At this point, the unknown position of the right soft-edges $a_k$ are fixed $\forall k$ by the occupation fraction $\epsilon_k$ via the condition
\begin{equation}\label{eq:ak_integral_def}
    \int_{b_k}^{a_k}\rho(x)dx=\frac{N_{P_k,\epsilon}}{N}\equiv \epsilon_k=\frac{\epsilon \mathbb{S}_{P_k\mathds{1}}(\mathds{1})}{N},
\end{equation}
where $P_k=(2k+1)\epsilon/2$ is the center of the $k$-th bin, connected to the energy variables appearing here via \eqref{eq:P_to_h}. 

Heuristically, from \cref{fig:inbin_density_ansatz}, we can understand that there are two regions where the general behavior of the eigenvalue density \eqref{eq:rho_ansatz} simplifies: a typically low-energy one where the bins are approximately empty, and one at higher energies where they are approximately full. The empty-bins approximation trivially reproduces the single-interval support results, so here, as well as in \cref{sec:meso_phys}, we will focus on the perhaps more interesting case of the high-energy region.
Now, we assume that $b_k-a_{k-1}\equiv d_{k-1}\ll\delta_k\equiv\delta h(P_k,\epsilon)$, where we recall that $\delta h(P_k,\epsilon)$ is related to $\epsilon$ via \eqref{eq:delta_eps}\footnote{notice that, in the high-energy region, all widths $\delta_k$ are parametrically equivalent to $\delta\equiv 2G$.}. This is the condition that the endpoints of the density cuts approximately pinch off, whose consistency we will need to check after solving \eqref{eq:ak_integral_def}. In this approximation, we notice that if $x$ belongs to the $k$-th bin, then $\forall j\neq k,\,k-1$ we have $x-b_j\gg d_j$, so that
\begin{equation}\label{eq:dj_approx_1}
    \frac{\sqrt{x-a_j}}{\sqrt{x-{b_{j+1}}}}\approx\biggr(1+\frac{d_j}{2(x-b_{j+1})}\biggr)\approx 1.
\end{equation}
Then, using the expression above, whenever $x$ belongs to the $k$-th microcanonical window, we can approximate the density to
\begin{equation}\label{eq:rho_inbin_approx}
\begin{aligned}
    &\rho(x)= \frac{1}{\pi\sqrt{x-b_0}\sqrt{b_{n+1}-x}}\prod_{j=0}^{n-1}\frac{\sqrt{x-a_j}}{\sqrt{x-{b_{j+1}}}}\approx C_k\frac{\sqrt{x-a_{k-1}}}{\sqrt{x-{b_{k}}}}\frac{\sqrt{x-a_k}}{\sqrt{x-{b_{k+1}}}},
\end{aligned}
\end{equation}
where we defined $C_k\equiv (\pi \sqrt{x-b_0}\sqrt{b_{n+1}-x})^{-1}$. The expression above holds when $x$ is in the $k$-th bin, and assuming this belongs to the high-region bulk of the spectrum, here we can approximate $C_k$ as a constant 
\begin{equation}\label{eq:Ck_binextrema}
    C_k\approx \frac{1}{\pi}\frac{1}{\sqrt{h_k-(c-1)/24)}\sqrt{E-h_k}}\quad\mathrm{and}\quad h_k=P_k^2+(c-1)/24,
\end{equation}
which we notice to be of the order $\mathcal{O}(E^{-1})$ in the center of the spectrum. In \eqref{eq:rho_inbin_approx}, we used that \eqref{eq:dj_approx_1} holds $\forall x,j$ such that $x-b_j\gg d_j$, in particular we cannot use the approximation only for the terms corresponding to the $k$-th bin where $x$ is localized and its closest soft and hard walls. Then, we notice that \eqref{eq:rho_inbin_approx} is a sort of `first neighbor approximation', where eigenvalues in a bin approximately feel only the effect coming from the closest walls/edges terms.

Now, we proceed to solve integral for the $k$-th filling fractions \eqref{eq:ak_integral_def} in the cut-pinching approximation. Using the approximate expression in the $k$-th bin for the eigenvalue density \eqref{eq:rho_inbin_approx}, we have that \eqref{eq:ak_integral_def} becomes:
\begin{equation}\label{eq:epsk_integralCk}
    \int_{b_k}^{a_k}C_k\frac{\sqrt{x-a_{k-1}}}{\sqrt{x-{b_{k}}}}\frac{\sqrt{x-a_k}}{\sqrt{x-{b_{k+1}}}}dx= \epsilon_k.
\end{equation}
As discussed in detail in \cref{app:elliptic_int}, in the pinching limit, that is for small $d_{k}/\delta_{k}$ and $d_{k-1}/\delta_{k}$, we can obtain the following expression for the elliptic integral above:
\begin{equation}\label{eq:dk_expression}
\begin{aligned}
\frac{\epsilon_k}{C_k}\approx &\,\delta_k\biggr(1-\frac{d_k}{2\delta_k}\log\left(\frac{\delta_k}{d_k}\right)+\frac{d_{k-1}}{2\delta_k}\log\left(\frac{\delta_k}{d_{k-1}}\right)\biggr)
\end{aligned}
\end{equation}
We can reorder the above equation as:
\begin{equation}\label{eq:dk_expression_reorder}
\begin{aligned}
\frac{d_{k-1}}{\delta_k}\log\left(\frac{\delta_k}{d_{k-1}}\right)\approx2\biggr(\frac{\epsilon_k}{\delta_k C_k}-1+\frac{d_{k}}{2\delta_k}\log\left(\frac{\delta_k}{d_{k}}\right)\biggr)\equiv R_k,
\end{aligned}
\end{equation}
in particular we notice that if we wish $d_{k-1}/\delta_k$ to be a small real number, then $R_k$ defined above needs to be small and positive.
From \eqref{eq:dk_expression_reorder}, we can find $d_{k-1}$ as a function of the occupation fraction $\epsilon_k$ and the pinching parameter $d_{k}/\delta_k$ of the previous bin:
\begin{equation}\label{eq:dk_sol_approx}
    \begin{aligned}
        \frac{d_{k-1}}{\delta_k}\approx \frac{-R_k}{W_{-1}\left(-R_k\right)}\approx -\frac{R_k}{\log\left(R_k\right)},
    \end{aligned}
\end{equation}
where $W_{j}(x)$ is the Lambert W-function in the $j=-1$ branch, and we expanded it for $R_k\ll1$. Notice that we chose the $j=-1$ branch branch precisely because we are searching for a real small $d_{k-1}/\delta_k$ that corresponds to a small and positive argument $R_k$ of the function. We remark that \eqref{eq:dk_sol_approx} takes the form of a descending recursion relation, that given all $d_j$ for $j\geq k$ determines $d_{k-1}$. At this point, we have solved \eqref{eq:ak_integral_def}. However, it remains possible that there is no region of the spectrum in which the approximations used above are valid. In \cref{app:cut_pinching_consistency}, we perform the corresponding consistency check and show that the cut-pinching approximation is valid, at least within a parametrically large high-energy window whose characteristic width scales as $E^{1/4}$. Therefore, the picture we have in mind is the following. We initialize the descending recursion \eqref{eq:dk_sol_approx} with the pinching parameter of the right-most bin inside the consistency region. Because this region is parametrically large, bins deep in its interior are insensitive, up to parametrically small corrections, to the physics outside the window. Consequently, provided the initial datum of the recursion is small, so are the pinching parameters it recursively determines throughout the interior, where the cut-pinching approximation is then self-consistent.\\

Before concluding the section, we comment on the relation between the regularization procedure introduced in \cref{sec:cardy_matrix} and the eigenvalue density inside the bins. In particular, notice that minimizing the potential \eqref{eq:cardy_potential} fixes the global structure to the Cardy profile, however the Vandermonde-driven physics of the bins pertains to a smaller mesoscopic scale controlled by $\epsilon$, which, by the prescriptions of \cref{sec:bins_width}, is such that $e^{-S}\ll \epsilon\ll 1$. Hence $\epsilon$ is small on macroscopic scales, but still parametrically larger than the microscopic level spacing $1/N\sim e^{-S}$, the scale at which individual eigenvalues become resolvable. It is in this sense that we refer to $\epsilon$ in the chosen scaling as mesoscopic.

Notice that the density of states in the regime where we have separation of scales between potential and Vandermonde strongly depends on the bins' positions, introduced upon regularization of the measure on the distributions entering the modular constraint. If we really wanted to build a `maximal ignorance' ensemble then the only physical information we have to inject in our model is the bin-regularized modular invariance. In particular then, we do not expect the dependence of the density on the bin position to be physically meaningful, but rather a completion of our model at mesoscopic scales, which is consistently provided by the random matrix behavior. A possibility of eliminating this dependence a posteriori, is performing a microcanonical-style averaging over the bins' center positions.
Such an averaging may be performed either before or after an observable is computed (for example via topological recursion), and the two
prescriptions are not equivalent. The first possibility is to average first, at the level of the spectrum: the
entire dependence on the bin positions enters through the mesoscopic density \eqref{eq:rho_ansatz},
so averaging there removes it everywhere, and any observable computed afterwards is free of any regularization imprints.
The alternative is to perform  computations in each regularization scheme and average the results afterwards.
The discrepancy between these two possibilities
measures how much of an answer is fixed by modular invariance and how much by the choice of
regulator, and whether it should be discarded as scheme noise or retained as some physical feature, as it would be if
one regarded the coarse-graining as itself part of the ensemble. We leave a more thorough study of these averaging procedures for the future.
In what follows, we adopt mainly the first prescription throughout.

At the level of the eigenvalue density, we can imagine that averaging over bins' centers effectively removes the mesoscopic features. Indeed, in the cut-pinching approximation,
the only dependence on the bin center is through $d_k$, hence through the occupation fraction
$\epsilon_k$. However, \eqref{eq:eps_consistency_subexp_acc} already demands that the crossing kernel be approximately
constant across a bin, so that, by construction, $\epsilon_k$, and thus $d_k$, depend very weakly on where the bin is centered. Therefore, to
first approximation, the averaging interpolates smoothly between the occupation
fractions, following the Cardy profile and erasing the scale $\epsilon$. We confirm this numerically, more generally than the cut-pinching limit, in \cref{fig:inbin_density_ansatz}.

\section{The scale of level repulsion}\label{sec:top_recursion}
In this section, we point out some peculiarities of the matrix theory we are describing. The origin of these special features lies in the potential \eqref{eq:cardy_potential} which depends on occupation numbers only and is indifferent to the internal eigenvalue configuration within the bins. This fact, together with the scale separation between potential and Vandermonde implied by the limits introduced in \cref{sec:saddle_point}, was used in \cref{sec:mesoscopic_inbin} to obtain the `mesoscopic density' \eqref{eq:rho_ansatz}. If we imagine being unable to resolve momentum scales of $\mathcal{O}(\epsilon)$, the density of states will be given by a piecewise constant approximation of Cardy, which will look smooth at the macroscopic scale. However, this profile is obtained simply via minimization of the potential \eqref{eq:cardy_potential}, and consistently at these scales the Vandermonde contribution is parametrically suppressed \eqref{eq:saddle_point_eq}. So, macroscopically it looks like the Cardy RMT is just trying to satisfy the S-modular invariance part of the bootstrap constraints, and, in this sense, the model is not a random matrix theory at all, as it does not (yet) manifest level repulsion. As the reader may already have surmised, the RMT nature becomes manifest at scales $\lessapprox\mathcal{O}(\epsilon)$, when the internal bin structure is resolved and determined by the Vandermonde repulsion between eigenvalues. Let us emphasize the logic further: the repulsion itself is of course built into the measure \eqref{eq:part_funct_cardy}, and is in this sense an input. What the model determines is the scale at which it becomes operative. Heuristically speaking, we attempt to implement certain properties of a CFT into random matrix theory, but any putative chaotic CFT is not expected to be described by RMT before its Thouless time. Thus, it is natural to identify the energy-width of the bin to the inverse of the Thouless time, the scale at which the underlying theory is expected to show the features of a RMT.

The following sections are dedicated to placing this heuristic discussion on more solid ground. In \cref{sec:macro_phys}, we will compute the leading disconnected components of correlation functions at macroscopic scales. On the other hand, \cref{sec:meso_phys} uses topological recursion to derive the connected components which become important at late times. In particular, we will focus on the computation of the connected spectral form factor, which only requires the initial data of the recursion: we compute the Thouless time, defined at the onset of its linear ramp behavior, and confirm it is connected to the energy-width of the bin, as we anticipated. The recursion itself, which we describe in \cref{app:g=1_top_rec_details}, becomes necessary only for higher correlators and higher-genus corrections. What the elliptic analysis in \cref{sec:meso_phys} actually provides is the correct $\omega_{0,2}$: it derives the effective local one-cut description from the multi-cut mesoscopic curve at frozen filling fractions, with corrections controlled by the pinching parameters.\\

\subsection{The macroscopic physics}\label{sec:macro_phys}

As discussed in \cref{sec:saddle_point}, at momentum scales $\gg\epsilon$, we can assume that the physics of the model is approximately determined by the minimization of the potential \cref{eq:cardy_potential}, without considering the Vandermonde repulsion. A strictly continuous density implies the absence of the plateau in the spectral form factor, and the absence of level correlations translates into the absence of the ramp in that same quantity. As a direct consequence of these two facts, we expect the spectral form factor to decay at large times\footnote{This statement refers to the contribution of the heavy continuum above the black-hole threshold, which is what we compute below. In the Cardy phase, the vacuum is also occupied and, being an isolated level, it contributes to $|\langle Z\rangle|^2$ with a bounded, quasi-periodic and non-decaying term
$|\chi_{\mathds{1}}\bigl(\frac{t+i\beta}{2\pi}\bigr)|^2$. This gives a trivial `plateau' of a single discrete state, together with vacuum-continuum cross
terms decaying more slowly than \eqref{eq:sff_disc_latetimes}. Since these
pieces carry no information about the statistics of the heavy spectrum, we
omit them in this section.}. In this section we verify these two statements and, in doing so, we determine the exact decay rate.

We start by computing the resolvent assuming the high-energy Cardy behavior \eqref{eq:cardy_formula} in the full spectrum\footnote{notice that, as per the definition \eqref{eq:cardy_formula}, here $\rho_{\rm Cardy}(h)$ is just the primaries' momenta density expressed in energy variables. In order to get the proper eigenvalue density we would have to reabsorb in the definition of $\rho_{\rm Cardy}$ the Jacobian factor for the change of variables.},
\begin{equation}\label{eq:resolvent_def}
    W(z)=\int_{(c-1)/24}^E\frac{1}{2N}\frac{\rho_{\rm Cardy}(h)}{(z-h)\sqrt{h-(c-1)/24}}dh\approx\int_{0}^\mathcal{E}\frac{1}{N}\frac{\sqrt{2}e^{2\pi Q P}}{l^2-P^2}dP,
\end{equation}
where we (re)-defined the variables
$$P^2+(c-1)/24=h\,,\quad l(z)=\sqrt{z-(c-1)/24}$$
and we recall that $\mathcal{E}=\sqrt{E-(c-1)/24}$. We can express the above integral as
\begin{equation}\label{eq:resolvent_macro}
\begin{aligned}
    W(l(z))=&-\frac{\sqrt{2}e^{2\pi Q l}}{2lN}\biggr(\mathrm{Ei}(2\pi Q(\mathcal{E}-l))-\mathrm{Ei}(-2\pi Q l)\biggr)\\+&\frac{\sqrt{2}e^{-2\pi Q l}}{2lN}\biggr(\mathrm{Ei}(2\pi Q(\mathcal{E}+l))-\mathrm{Ei}(2\pi Q l)\biggr),
\end{aligned}
\end{equation}
where $\mathrm{Ei}(x)=\int_{-\infty}^x dt \;e^t/t $ is the exponential integral. Using its large-$x$ behavior $\mathrm{Ei}(x)\approx e^x/x$, and remembering that to be in the Cardy phase we need $N\approx M_{\mathds{1}}$, given by \eqref{eq:ntildeprime_expression}, we can obtain the large-$z$ asymptotics of the resolvent
\begin{equation}\label{eq:resolvent_approxed_macro}
     W(z)\approx\frac{\sqrt{2}e^{2\pi Q \mathcal{E}}}{2lM_{\mathds{1}}(2\pi Q)}\biggr(\frac{1}{\mathcal{E}+l}+\frac{1}{l-\mathcal{E}}\biggr)+\dots\approx_{z\to\infty}\frac{1}{z}+\mathcal{O}(z^{-2})\,.
\end{equation}
We notice that it has the expected $1/z$ asymptotic behavior.

Now, we proceed to compute the spectral form factor, which at the scales considered in this section, is not expected to show signs of the tell-tale chaotic RMT features, such as level repulsion. The partition function can be expressed as \eqref{eq:Z_characters}, and under the assumption that the mean eigenvalue density is given everywhere by Cardy \eqref{eq:cardy_formula} we obtain
\begin{equation}\label{eq:<part_funct>_def}
    \langle Z(\tau)\rangle =\int dh \frac{\rho_{\rm Cardy}(h)}{2\sqrt{h-(c-1)/24}}\frac{e^{2\pi i \tau (h-(c-1)/24)}}{\eta(\tau)}.
\end{equation}
We set $\tau=(t+i \beta)/(2\pi)$, and in particular we notice that when $\beta>0$ the high-energy contribution will be exponentially suppressed. As a consequence, \eqref{eq:<part_funct>_def} also depends on the low-energy spectrum where, in general, the vacuum character doesn't strictly dominate as in \eqref{eq:high_energy_potential_term_approx}, and we can expect deviations from the Cardy behavior. Here, we perform our computations assuming Cardy behavior \eqref{eq:cardy_formula} down to the edge of the spectrum, and ignore this complication. This can be made more rigorous if we also consider the limit $c\to\infty$ in order to extend the Cardy regime \cite{Hartman:2014oaa}.

From the definition of the disconnected spectral form factor we have
\begin{equation}\label{eq:sff_disc_def}
\begin{aligned}
      &SFF_{\rm disc}(\beta,t)=|Z(\beta-it)|^2=\\&=\frac{4}{\bigr|\eta\bigr(\frac{t+i \beta}{2\pi}\bigr)\bigr|^2}\biggr|\int_{\frac{c-1}{24}}^\infty dh \frac{\sqrt{2}\sinh{2\pi b \sqrt{h-\frac{c-1}{24}}}\sinh{2\pi \sqrt{h-\frac{c-1}{24}}/b }}{\sqrt{h-(c-1)/24}}e^{-(\beta-it)(h-(c-1)/24)}\biggr|^2.
\end{aligned}
\end{equation}
The integral appearing inside the absolute value of the above expression is
\begin{equation}\label{eq:sff_disc_integral}
\begin{aligned}
    \int_{\frac{c-1}{24}}^\infty dh& \frac{\sqrt{2}\sinh{2\pi b \sqrt{h-\frac{c-1}{24}}}\sinh{2\pi \sqrt{h-\frac{c-1}{24}}/b }}{\sqrt{h-(c-1)/24}}e^{-(\beta-it)(h-(c-1)/24)}=\\&=\sqrt{2}\int_0^\infty du e^{-(\beta-it)u^2}(\cosh(2\pi (b+1/b)u)-\cosh(2\pi (b-1/b)u)),
\end{aligned}
\end{equation}
where we changed variables to $u=\sqrt{h-(c-1)/24}$ and used the product formula for hyperbolic sines. We have the following integral result
\begin{equation}
    \int_0^\infty du e^{-su^2} \cosh{ k u}=\frac{e^{k^2/4s}\sqrt{\pi}}{2\sqrt{s}},
\end{equation}
so that if we insert the above expression in \eqref{eq:sff_disc_integral} we obtain:
\begin{equation}
\begin{aligned}
    \int_{\frac{c-1}{24}}^\infty dh& \frac{\sqrt{2}\sinh{2\pi b \sqrt{h-\frac{c-1}{24}}}\sinh{2\pi \sqrt{h-\frac{c-1}{24}}/b }}{\sqrt{h-(c-1)/24}}e^{-(\beta-it)(h-(c-1)/24)}=\\&=\sqrt{\frac{\pi}{2(\beta-it)}}\left(e^{\frac{\pi^2(b+1/b)^2}{\beta-it}}-e^{\frac{\pi^2(b-1/b)^2}{\beta-it}}\right)\approx_{t\to\infty}2\sqrt{2} \pi^{5/2}(\beta-it)^{-3/2}.
\end{aligned}
\end{equation}
In particular, by inserting the above expression in \eqref{eq:sff_disc_def}, we find that at late times the disconnected spectral form factor behaves as
\begin{equation}\label{eq:sff_disc_latetimes}
\begin{aligned}
      SFF_{disc}(\beta,t)\approx_{t\to\infty}\frac{32\pi^5}{\bigr|\eta\bigr(\frac{t+i \beta}{2\pi}\bigr)\bigr|^2}\frac{1}{(\beta^2+t^2)^{3/2}}
\end{aligned}
\end{equation}
The function $\eta(z)$ satisfies $|\eta(z)|=|\eta(z+1)|$, so that the first factor $1/\bigr|\eta\bigr(\frac{t+i \beta}{2\pi}\bigr)\bigr|^2$ in \eqref{eq:sff_disc_latetimes} is a bounded function with periodicity $t\to t+2\pi$, and, at late times such that $t\gg\beta$, the disconnected spectral form factor decays as $\propto t^{-3}$. Let us note that this decay is typical for a continuous density with square-root edge, and indeed, written in the more physical $h$ variables, as in Eq. \eqref{eq:sff_disc_integral}, the Cardy density starts with a square-root edge, which in our Cardy RMT sits at the black-hole threshold. For completeness, we remark that here we considered the spectral form factor built from the partition function including also descendants states, accounted for by the aforementioned $\eta$-function. In the following section, we will instead focus directly on the one containing only primary states, whose disconnected component then only showcases the $t^{-3}$ time-decay without the bounded $\eta$-dependence.

Let us now explain why, at the macroscopic resolution of this subsection the leading part of the spectral form factor is captured by its disconnected component. Consider $\langle Z(\tau)Z(\Bar{\tau})\rangle$ in the Cardy RMT. In the limits discussed in \cref{sec:vac_occ_gap}, the potential \eqref{eq:cardy_potential} becomes a parametrically deep well which freezes the occupation number of every bin to its Cardy value. The only remaining freedom is the arrangement of the eigenvalues within each bin, governed by the Vandermonde, and this affects correlations only at momentum separations of $\mathcal{O}(\epsilon)$. Probed at any coarser resolution, the density $\rho(h)$ is therefore effectively a fixed, non-fluctuating function, specifically the coarse-grained Cardy profile. This means that $\langle\rho(h)\rho(h')\rangle\approx\langle\rho(h)\rangle\langle\rho(h')\rangle$,
so that $\langle Z(\tau)Z(\bar\tau)\rangle$ reduces to the disconnected piece
computed above, decaying as $t^{-3}$, while the diagonal self-correlation of each
eigenvalue only contributes the time-independent constant
$\langle Z(2\beta)\rangle$. At this resolution the spectral form factor thus decays as $\propto t^{-3}$ and never develops a ramp. The missing connected correlations reside entirely in the in-bin structure. As we show in the next subsection, bin-scale separated eigenvalues are correlated by the Vandermonde repulsion, and it is precisely these correlations that restore the characteristic random matrix ramp at times $t\gtrsim 1/G$.

\subsection{The mesoscopic RMT physics}\label{sec:meso_phys}
In this section, we examine the physics of the Cardy matrix model at momenta mesoscopic scales of $\mathcal{O}(\epsilon)$, at which we are able to discern the internal structure of the bins, determined by the Vandermonde repulsion as discussed in \cref{sec:mesoscopic_inbin}, as well as its interaction with the average Cardy density that now gets modulated by these effects. At the scale of the mean level spacing, the Cardy RMT exhibits the level repulsion built in from the Vandermonde factor. At mesoscopic scales, remarkably, the a priori intractable spectral curve of parametrically large genus underlying the mesoscopic density reduces, in a controlled cut-pinching limit derived from the saddle point itself, to a local genus-one problem solvable with standard topological recursion technology \cite{Eynard:2007kz, Bonnet_2000,Mari_o_2009}. This gives us in-principle access to the complete smoothed correlation structure beyond the two-point function. The genuinely microscopic features, namely the level-spacing oscillations and the plateau of the spectral form factor, are instead nonperturbative in the genus expansion, as emphasized in \cite{Altland:2020ccq}, and in the present model they are exhibited by the numerics of \cref{sec:numerics}.

In this section, our goal will be to describe, using a topological recursion procedure \cite{Eynard:2007kz}, how to compute the first resolvent correlators and the spectral form factor, where the small energy difference limit translates in long time effects.

The topological recursion is a procedure that, given a certain spectral curve as initial data, computes all correlators of the resolvent to arbitrary orders in a genus expansion. A spectral curve $\mathcal{S}=(\Sigma, w_{0,1}, B)$ is determined by a (compact) Riemann surface, a meromorphic 1-form $w_{0,1}$ and the Bergman kernel $B$, which
also goes under the name of fundamental differential of the second kind on $\Sigma$. Determining $w_{0,1}$ is equivalent to specifying two functions $y,x:\Sigma\to \mathbb{P}^1$, where $\mathbb{P}^1$ is the complex plane plus the point at infinity, such that $w_{0,1}=y dx$. The function $y(x)$ is found using the fact that discontinuity along the cut of $w_{0, 1}$ gives back the density of states, while $x$ is a multi-branched covering of the complex plane depending on the genus of $\Sigma$ and the position of the branch cuts. In recent years, topological recursion has been crucial to interpretations of low dimensional quantum gravity as a random matrix theory \cite{Saad:2019lba,Jafferis:2022wez}. The majority of the literature examines the genus-zero spectral curve, where $\Sigma$ is a sphere and the Bergman kernel is universal. However, for the case at hand, as we will see momentarily, we will need higher-genus spectral curves, because of the multi-cut structure of the mesoscopic eigenvalue density.
We discuss the relevant recursion formulas for our analysis, together with the specific case of the genus-1 topological recursion in \cref{app:g=1_top_rec_details}. For a more in depth review of the topological recursion, please refer to \cite{eynard2018randommatrices, Eynard:2016yaa}.

As discussed in \cref{sec:mesoscopic_inbin}, the mesoscopic density \eqref{eq:rho_ansatz}, which is sensitive to the internal bin structure, outlines a multi-cut matrix model with eigenvalues supported in the intervals $[b_k,a_k],\; k=0,\dots \mathcal{E}/\epsilon-1$. The $\Sigma$ corresponding to this cut structure is a Riemann surface of (parametrically infinite) genus $n=\mathcal{E}/\epsilon-1$, and this is already a manifestation of the fact that the mesoscopic structure is considerably richer than the macroscopic one. However in the parameter limit we are considering \eqref{eq:limits}, as shown in \cref{sec:mesoscopic_inbin}, we have a high-energy shell where $a_k\to b_{k+1}$ and it is licit to consider the cut-pinching approximation $d_k/\delta_k\ll 1$. This approximation collapses the generic infinite genus structure to a more tractable case of a sphere with $n$ punctures, similarly to the spectral curves analyzed in \cite{Collier_2024, Collier_2025}. We will consider a further simplification and restrict our attention to the local behavior of correlation functions, for energies typically in the same bin. As discussed in \cref{sec:mesoscopic_inbin}, at separation scales comparable with the bin size, we can consider a first neighbor approximation that simplifies the density in the $k$-th bin to \eqref{eq:rho_inbin_approx}, which translates into
\begin{equation}\label{eq:spectral_curve_ansatz}
     y^2(x)=\sum_k(i\pi C_k)^2\frac{(x-a_{k-1})(x-a_k)}{(x-b_k)(x-b_{k+1})}\chi_{I_k}(x),
\end{equation}
where $\chi_{I_k}(x)$ is the characteristic function of the $k$-th bin $I_k$, and we recall that $C_k$ is approximately constant \eqref{eq:Ck_binextrema} in $I_k$. The expression \eqref{eq:spectral_curve_ansatz} provides a family of local auxiliary spectral curves: the term associated to the $k$-th bin in \eqref{eq:spectral_curve_ansatz} can be locally seen as the spectral curve which subtends the support $[-\infty,x_1]\cup[x_2,x_3]\cup[x_4,\infty]$ on the real axis, where $x_1=a_{k-1}$, $x_2=b_{k}$, $x_3=a_{k}$ and $x_4=b_{k+1}$. Notice however, that this structure still describes a genus-one spectral curve, due to the inclusion of the point at infinity to the complex plane discussed above. In principle, in order to capture higher orders in $\epsilon$, we can imagine to extend the approximation \eqref{eq:rho_inbin_approx} to the generic $j$-th neighbor interactions. However, this would result in a genus $j$ spectral curve, so in practice we restrict to genus 1, which is still tractable following, for example, \cite{Bonnet_2000,Mari_o_2009}.
\begin{figure}
\centering
        \includegraphics[width=\textwidth]{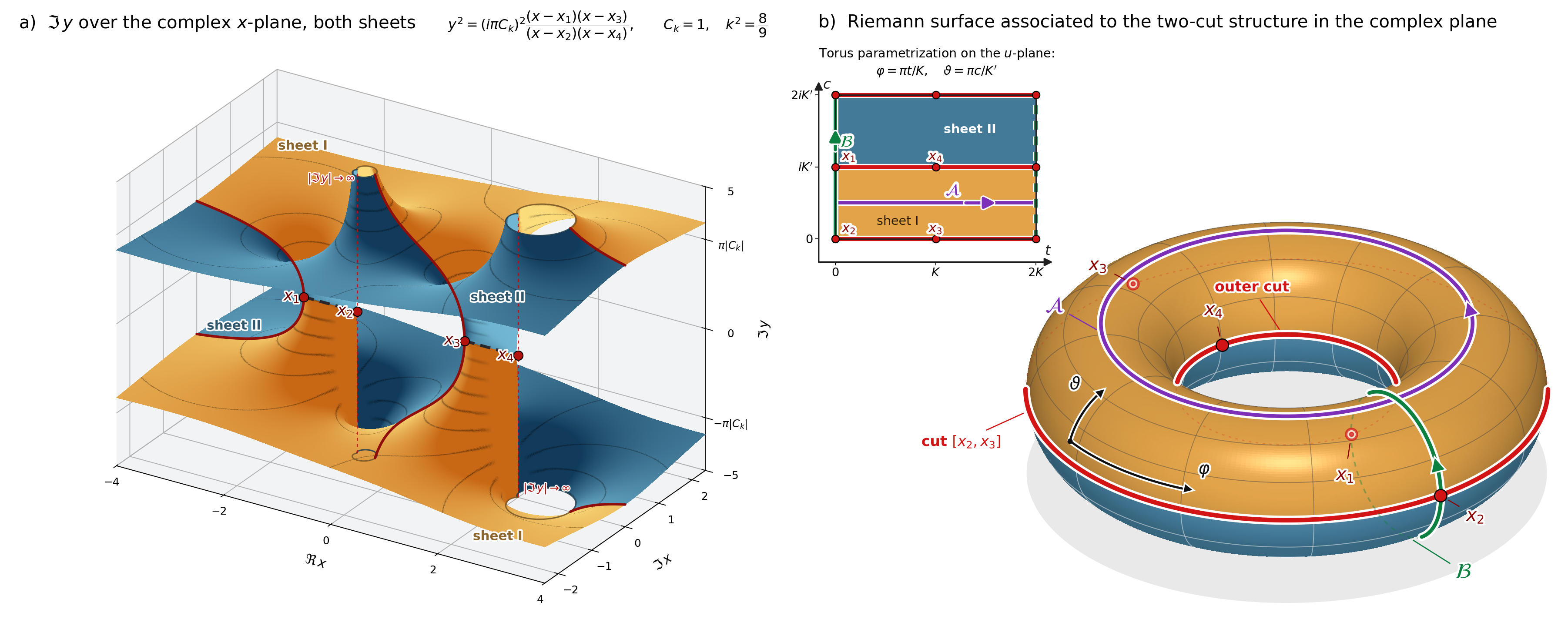}
    \caption{\emph{Left (a)}: Plot of $\Im y$ in the $k$-th bin from \eqref{eq:spectral_curve_ansatz}, on both sheets of the complex plane. On the cuts, highlighted in red, $y$ on sheet I is related to the eigenvalue density via \eqref{eq:resolvent_to_eigenv}. \emph{Right (b)}: Visualization of the uniformization of the two-sheeted complex plane on the torus. In the upper part of the figure, the parallelogram realization of the torus in coordinates $\rm u\equiv z/g$, with periods $\omega_1=2K(k)$ and $\omega_2=2i K(k')$, and its relation to the more common angular parametrization drawn under it. The cuts, as well as the images of the Riemann sheets, are colored correspondingly to (a). The fundamental cycles $\mathcal{A}$ and $\mathcal{B}$, needed to define the coordinates \eqref{eq:uniform_abel_map}, are chosen so that $\mathcal{A}$ encircles the $[x_2,x_3]$ cut.}
    \label{fig:two_cut_torus}
\end{figure}

Now we want to determine the spectral curve in the first-neighbor approximation discussed above. We argued that in this case $\Sigma$ is a genus 1 surface, that is a torus, while the function $y(x)$ is given by \eqref{eq:spectral_curve_ansatz}. In order to determine the 1-form $w_{01}$, we need the function $x(z):\Sigma\to\mathbb{P}^{1}$, where $z\in\Sigma$ is called the uniformization coordinate on the Riemann surface.
For genus one, a good parametrization is the Abel map
\begin{equation}\label{eq:uniform_abel_map}
    z(x)=\int_{b_k}^{x}\frac{dt}{\sqrt{(t-a_{k-1})(t-b_k)(t-a_k)(t-b_{k+1})}}=g F(\alpha(x),k)\,,
\end{equation}
where $F$ is the incomplete elliptic integral of the first kind and we defined
\begin{equation}\label{eq:k_torus}
\begin{aligned}
   \alpha(x)=\arcsin{\sqrt{\frac{(a_k-a_{k-1})(x-b_k)}{(a_k-b_k)(x-a_{k-1})}}},\quad g=\frac{2}{\sqrt{(b_{k+1}-b_k)(a_k-a_{k-1})}}\,.
\end{aligned}
\end{equation}
Notice that we defined the Abel map coordinates in \eqref{eq:uniform_abel_map} choosing the cycle encircling the interval $[x_2,x_3]$, which coincides with our eigenvalue support\footnote{but differs from the canonical choice of $[x_1,x_2]$ more common in the literature \cite{Mari_o_2009}.}. The definitions and properties of the elliptic functions considered here and in \cref{app:g=1_top_rec_details} can be found on the Mathematica functions site, while for a considerably deeper discussion we refer to \cite{lawden1989elliptic}.

The double covered complex plane identified by the cuts of our spectral curve is equivalent to a torus, which, in the uniformization coordinates \eqref{eq:uniform_abel_map}, is realized as the parallelogram $\frac{\mathbb{C}}{\omega_1\mathbb{Z}+\omega_2\mathbb{Z}}$ with periods $\omega_1=2\mathcal{K}$ and $\omega_2=2i\mathcal{K}'$ given by
\begin{equation}\label{eq:ks_torus}
\begin{aligned}
    &\mathcal{K}=\int_{b_k}^{a_k}\frac{dt}{\sqrt{(t-a_{k-1})(t-b_k)(t-a_k)(t-b_{k+1})}}=gK(k),\\&
    \mathcal{K}'=\int_{a_{k}}^{b_{k+1}}\frac{dt}{\sqrt{(t-a_{k-1})(t-b_k)(t-a_k)(t-b_{k+1})}}=gK(k')\,,
\end{aligned}
\end{equation}
where $k'^2=1-k^2$.

The Bergman kernel for a torus can be written explicitly using Weierstrass functions \cite{eynard2018randommatrices},
\begin{equation}\label{eq:berg_ker_torus}
    B(z,z')=\biggr(\wp(z-z';\omega_1,\omega_2) +2\frac{\eta_1}{\omega_1}\biggr)dzdz'\,,
\end{equation}
and $\eta_1$ is the quasi-periodicity of the Weierstrass zeta-function $\eta_1=\zeta\left(\frac{\omega_1}{2};\omega_1,\omega_2\right)$.
We refer to \cref{app:elliptic_int} for the computation of the elliptic parameters in the cut-pinching approximation and in particular to \cref{eq:alpha_k_elliptic}, which we can use to obtain the following expressions for \eqref{eq:uniform_abel_map}
\begin{equation}
\begin{aligned}\label{eq:K_Kp_pinched}
    K(k)\approx\frac{1}{2}\log\left(\frac{16\delta_k^2}{d_{k-1}d_k}\right) ,\quad K(k')\approx \frac{\pi}{2}+\frac{\pi d_k d_{k-1}}{8\delta_k^2}\,.
\end{aligned}
\end{equation}
We therefore see that, from the point of view of the torus, the cut-pinching approximation corresponds, because of \eqref{eq:ks_torus}, to sending $\omega_1\to\infty$ while fixing $\omega_2$, so that $\tau\equiv\omega_2/\omega_1\to 0$. In this limit, the spectral curve $(\Sigma, ydx, B)$ subtends a topological recursion on a pinched torus, which as expected reduces to the more standard genus 0 case. In particular, we can find a coordinate $Z$ that brings the degenerate torus uniformization into the standard Joukowski form expected when the Riemann surface associated to the topological recursion is a sphere \cite{eynard2018randommatrices}:
\begin{equation}\label{eq:jouk_coords_main}
    x(Z)=\frac{x_3+x_2}{2}+\frac{x_3-x_2}{4}\left(Z+\frac{1}{Z}\right)\,.
\end{equation}
Notice that \eqref{eq:jouk_coords_main} is the Joukowski parametrization for the surviving cut $[x_2,x_3]$ after the pinching procedure. As expected, in the coordinate $Z$ appearing in \eqref{eq:jouk_coords_main}, the degenerate Bergman kernel, in a compact region, can be brought in the universal genus-zero form,
\begin{equation}\label{eq:Bergman_deg}
    B(Z,Z')=\frac{dZdZ'}{(Z-Z')^2}.
\end{equation}
Please refer to \cref{app:torus_deg} for the specific coordinate changes from which we define $Z$, as well as the derivations of \eqref{eq:jouk_coords_main} and \eqref{eq:Bergman_deg} in the degeneration limit of the torus.

\begin{figure}
\centering
  \begin{minipage}[b]{0.495\textwidth}
    \centering
    \includegraphics[width=\textwidth]{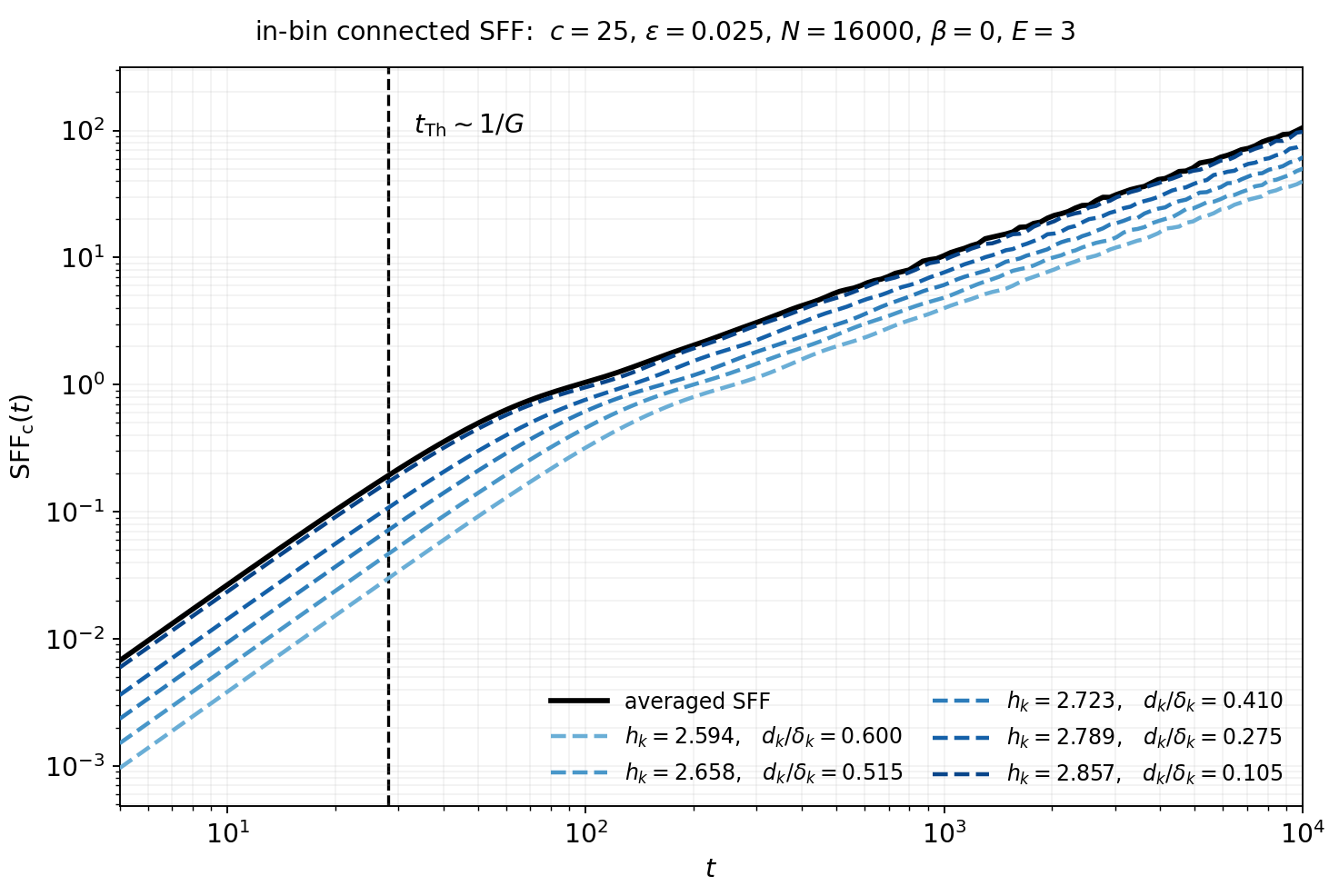}
  \end{minipage}
  \hfill
  \begin{minipage}[b]{0.495\textwidth}
    \centering
    \includegraphics[width=\textwidth]{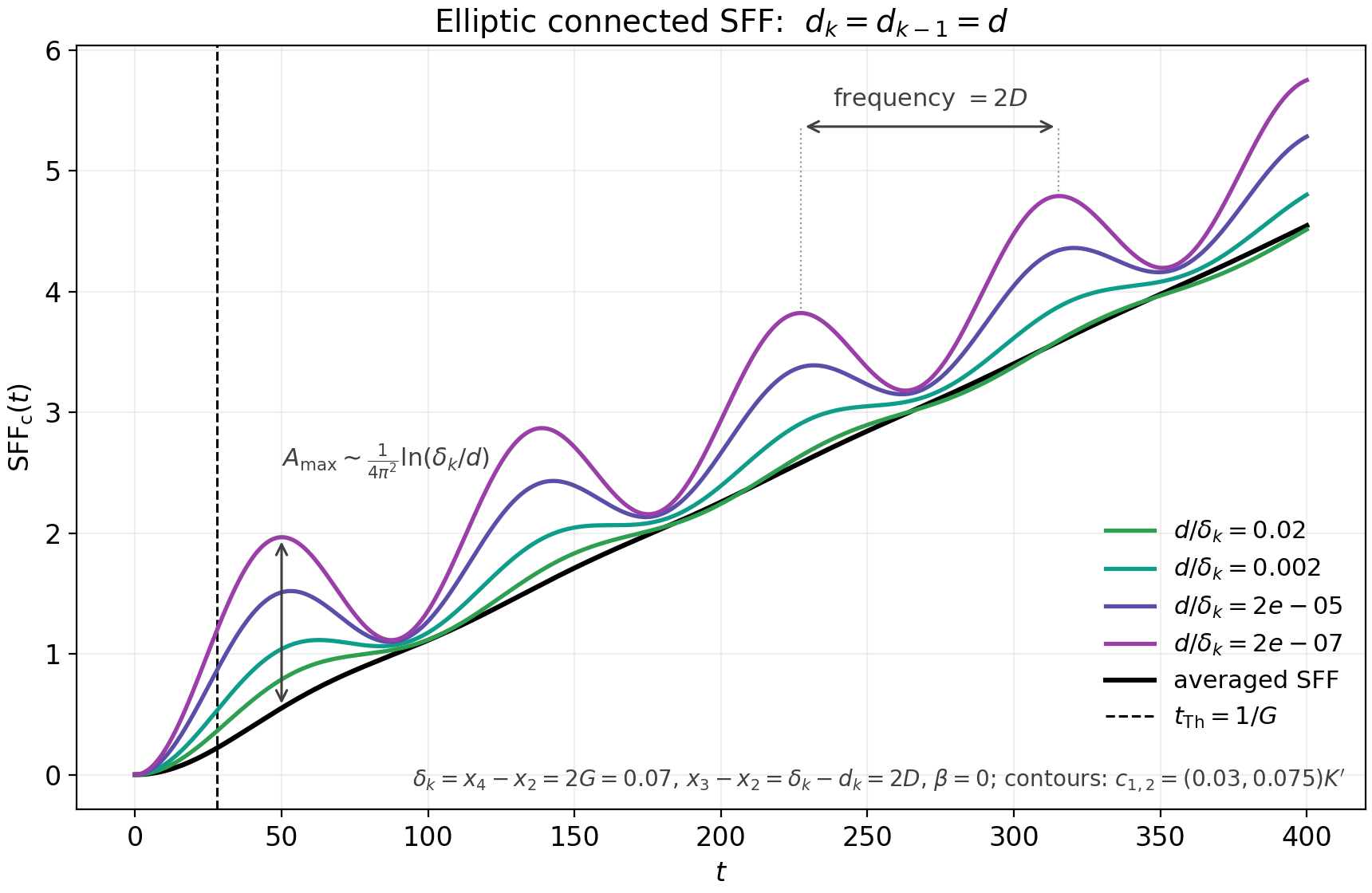}
  \end{minipage}
    \caption{\emph{Left}: connected in-bin spectral form factor (log--log) computed from the
frozen-occupations model of \cref{fig:inbin_density_ansatz}, for several bins of
a single Monte Carlo run, labelled by their centre $h_k$ and by the pinching
parameter $d_k/\delta_k$. As the pinching parameter decreases the curves
approach the black reference, the genus-0 answer evaluated on the full bin:
this is the result obtained if the average over the bins' positions is performed
on the density \emph{before} the two-point function is built, since the empty gap
sits at a different place in every scheme and averages away. The result
carries in addition some regularization-dependent features, invisible at this
resolution, which are displayed in the right panel, in the limiting case of the cut-pinching approximation, and which the same averaging
removes. \emph{Right}: the analytical SFF in the genus 1 cut-pinching approximation, computed numerically as \eqref{eq:sff_def} using the
exact elliptic functions, in the case where $d_{k-1}=d_k=d$. Notice that the aforementioned regularization-dependent features are expected to arise from the fact that we are computing a higher-genus elliptic SFF rather than the usual genus 0 result. To make such features evident, we restrict to a linear
and shorter time axis and push the pinching down to $d/\delta_k=2\times
10^{-7}$, far below the values realized in the Monte Carlo. On top of the universal ramp there is then a clearly visible oscillatory term, with frequency $2D$, that is the bin scale in the pinching limit, and amplitude growing logarithmically as $\tfrac{1}{4\pi^2}\ln(\delta_k/d)$. Its amplitude decays in time, so relative to the linearly growing ramp it becomes negligible at late times; for more details on its origin refer to \cref{app:elliptic-sff}.}
    \label{fig:sff_analytics_mcmc}
\end{figure}

If we run the genus-zero topological recursion procedure, which we summarize in \cref{app:top_rec_recap}, we can obtain the leading behavior in the pinching parameter of all-orders resolvent correlators at small energy distances. Here we will be interested mostly in one particular observable, obtained from the initial data of the recursion, that is the spectral form factor, which we define from the connected $2$-resolvent correlator $W_{02}$ as
\begin{equation}\label{eq:sff_def}
    SFF(\beta, t)\equiv Z_{02}(\beta_1,\beta_2)\biggr|_{\substack{\beta_1=\beta+i t \\ \beta_2=\beta-i t}},\quad Z_{02}(\beta_1,\beta_2)=\oint_\Gamma\oint_\Gamma W_{02}(Z_1,Z_2)\prod e^{-\beta_ix(Z_i)}\,,
\end{equation}
where $\Gamma$ encircles $[x_2,x_3]$ and $W_{02}(Z_1,Z_2)=(1-Z_1Z_2)^{-2}$. As explained in detail in \cref{app:sff_details}, we obtain the following spectral form factor:
\begin{equation}\label{eq:sff_finite_T}
    SFF(\beta,t)\equiv Z_{02}(\beta+it, \beta-it)\approx e^{-2\beta b_{k}}\frac{\sqrt{\beta^2+t^2}}{4\pi \beta},
\end{equation}
computed at $\beta G\gg 1 $ and late times $t\gg 1/G$, where we remind that $\delta=2G$ is the typical high-energy bin-width.
This is an analogous result to the spectral form factor of JT gravity, as computed in \cite{Saad:2019lba}, and in particular at times $t\gg\beta$ it shows the characteristic linear time-growth associated to the RMT nature. We remark that the computation of the SFF \eqref{eq:sff_def}, in the elliptic case, also gathers an oscillatory contribution, around the ramp behavior, from a contour region where the approximation reducing the elliptic functions to their sphere counterparts \cref{eq:jouk_coords_main,eq:Bergman_deg} does not hold (for more details please refer to \cref{app:elliptic-sff}). We plot the connected part of the spectral form factor in \cref{fig:sff_analytics_mcmc}: the aforementioned bin-scale oscillations can be considered a regularization-dependent effect that can be filtered away if we perform an averaging procedure over the bins' positions, analogously to \cref{fig:inbin_density_ansatz}, before computing the SFF.

The analytically computed in-bin ramp is consistent with the random-matrix behaviour observed numerically in \cref{fig:mcmc_sff} over a window containing several bins. The numerical form factor also reaches a plateau consistent with unity despite frozen occupations. This microscopic saturation lies beyond the perturbative, smoothed-correlator calculation presented here.
Notice that the exponential prefactor in \eqref{eq:sff_finite_T} is present because the left branch point in our case coincides with the left edge of a high energy bin, in contrast with \cite{Saad:2019lba}, where it sits in the zero-energy position. As explicitly checked in \cref{app:sff_details}, this constant disappears when we take infinite temperature limit $\beta\to0$ \eqref{eq:sff_inf_T}.

Now let us put together \eqref{eq:sff_finite_T} and \eqref{eq:sff_disc_latetimes}: the leading behavior of the spectral form factor at early-to-intermediate times is given by its disconnected component \eqref{eq:sff_disc_latetimes}, which is decaying $\propto t^{-3}$. However, at times $\mathcal{O}(1/G)$, the connected contribution becomes non-negligible, and in the case where $G\ll1$, it is actually the dominant contribution, reinstating the late-time linear ramp behavior typical of random matrix theories\footnote{The Thouless time-scale is determined by the ramp onset in the connected spectral form factor, which is given by $1/G$ from \eqref{eq:Z02_besselsum}. Here we are saying that, if $G\ll1$, we expect this scale to be parametrically of the same order as the onset of the ramp in the full spectral form factor, obtained by summing its connected and disconnected components.}.
Analogously, one can run the topological recursion as reviewed in \cref{app:top_rec_recap}, to compute the $n$-point resolvent correlators $W_{gn}$ for all genus $g$, as long as all the $n$ energies lie inside a single bin. Then, similarly to what happened for the spectral form factor, the small-time leading behavior will be the disconnected component \cref{sec:macro_phys}, while, at later times $\mathcal{O}(1/G)$, the connected component computed from the topological recursion will start to contribute. From the behavior of these observables we understand that, in this bin-regularized model, the parameter $\delta$ takes on the role of the inverse of the Thouless time, at which we first see the onset of the random matrix theory features. This is a direct consequence of the limits \eqref{eq:limits}, which separate the scales at which the Cardy potential and Vandermonde act: as we expected heuristically the RMT nature of the model becomes evident only at timescales $t_{Th}\sim 1/G$ when we are able to discern the mesoscopic structure inside the bins. The interpretation of the bins' width determining the Thouless time-scales also justifies the parameter regime identification to describe the spectral properties of a chaotic CFT that we proposed in \cref{sec:vac_occ_gap}. Indeed, as recently argued for example in \cite{Chen:2024oqv}, one expects that a strong chaotic nature is trademarked by an $\mathcal{O}(1)$ Thouless time, and this can be achieved via the limits \eqref{eq:limits} while scaling $G$ with a constant independent of $N$.

We end this section with a more technical remark. If we perform the \textit{a-priori} averaging of \cref{fig:inbin_density_ansatz}, this collapses the bin-localized spectral curve to the genus 0 case, which retrieves the RMT nature of the Cardy matrix model. However, as explained in this section, in the nearest neighbor approximation introduced in \cref{sec:mesoscopic_inbin}, we can also consistently obtain the first order regularization-dependence in the pinching parameters from the genus 1 topological recursion. As anticipated, averaging \textit{a posteriori} over the bins' positions retains the imprint of the regularization instead of removing it, and it would be interesting to understand what physics it encodes. The procedure is, however, rather cumbersome to carry out in general. We report that, if one wishes to obtain higher order effects in the pinching parameters, then the procedure to carry out would be to generalize the approximation in \cref{sec:mesoscopic_inbin} to more than nearest-bin interactions, and run the appropriate local genus $g>1$ recursion. It is also very hard to consider more general limits than the ones discussed in \cref{sec:saddle_point}: indeed, in the case where the Cardy potential and the Vandermonde can act at comparable scales, one should modify the topological recursion procedure to account for the multi-trace nature of the model \cite{BOROT_2016}.

\section{Discussion and outlook}\label{sec:discussion}
Let us begin by summarizing what we achieved in this paper. We introduced the Cardy random matrix theory, a regulated `constraint$^2$' matrix model implementing coarse-grained $\mathbb{S}$-modular invariance, which we take to describe the spectrum of a CFT. We identified a scaling limit (large-N and associated scaling of regulators) in which the vacuum drives a Cardy phase: the density develops the expected exponential growth: the preferred number of states scales exponentially with the cutoff. Numerical and analytic results reveal a separation of scales: the modular potential fixes the macroscopic density, while Vandermonde repulsion governs the finer in-bin structure. Consequently, the spectrum exhibits level repulsion and a linear spectral ramp at mesoscopic scales, with the bin width determining an effective inverse Thouless time in our scaling limit. The numerical connected form factor further reaches a plateau consistent with unity, demonstrating that frozen coarse occupations do not prevent microscopic plateau saturation. Now, we move to discuss possible future directions of inquiry. 

\subsection*{Corrections to Cardy, and constraint$^2$ potentials as frustrated systems}

The peculiar hierarchy of scales, determined in \cref{sec:saddle_point}, could suggest a more general use of these `constraint$^2$' matrix models. Indeed, as discussed in \cref{sec:top_recursion}, the system is fully integrable at a certain macroscopic scale, while the chaotic features of a random matrix theory emerge at the smaller mesoscopic length set by the bin regularization, well before we are able to discern the microscopic eigenvalue structure. We can imagine that analogous `constraint$^2$' matrix potentials can be used to model systems where there is a `tension' between an ordered and disordered phase, as it happens for example in glasses, where full ergodicity is broken.

If we consider the corrections to the Cardy behavior given by the $\mathbb{S}$-kernels of heavy-states we can observe such effects also in the Cardy RMT. Modular invariance is a self-duality condition, and no positive atomic
spectrum is exactly self-dual under the regulated $\mathbb{S}$-transform: the
unavoidable misfit can be distributed in exponentially many nearly
degenerate ways, which is the disorder-free frustration mechanism typical of some deterministic
glass models \cite{Marinari_1994, Parisi_1995}. The prolate
functions that diagonalize the windowed kernel are the natural tool to
quantify this degeneracy and the role of the vacuum in possibly relieving it. We remark that a
genuine glassy phase would make quenched and annealed ensemble averages
differ, with direct implications for the averaging procedures of chaotic
CFTs. We leave this intriguing direction for future work.

  \subsection*{From Cardy RMT to a 2d CFT matrix model}

The main goal of this paper was to make progress on the tensor model for 3D gravity introduced in \cite{Belin:2023efa}. To improve on the toy model Cardy RMT described in this paper, the first step is to introduce a full 2d CFT spectrum. In our model, we consider only S-modular invariance and treat all states to be scalar. In a more realistic model, we should have both left and right moving conformal weights (the eigenvalues of 2 commuting matrices) and additionally impose $\mathbb{T}$-modular invariance.

To construct a random matrix model that takes into account both approximate $\mathbb{S}$ and $\mathbb{T}$-modular invariance,\footnote{In \cite{Belin:2023efa}, it was also discussed of taking a hard $\mathbb{T}$-invariant constraint by considering directly quantized spin. This could be studied from the model we describe now, by taking 2 different constraint enforcing parameters, and sending the one for $\mathbb{T}$ invariance to zero much faster than that for $\mathbb{S}$ invariance.} we can generalize the regularization procedure described in this paper for the non-scalar $\mathbb{S}$-modular invariance, and obtain, after integrating in a square window of width $\epsilon$ around $P$ and $\Bar{P}$, the following $\epsilon$-regularized constraints:
\begin{equation}\label{eq:constraints_anti_hol}
    \mathcal{C}_\mathbb{S}\equiv\frac{1}{\epsilon^2}\left( N_{P,\Bar{P},\epsilon}-\epsilon^2\sum_i\mathbb{S}_{PP_i}(\mathds{1})\mathbb{S}_{\Bar{P}\Bar{P}_i}(\mathds{1})\right)=0,
\end{equation}
where $N_{P,\Bar{P},\epsilon}=\sum_i \chi_{P,\epsilon}(P_i)\chi_{\Bar{P},\epsilon}(\Bar{P}_i)$ indicates the number of eigenvalues in a square cell of the two dimensional $(P,\Bar{P})$-histogram. Now, with the same considerations as in \eqref{eq:cardy_potential}, we can build a (constraint$^2$) potential that this time is associated to a two-matrix theory, due to the extra degrees of freedom associated to the anti-holomorphic sector.  
The novel complication is introduced by the $\mathbb{T}$-modular invariance condition, which can be approximately implemented, without extra-regularizations, via the following (constraint)$^2$ potential \cite{Belin:2023efa}:
\begin{equation}
    V_{\mathrm{spin}}=\sum_i \sin^2(\pi(h_i-\Bar{h}_i)),
\end{equation}
which softly imposes spin quantization.
The model would then take the following form:
 \begin{equation}\label{eq:full_model}
\begin{aligned}
    Z\propto\int D[h,\Bar{h}] \exp{\left(-\frac{1}{4a_\mathbb{S}}C_\mathbb{S}^2-\frac{1}{4a_\mathrm{spin}}V_\mathrm{spin}\right)},
\end{aligned}
\end{equation}
 where $D[h,\Bar{h}]$ are the integration measures over the eigenvalues. Notice in particular that we absorbed the factor coming from the simultaneous diagonalization of commuting matrices $L_0$ and $\Bar{L}_0$ in $D[h,\Bar{h}]$ so that, as discussed for example in \cite{Wiegmann:2002tk, Berenstein_2006, Filev:2014jxa}, $D[h,\Bar{h}]=\prod_{i<j}\left((h_i-h_j)^2+(\Bar{h}_i-\Bar{h}_j)^2\right)dhd\Bar{h}$. This model can be analyzed numerically by a Monte-Carlo analysis similar to how we proceeded in this paper.\footnote{It is interesting to note that a model where spin quantization is \textit{exact} is actually harder to treat. This follows from the fact that the size of the Hilbert space at each spin becomes a dynamical quantity, that can undergo jumps as parameters are varied. The approximate spin quantization potential is thus in some sense smoother.} We leave a more in-depth study to future work.

 In the low spin and high energy part of the model, we expect very few deviations from the physics we found in this paper. At the level of non-perturbatively small corrections, there should still be signatures, and we should recover the full $SL(2,\mathbb{Z})$ sin-kernel \cite{Boruch:2025ilr}. But there are regions of parameter space that would display new physics, the most interesting being the near-extremal regime. A theory with only an identity operator below the black hole threshold, and modular invariance leads to a negative and exponentially large density of states \cite{Benjamin:2019stq}. In our model, in general, we do not enforce to have only the identity as the below threshold state. The low-energy spectrum is populated dynamically as we increase the number of eigenvalues. There have been proposed cures for the negativity of pure gravity \cite{Maxfield:2020ale,Benjamin:2020mfz,DiUbaldo:2023hkc}, and it would be very interesting to see what our random matrix model dynamically tends to.

\subsection*{Incorporating crossing and OPE coefficients}

The next step is to incorporate physics beyond the spectrum, and add in OPE coefficients. These are then constrained by modular covariance of
torus one-point functions and sphere four-point crossing. We start by considering the insertion of these last two constraints, for simplicity limiting the discussion to the toy model Cardy RMT (i.e. without spin as a second quantum number). We need to introduce (constraint)$^2$ potentials pertaining to a random tensor model for the OPE coefficients, appropriately coupled to the eigenvalue sector we have been discussing. Following \cite{Belin:2023efa, Jafferis:2025vyp}, the regularized version for the constraint of modular covariance of
torus one-point functions reads
\begin{equation}\label{eq:onept_mod_inv}
    \mathcal{C}_{1-pt.}\equiv\frac{1}{\epsilon}\left(N_{P,\epsilon}\overline{C_{ii\mathcal{O}}}-\epsilon\sum_i C_{ii\mathcal{O}}\mathbb{S}_{PP_i}(\mathcal{O})\right)=0\,,
\end{equation}
where $\Tilde{\rho}_\mathcal{O}(P)\equiv \sum_i C_{ii\mathcal{O}}\delta(P-P_i)$, the indices $i$ label the primary operator with momentum $P_i$ and $\overline{\cdot}$ indicates an average $\overline{C_{ii\mathcal{O}}}\equiv \int_{P-\epsilon/2}^{P+\epsilon/2}dP'\Tilde{\rho}_\mathcal{O}(P')/N_{P,\epsilon}$ over the bin centered around $P$ of width $\epsilon$. Similarly, the regularized version of sphere four-point crossing can be recast as
\begin{equation}\label{eq:crossing_eq_rhos}
    \mathcal{C}_{cross.}\equiv\frac{1}{\epsilon}\left(N_{P,
    \epsilon}\overline{C_{12P}C_{34P}}-\epsilon\sum_i \mathbb{F}_{P P_i}C_{41i}C_{23i}\right)=0\,,
\end{equation}
where $\rho_s(P)=\sum_sC_{12s}C_{34s}\delta(P-P_s)$, $\rho_t=\sum_t C_{41t}C_{23t}\delta(P-P_t)$, $\overline{C_{12P}C_{34P}}=\int_{P-\epsilon/2}^{P+\epsilon/2}dP'\rho_s(P')/N_{P,\epsilon}$  and $\mathbb{F}_{P_s P_t}$ is the fusion kernel \cite{ponsot1999liouvillebootstrapharmonicanalysis,Ponsot_2001, Teschner_2001}. 

Now, if we perform the scaling limits similarly to what we have done in this paper, we expect to find as a saddle point:

\begin{equation}\label{eq:C0_formula}
     \overline{C_{12P}^2} N_{P,\epsilon}\approx\epsilon \mathbb{F}_{P\mathds{1}}+\mathrm{non-vac.\;corrections}.
\end{equation}
Note that we have restricted here to the symmetric configuration $P_1=P_4$ and $P_3=P_2$, so that the vacuum is included, and dominates, in the T-channel OPE. The full (left and right) version of \eqref{eq:C0_formula}, which may be interpreted as the Heavy-Heavy-Heavy analogue of the Cardy formula for OPE coefficients, is the $C_0$ formula derived in \cite{Cardy:2017qhl,Collier:2019weq}. The saddle-point equation would thus produce the coarse-grained $C_0$ formula much like we obtained Cardy's formula in this paper. Note that we expect both $N_{P,\epsilon}$ and $\overline{C_{12P}^2}$ to be driven towards the Cardy/$C_0$ formula simultaneously. Macroscopically, we do not expect any interaction between the two as the spectrum and OPE coefficients can arrange themselves quasi-independently.

However, as a final remark, we notice that even though \eqref{eq:C0_formula} imposes conditions at the macroscopic level, it can leave an imprint via a residual potential on the in-bin mesoscopic physics. In this case, even with the approximations of \cref{sec:mesoscopic_inbin} becoming less controlled, we expect that the intuitive picture of the emergence-scale of universal RMT behavior, which is dependent only on the eigenvalues cut structure, is maintained. However, the computation of higher-point functions will be heavily affected by the detailed form of the mesoscopic density determined by the residual potential, as well as crossing/modular-invariance interaction.

\section*{Acknowledgments}
We thank Jordan Cotler, Gabriele Di Ubaldo, Tom Hartman, Daniel Jafferis, Diego Liska, Alex Maloney, Eric Perlmutter, Steve Shenker, Douglas Stanford and Diandian Wang for discussions. We thank Bharath Radhakrishnan for collaboration in the initial stage of this work. We also thank Alexander Hock and Marcos Mariño for enlightening assistance regarding some technical aspects of the matrix topological recursion and the use of elliptic functions. The MCMC codes in this paper were developed in python with Anthropic's Claude Code (Fable 5) and OpenAI's Codex (ChatGPT-Pro 5.6), and checked and verified by human authors. Both Labs' frontier models were used for adversarial review of the material in this paper.  JS and JdB thank the Aspen Center for Physics for hospitality during the final stages of this project. This research is supported in part by the Fonds National Suisse de la Recherche Scientifique (Schweizerischer Nationalfonds zur Förderung der wissenschaftlichen Forschung) through the Project Grant 200021\_215300 and the NCCR51NF40-141869 The Mathematics of Physics (SwissMAP). The work was performed in part at the Aspen Center for Physics, which is supported by the National Science Foundation grant PHY-2210452.

\appendix

\section{Details on the orthogonality of Virasoro characters}\label{app:f_details}

In this appendix we review the construction of the function $f(\tau)$ used in \cref{sec:S_modinv_reg} to project the $S$-modular invariance constraint to a single conformal block. For a more in depth and technical discussion you can refer to \cite{Verlinde:1989ua, Collier_2023}.

We consider the setting were we have already quotiented by reflection, so that we can limit our analysis to heavy states with positive momenta. So, we want the following orthogonality relation between characters with $P,P'\geq0$:
\begin{equation}\label{eq:ort_rel}
\int \frac{d^2\tau}{\sqrt{\Im(\tau)}} f(\tau) \chi_P(\tau) \chi_{P'}(\tau)^* \sim \delta(P-P') \,,
\end{equation}
where $\chi_{P'}(\tau)^*=\Bar{\chi}_{P'}(\Bar{\tau})$, as $e^{\overline{2\pi i \tau}}=e^{-2\pi i \Bar{\tau}}$, and $\int d^2\tau$ is the integral in the complex half-plane. Notice that the $\Bar{\tau}$ here has nothing to do with the $\tau$-integral to be performed for the anti-holomorphic sector in a non-chiral CFT, and they should be treated as independent.

Let us start by considering the case of non-degenerate characters. We give the following ansatz for the function $f(\tau)$:
\begin{equation}\label{eq:ansatz_f_proj}
    f(\tau)=\frac{|\eta(\tau)|^2}{\Im(\tau)}.
\end{equation}
In order to verify this ansatz,  we insert the above expression in \eqref{eq:ort_rel}. Using \eqref{eq:character_P_eta_expr} for the characters, and with $\tau= x + i y$, we obtain the following integral:
\begin{equation}\label{eq:inner_prod_def}
\int \frac{d^2\tau}{\sqrt{\Im(\tau)}} f(\tau) \chi_P(\tau) \chi_{P'}(\tau)^* = \int \frac{dx d y}{y\sqrt{y}} e^{2\pi i x(P^2-P'^2)}  e^{-2 \pi y (P^2+P'^2)} \,.
\end{equation}
The $x$-integral gives a delta function $\delta(P^2-P'^2)=\frac{\delta(P-P')}{2P}$, so we can restrict the analysis of the $y$-integral on the support $P'=P>0$. Then, if we change variables $y\to 4yP^2$ we obtain:
\begin{equation}
    \int_0^\infty dy  \frac{e^{-4\pi y P^2}}{y\sqrt{y}}=2P\int_0^\infty dy  \frac{e^{-\pi y}}{y\sqrt{y}}
\end{equation}
So in total, we obtain the following result:
\begin{equation}
    \int \frac{d^2\tau}{\sqrt{\Im(\tau)}} f(\tau) \chi_P(\tau) \chi_{P'}(\tau)^* =\frac{\delta(P-P')}{\cancel{2P}}\cancel{2P}\int_0^\infty dy  \frac{e^{-\pi y}}{y\sqrt{y}}.
\end{equation}
As it turns out, the $y$ integral diverges due to the contribution near $y\sim 0$, so we introduce a regularization parameter $\varepsilon$ and obtain
\begin{equation}
    \int_\varepsilon^\infty dy  \frac{e^{-\pi y}}{y\sqrt{y}}=\frac{2e^{-\pi\varepsilon}}{\sqrt{\varepsilon}}-2\pi \,\mathrm{Erfc}(\sqrt{\pi\varepsilon})\approx_{\varepsilon\to0}\frac{2}{\sqrt{\varepsilon}}-2\pi+\mathcal{O}(\sqrt{\varepsilon}).
\end{equation}
So regularizing the integral boils down to discarding an overall constant \cite{Collier_2023}, which we can reinterpret as a renormalization procedure which in the end leaves us with an inner product $\propto \delta(P-P')$, thus proving the ansatz \eqref{eq:ansatz_f_proj}.

Now we move to discuss what happens if we want to project to the vacuum character \eqref{eq:chi_identity}. Using naively the same inner product \eqref{eq:ansatz_f_proj} we would obtain
\begin{equation}\label{eq:vac_noninnerprod}
     \int \frac{d^2\tau}{\sqrt{\Im(\tau)}} f(\tau) \chi_P(\tau) \chi_{\mathds{1}}(\tau)^* \overset{?}{=}\delta\left(P-i\frac{Q}{2}\right)-\delta\left(P-\sqrt{\frac{25-c}{24}}\right)\equiv \delta(P-\mathds{1}),
\end{equation}
which corresponds to the characters of conformal dimension $h=0,\,1$. Notice that, as it stands, \eqref{eq:vac_noninnerprod} is quite problematic, as complex argument in the Dirac deltas are associated to non-converging integrals in \eqref{eq:inner_prod_def}\footnote{the divergence here is due to the $y\to\infty$ region, so it is separate from the $y\to0$ one we discussed, and in particular is not cured by the $\varepsilon$ regularization.}, and consistently the vacuum is not normalizable. With more care, it is nonetheless possible to define such distributions $\delta(P-\mathds{1})$, and for more details we refer for example to \cite{Maxfield_2019, Hartman:2025cyj}. Here we limit to report that, even though \eqref{eq:vac_noninnerprod} is not a standard inner product, these complex argument deltas behave analogously to their heavy-states counterparts as long as we integrate over characters, which is all we need in \cref{sec:cardy_matrix}. In any case, in the main text we will only use that the vacuum character has a non-null overlap with the $h=1$ heavy-state when $c<25$, and this conclusion is independent of the above discussion. So, to summarize, the effect of the vacuum being degenerate is that we get an extra term in the orthogonality relations, which kicks in if the state has dimension $h=1$. We discuss how to modify the Cardy RMT potential to take into account this extra piece at the end of \cref{sec:S_modinv_reg}, however in the rest of the paper we will neglect this correction. Indeed, we will be mainly interested in the large $c\geq 25$ case with no light states aside from the vacuum, so that this $h=1$ term is never active.

We thus have a strategy for \cref{sec:S_modinv_reg}. We can take the modular crossing equation, multiply it by $f$ and a character of our choice in one channel, which we call $P'$, and this will project the equation onto the character with the chosen momenta:
\begin{equation}
\int \frac{d^2\tau}{\sqrt{\Im(\tau)}} Z(\tau) f(\tau) \chi_{P'}(\tau)^* = \sum_{i}\delta(P_i-P') \,.
\end{equation}

\section{Details on the elliptic integral computations}\label{app:elliptic_int}
In this appendix, we report the explicit computation of the elliptic integral appearing in \eqref{eq:epsk_integralCk}, and discuss its cut-pinching limit.
If we denote $b_{k+1}=a$, $a_k=b$, $b_k=c=y$, $a_{k-1}=d$, we can compute the left hand side of elliptic \eqref{eq:epsk_integralCk} as\footnote{refer to formulas $254.21$ and $362.16$ in \cite{elliptic_handbook}.}:
\begin{equation}
\begin{aligned}
     &\int_{b_k}^{a_k}\frac{\sqrt{x-a_{k-1}}}{\sqrt{x-{b_{k}}}}\frac{\sqrt{x-a_k}}{\sqrt{x-{b_{k+1}}}}dx=\\&=\frac{(b-c)(c-d)g}{2\alpha^2 (k^2-\alpha^2)}(\alpha^2 E(k^2)+(k^2-\alpha^2)K(k^2)+(2k^2\alpha^2-\alpha^4-k^2)\Pi(\alpha,k^2)),
\end{aligned}
\end{equation}
where $K$, $E$ and $\Pi$ are respectively the complete elliptic integrals of the first, second and third kind and
\begin{equation}\label{eq:elliptic_param_def}
\begin{aligned}
    g\equiv \frac{2}{\sqrt{(a-c)(b-d)}}&=\frac{2}{\sqrt{(b_{k+1}-b_k) (a_k-a_{k-1})}},\quad \alpha^2\equiv\frac{b-c}{b-d}=\frac{a_k-b_k}{a_k-a_{k-1}}<k^2,\\&k^2\equiv \frac{(b-c)(a-d)}{(a-c)(b-d)}=\frac{(a_k-b_k)(b_{k+1}-a_{k-1})}{(a_k-a_{k-1})(b_{k+1}-b_k)}.
\end{aligned}
\end{equation}
If we insert the expressions above in the result of the elliptic integral, we obtain:
\begin{equation}\label{eq:epsk_elliptic}
\begin{aligned}
    &\int_{b_k}^{a_k}\frac{\sqrt{x-a_{k-1}}}{\sqrt{x-{b_{k}}}}\frac{\sqrt{x-a_k}}{\sqrt{x-{b_{k+1}}}}dx=\\&=\frac{(a_k-a_{k-1})^{3/2}\sqrt{b_{k+1}-b_k}}{(a_k-b_k)}(\alpha^2 E(k^2)+(k^2-\alpha^2)K(k^2)+(2k^2\alpha^2-\alpha^4-k^2)\Pi(\alpha^2,k^2)).
\end{aligned}
\end{equation}
Now we use the cut-pinching approximation so that the elliptic parameters defined in \eqref{eq:elliptic_param_def} become:
\begin{equation}\label{eq:alpha_k_elliptic}
\begin{aligned}
    &g=\frac{2}{\sqrt{\delta_k(\delta_k-d_k+d_{k-1})}}\approx \frac{2}{\delta_k}\biggr(1+\frac{d_k}{2\delta_k}-\frac{d_{k-1}}{2\delta_k}\biggr),\quad\alpha^2=\frac{\delta_k-d_k}{\delta_k-d_k+d_{k-1}}\approx 1-\frac{d_{k-1}}{\delta_k},\\& k^2=\frac{(\delta_k-d_k)(\delta_k+d_{k-1})}{\delta_k (\delta_k-d_k+d_{k-1})}\approx 1-\frac{d_kd_{k-1}}{\delta_k^2},\quad k'^2\equiv 1-k^2\approx\frac{d_kd_{k-1}}{\delta_k^2}.
\end{aligned}
\end{equation}
If we insert the expressions above in \eqref{eq:epsk_elliptic}, we obtain, from \eqref{eq:epsk_integralCk}, equation \eqref{eq:dk_expression}.

\subsection{Region of validity of the cut-pinching approximation}\label{app:cut_pinching_consistency}
In this appendix, we prove that the cut-pinching approximation used to solve \eqref{eq:epsk_integralCk} holds in a parametrically large high-energy region of the spectrum. From \eqref{eq:dk_expression_reorder} and \eqref{eq:dk_sol_approx}, we have that the pinching condition on $d_{k-1}/\delta_k$, assuming that $d_k/\delta_k$ is already parametrically small, at leading logarithmic order becomes:
\begin{equation}\label{eq:dk_dkk_ll1}
   \frac{d_{k-1}}{\delta_k}\log\left(\frac{\delta_k}{d_{k-1}}\right)-\frac{d_{k}}{\delta_k}\log\left(\frac{\delta_k}{d_{k}}\right)\approx2\biggr(\frac{\epsilon_k}{\delta_k C_k}-1\biggr)\ll1,
\end{equation}
which implies, by \eqref{eq:dk_expression_reorder}, that the consistency condition across the single bin should have $\frac{\epsilon_k}{\delta_k C_k}\approx 1$.
In order to translate this condition into a bound on a spectral region, we insert the explicit expression for the occupation fraction \eqref{eq:ak_integral_def}, and define $h_*$ as the energy for which we have:
\begin{equation}
\begin{aligned}
     \frac{\epsilon_k}{\delta_k C_k}_{\bigr|_{h_k=h_*}}&=\pi^2 Q\sqrt{E-h_*}e^{2\pi Q \biggr(\sqrt{h_*-\frac{c-1}{24}}-\sqrt{E-\frac{c-1}{24}}\biggr)}=1\implies_{h_*, E\gg c}\\&\implies \sqrt{h_*}-\sqrt{E}+\frac{1}{4\pi Q}\log\bigr(\pi^4Q^2(E-h_*)\bigr)=0.
\end{aligned}
\end{equation}
Now if we define $\Delta=E-h_*$ and expand the equation above for $\Delta\ll E$, we can find the following solution\footnote{for large $E$ the principal branch supplies a second solution, lying closer to the UV cutoff.}:
\begin{equation}
     h_*\approx E+ \frac{\sqrt{E}}{2\pi Q}W_{-1}\biggr(-\frac{2}{\pi^3Q\sqrt{E}}\biggr)\approx E- \frac{\sqrt{E}}{4\pi Q}\log(E)+\dots
\end{equation}
We remark that such a branch choice for the Lambert $W$-function captures the case $d_k<d_{k-1}$, that is when the LHS of \eqref{eq:dk_dkk_ll1} is positive. In particular then, we can interpret this $h_*$ as the leftmost energy of the consistency window for the pinching approximation appropriate to describe the case where the parameters $d_k$ are monotonously decreasing with $k$.
Now if we expand the occupation fraction around $h_*$:
\begin{equation}\label{eq:dk_val_bulk_shell}
    \frac{\epsilon_k}{\delta_k C_k}\approx_{h_k\approx h_*}1+(h_k-h_*)\frac{\pi Q}{\sqrt{h_*-\frac{c-1}{24}}}\gtrapprox 1\quad \mathrm{when}\quad 0<h_k-h_*\ll\frac{\sqrt{h_*-\frac{c-1}{24}}}{\pi Q}\approx \frac{\sqrt{E}}{\pi Q}
\end{equation}
So if we choose the high-energy shell of width $W$ according to the condition above, we have that for every bin $d_k/\delta_k\ll1\implies d_{k-1}/\delta_k\ll 1$. However, this procedure needs to be recursively iterated for all the $\mathcal{O}(W/\delta)$ bins inside the window\footnote{we used that at high energies the bins' width are $\delta_k\approx \delta=2G$.}. So the consistency condition for the full high-energy shell is the following:
\begin{equation}
    W^2\frac{\pi Q}{\delta\sqrt{E}}\ll 1\implies W\ll \sqrt{\frac{\delta}{\pi Q}} E^{1/4}
\end{equation}
So, provided the initial datum of the recursion, that is the pinching parameter of the rightmost bin, is small, we have shown that the pinching approximation is self-consistent at least within a
parametrically large high-energy region, of characteristic width $\propto E^{1/4}$. We expect
this condition to be met at sufficiently high energies, and the numerics of
\cref{fig:inbin_density_ansatz} bear this out at the upper end of the window. We remark that the
region in which $d_k/\delta_k\ll1$ may in principle extend beyond the characteristic width found
above; determining how far would require explicit knowledge of the earlier recursive data, that
is of the pinching parameters of the bins at higher energy, which are themselves fixed by the
eigenvalue distributions there.

\section{Details on the topological recursion}\label{app:g=1_top_rec_details}
In this appendix, we provide some details regarding the topological recursion procedure used in \cref{sec:top_recursion}. We start with a fast review of the generic topological recursion procedure, to be applied to the genus $1$ case of interest in this paper, in \cref{app:top_rec_recap}. Then, in \cref{app:torus_deg}, we discuss how, in the cut-pinching limit introduced in \cref{sec:mesoscopic_inbin}, the genus $1$ spectral curve degenerates to the more standard genus $0$ case. In particular, in this case we will focus on the computation of the spectral form factor, the details of which we report in \cref{app:sff_details}.
\subsection{Recursion formulae}\label{app:top_rec_recap}
In this appendix we want to give a self-contained, albeit brief, report of the main formulas to compute higher-point correlations using the topological recursion. For more details and derivations please refer to \cite{eynard2018randommatrices, Eynard:2016yaa}.

Let us consider a random matrix theory with (single-trace) potential $V$, and eigenvalue density $\rho(x)$ in the $N\to\infty$ limit. 
We define the connected $n$-point correlation function $W_n$, and its formal asymptotic expansion in powers of $N$:
\begin{equation}\label{eq:n_resolvent_def}
    W_n(x_1,\dots x_n)\equiv \frac{1}{N^n}\biggr\langle\mathrm{tr}\frac{1}{x_1-M}\dots\mathrm{tr}\frac{1}{x_n-M}\biggr\rangle_c=\sum_{g=0}^{\infty}N^{2-2g-2n}W_{gn}(x_1,\dots,x_n).
\end{equation}
Notice that, in the $N\to\infty$ limit, we can determine the planar $W_{01}$ from the eigenvalue density, and hence from the potential $V$, using the following standard relation \cite{Livan_2018}:
\begin{equation}
    \rho(x)=-\frac{1}{\pi}\Im W_{01}(x).
\end{equation}
Now, we assume that the eigenvalue density $\rho(x)$ has a support made by the union of $n+1$ disjoint intervals. Such a density implies in the complex plane a branch cut structure which can be unraveled on a genus $n$ Riemann surface $\Sigma$, via a uniformization coordinate $z:\Sigma\to\mathbb{P}^1$, where $\mathbb{P}^1 \simeq \mathbb{C}\cup \{\infty\}$. At this point we define the following:
\begin{equation}
   y(z)\equiv W_{01}(x(z)),\quad\mathrm{and}\quad \omega_{01}=y(z)dx(z).
\end{equation}
Analogously, we can repackage the information contained in the higher-point resolvent correlations \eqref{eq:n_resolvent_def}, in the differential forms defined as \cite{eynard2018randommatrices}:
\begin{equation}\label{eq:n_wgn_def}
     \omega_{gn}(z_1,\dots,z_n)\equiv W_{gn}(x(z_1),\dots x(z_n))dx(z_1)\dots dx(z_n)+\delta_{0g}\delta_{2n}\frac{dx(z_1)dx(z_2)}{(x(z_1)-x(z_2))^2}.
\end{equation}
In particular, notice that the Kronecker deltas $\delta_{0g}, \delta_{2n}$ add a term to the definition of $\omega_{02}$. Indeed, we can define, after fixing a basis of closed A-cycles $A_i$, for a Riemann surface $\Sigma$, a fundamental second kind differential as a meromorphic symmetric bilinear differential form $B(z_1,z_2)$ whose only singularity is a double pole and such that:
\begin{equation}
    B(z_1,z_2)\approx_{z_1\to z_2}\frac{dz_1 dz_2}{(z_1-z_2)^2}+\mathrm{reg.},\quad \forall i\; \oint_{z_1\in A_i}B(z_1,z_2)=0.
\end{equation}
Given that $\omega_{02}\equiv B$, which we refer to as the Bergman kernel, the last term of \eqref{eq:n_wgn_def} ensures that $B$ only has the required diagonal singularity.

Now, we can use the fact that the resolvent correlation function must respect the so-called loop equations \cite{eynard2018randommatrices}. In particular, for a single-trace potential $V$, if we insert the topological expansion of \eqref{eq:n_resolvent_def}, we can recast these loop equations as a solvable triangular system of equations. Then, in terms of the differential forms \eqref{eq:n_wgn_def}, the loop equations are solved by the following recursive expressions:
\begin{equation}\label{eq:top_rec_wgn}
    \begin{aligned}
\omega_{g, n}\left(z_1, \cdots z_n\right)=\sum_a \operatorname{Res}_{z=a} K_a\left(z_1, z\right) & {\Bigg[\omega_{g-1, n+1}\left(z, \sigma_a(z), z_2, \cdots z_n\right)} \\
& +\sum_{\substack{h+h^{\prime}=g \\
I \sqcup I^{\prime}=\left\{z_2, \cdots z_n\right\}}}^{\prime} \omega_{h, 1+|I|}(z, I) \omega_{h^{\prime}, 1+\left|I^{\prime}\right|}\left(\sigma_a(z), I^{\prime}\right)\Bigg],
\end{aligned}
\end{equation}
where $\sum_a$ is performed over all branch points such that $x'(a)=0$ and the notation of the primed summation $\sum^\prime$ indicates that we have to exclude the terms $(h,I)=(0,\emptyset)$ and $(h,I)=(g,\{z_2,\dots z_n\})$.
In the formula above, $\sigma_a(z)$ is the local Galois involution around $a$, determined by $x(\sigma_a(z))=x(z)$ and $\sigma_a(a)=a$, and we defined the following recursion kernel:
\begin{equation}
    K_a(z_1,z)\equiv \frac{1}{2}\frac{\int_{\sigma_a(z)}^z\omega_{02}(z_1,z')dz'}{\omega_{01}(z)-\omega_{01}(\sigma_a(z))}.
\end{equation}
To summarize, we reviewed a computational procedure that, given the data contained in the spectral curve $(\Sigma, \omega_{01}, B)$, determines all the topological invariants $\omega_{gn}$. We now proceed to give example computations in the genus $0$ case, the most common in the literature, as well as the genus $1$ of interest in \cref{sec:top_recursion}. In particular, we will focus on $\omega_{03}$, the simplest non-trivial observable, given from \eqref{eq:top_rec_wgn} by:
\begin{equation}\label{eq:w03_general}
\begin{aligned}
   \omega_{03}(z_1,z_2,z_3)&=2\sum_a\mathrm{Res}_{z=a}K_a(z_1,z)B(z,z_2)B(\sigma_a(z),z_3)=\\&= \sum_{a}\mathrm{Res}_{z=a}\frac{w_{0,2}(z,z_1)w_{0,2}(z,z_2)w_{0,2}(z,z_3)}{dx(z) dy(z)}.
\end{aligned}
\end{equation}

\subsection{Computations for the genus $0$ and genus $1$ cases}
We briefly report some example computations using the genus $0$ topological recursion, in the case where the eigenvalue density has support on a single interval $[a,b]$\footnote{It is going to be useful to have this case in mind when we discuss the degeneracy, in the pinching limit, of the genus $1$ case examined in this paper (see \cref{app:torus_deg}).}.  For a more extended discussion on the genus $0$ recursion you can refer for example to \cite{Eynard:2007kz, Eynard:2016yaa, eynard2018randommatrices}. Then, we move to examine the topological recursion for the genus $1$ case encountered in \cref{sec:top_recursion}, already discussed for example in \cite{Bonnet_2000,Mari_o_2009}. 
\subsubsection*{The genus $0$ case recap}
For an eigenvalue density supported on an interval $[a,b]$, the associated Riemann surface $\Sigma$ is a sphere, and the uniformization coordinate can be chosen to be the Joukowski map:
\begin{equation}\label{eq:jouk_def}
    x(z)=\frac{a+b}{2}+\frac{b-a}{4}\left(z+\frac{1}{z}\right).
\end{equation}
Notice that the only ramification points are in $z_i=\pm 1$, and that changing sheets at fixed $x$ corresponds to the involution $z\to1/z$.

In the genus $0$ case, the Bergman kernel is uniquely determined, independently of the details of $\Sigma$, as:
\begin{equation}
    B(z,z')=\frac{dzdz'}{(z-z')^2}=\omega_{02}(z,z').
\end{equation}
At this point, we can already determine the $2$ point function resolvent, defined from $\omega_{02}$ using \eqref{eq:n_wgn_def}, noticing that:
\begin{equation}\label{eq:W02_jouk}
    \frac{dxdx'}{(x-x')^2}=B(z,z')+B\left(z,\frac{1}{z'}\right)\implies W_{02}(x_1,x_2)dx_1dx_2=-B\left(z_1,\frac{1}{z_2}\right)
\end{equation}
Now, we should specialize to the specific eigenvalue density of interest and find the appropriate resolvent that completes the full set of data of the spectral curve needed to determine all its topological invariants $\omega_{gn}$ is \eqref{eq:w03_general}.

We finish this brief summary of the genus $0$ recursion with the following remark. If we expand around $z\approx1$, the uniformization coordinates \eqref{eq:jouk_def} become:
\begin{equation}
    x(z)\approx b+\frac{b-a}{4}(z-1)^2+\dots,
\end{equation}
which can be rescaled to define a new uniformization coordinate $\Tilde{x}(\xi)=\xi^2$. This regime can be reached via a double-scaling procedure, where we zoom in into one endpoint and send to infinity the other. In particular, this is the setting where the Airy model as well as the JT gravity topological recursions are defined, which are determined respectively by the spectral curves $\{y(\xi)=\xi,x(\xi)=\xi^2\}$ and $\{y(\xi)=\sin(2\pi\xi)/(2\pi),x(\xi)=\xi^2\}$ \cite{Saad:2019lba}.

\subsubsection*{The genus $1$ case of interest}
Now, we move to discuss the considerably less common case of a genus $1$ topological recursion. Let us assume that the branch points subtending the spectral curve of interest are $x_1<x_2<x_3<x_4$, and consider the uniformization coordinates \eqref{eq:uniform_abel_map} for the torus, reported here for the reader's convenience:
\begin{equation}
\begin{aligned}
    z(x)=\int_{b_k}^{x}&\frac{dt}{\sqrt{(t-x_1)(t-x_2)(t-x_3)(t-x_4)}}=g F(\alpha(x),k),\\&\qquad\mathrm{where}\qquad \alpha(x)=\arcsin{\sqrt{\frac{(x_3-x_1)(x-x_2)}{(x_3-x_2)(x-x_1)}}}.
\end{aligned}
\end{equation}
Using the definition of the Jacobi sine function $\sin\alpha=\mathrm{sn}(F(\alpha),k)$, we can invert the map and obtain:
\begin{equation}\label{eq:sn_to_x_map}
    \mathrm{sn}(\mathrm{u}(x))=\sqrt{\frac{(x_3-x_1)(x-x_2)}{(x_3-x_2)(x-x_1)}}\implies x(\mathrm{u})=\frac{\left(\mathrm{sn(u)}^2-1\right) x_1 x_2-\mathrm{sn(u)}^2 x_1 x_3+x_2 x_3}{\mathrm{sn(u)}^2 (x_2-x_3)-x_1+x_3},
\end{equation}
where we defined $\mathrm{u}=z/g$\footnote{please notice that the $\mathrm{u}$ coordinate defined here, and appearing exclusively throughout this appendix, is different from the $u$ of \eqref{eq:jacobi_unif}, used in \cref{app:torus_deg}.} and from now on we will often omit the elliptic $k$ in the arguments of the Jacobi functions for brevity. If we define $\mathrm{u}(\infty)=\mathrm{u}_\infty$, and take the $x\to\infty$ limit in the equation above, we can find the following useful relation:
\begin{equation}\label{eq:sn_x_to_infty}
    \mathrm{sn}^2(\mathrm{u}_\infty)=\frac{x_3-x_1}{x_3-x_2}.
\end{equation}
If we use the Jacobi elliptic trigonometric relations \cite{akhiezer1990elements}, we can find analogously the following from \eqref{eq:sn_x_to_infty}:
\begin{equation}
\begin{aligned}
     & \mathrm{cn}^2(\mathrm{u})+\mathrm{sn}^2(\mathrm{u})=1\implies \mathrm{cn}^2(\mathrm{u}_\infty)=\frac{x_1-x_2}{x_3-x_2},\\&k^2\mathrm{sn}^2(\mathrm{u})+\mathrm{dn}^2(\mathrm{u})=1\implies \mathrm{dn}^2(\mathrm{u}_\infty)=\frac{x_1-x_2}{x_4-x_2}.
\end{aligned}
\end{equation}
From these results, we can find the following ratios:
\begin{equation}
    \frac{\mathrm{sn}^2(\mathrm{u})}{\mathrm{sn}^2(\mathrm{u}_\infty)}=\frac{x-x_2}{x-x_1},\quad \frac{\mathrm{cn}^2(\mathrm{u})}{\mathrm{cn}^2(\mathrm{u}_\infty)}=\frac{x_3-x}{x_1-x},\quad\frac{\mathrm{dn}^2(\mathrm{u})}{\mathrm{dn}^2(\mathrm{u}_\infty)}=\frac{x_4-x}{x_1-x}.
\end{equation}
The goal of the manipulations above is to express the $(x-x_i)$ factors as a function of $u$ in a convenient manner, and indeed we obtain:
\begin{equation}
    \begin{gathered}
x-x_3=\left(x_2-x_3\right) \frac{\mathrm{cn}^2(u)}{1-\frac{\mathrm{sn}^2(\mathrm{u})}{\mathrm{sn}^2\left(\mathrm{u}_{\infty}\right)}}, \quad x-x_4=\left(x_2-x_4\right) \frac{\mathrm{dn}^2(u)}{1-\frac{\mathrm{sn}^2(\mathrm{u})}{\mathrm{sn}^2\left(\mathrm{u}_{\infty}\right)}}, \\
x-x_1=\left(x_2-x_1\right) \frac{1}{1-\frac{\mathrm{sn}^2(\mathrm{u})}{\mathrm{sn}^2\left(\mathrm{u}_{\infty}\right)}}, \quad x-x_2=\left(x_3-x_2\right) \frac{x_2-x_1}{x_3-x_1} \frac{\mathrm{sn}^2(\mathrm{u})}{1-\frac{\mathrm{sn}^2(\mathrm{u})}{\mathrm{sn}^2\left(\mathrm{u}_{\infty}\right)}}.
\end{gathered}
\end{equation}
Now we can specialize to the model encountered in this paper, and make use of the equations above to conveniently express $y(x)$ from \eqref{eq:spectral_curve_ansatz} in terms of the uniformization coordinates. As in the rest of this appendix, we restrict to the $j$-th bin, so that we obtain:
\begin{equation}\label{eq:sp_curve_y}
     y_j(\mathrm{u})=(i\pi C_j)\sqrt{\frac{(x(\mathrm{u})-x_1)(x(\mathrm{u})-x_3)}{(x(\mathrm{u})-x_2)(x(\mathrm{u})-x_4)}}=(i\pi C_j)\frac{\mathrm{cn(u,k)}}{\mathrm{sn(u},k)\mathrm{dn(u},k)}\sqrt{\frac{x_3-x_1}{x_4-x_2}},
\end{equation}
where we notice that the last factor is such that $(x_3-x_1)/(x_4-x_2)\approx 1$ to leading order in the pinching limit.

\begin{figure}
\centering
        \includegraphics[width=\textwidth]{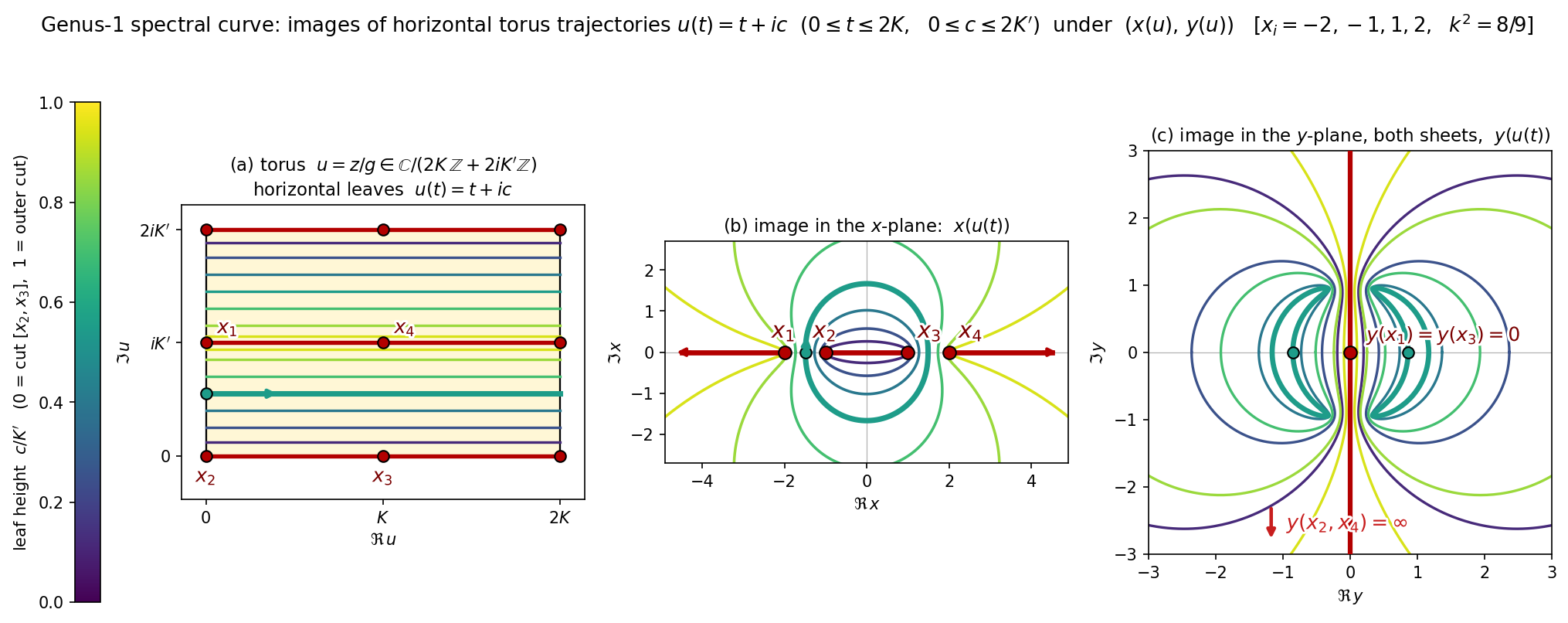}
    \caption{Visualization of the genus-1 spectral curve $(\Sigma, y dx, B)$.  \emph{Left (a)}: $\Sigma$ is a torus, obtained in the uniformization coordinate $\rm u$ as a parallelogram in a complex plane with periods $\omega_1=2K(k)$ and $\omega_2=2iK(k')$. For visualization purposes, we performed a horizontal foliation of this surface with the indicated color grading on $\Im \rm u=c$. Notice that we highlighted the positions of  $(x_1,x_2,x_3,x_4)$ as well as an example distinguished point in green. In plots (b) and (c) we indicate analogously their images respectively under $x(\rm u$ and $y(\rm u)$. \emph{Center (b)}: Trajectories in the complex $x$-plane given by $x(\rm u(t))$, given by \eqref{eq:sn_to_x_map}, with the aforementioned color-grading to distinguish different values of $c$. \emph{Right (c)}: Trajectories (on both sheets) in the complex $y$-plane given by \eqref{eq:sp_curve_y}, with the same color-grading convention explained above.}
    \label{fig:curve_unif}
\end{figure}

At this point, once we insert $\omega_{01}$ and $\omega_{02}$, the machinery of the topological recursion is going to determine all the $\omega_{gn}$. In particular, we can use the result above, together with \eqref{eq:sn_to_x_map}, to obtain $\omega_{01}(\mathrm{u})=y(x(\mathrm{u}))dx(\mathrm{u})=y(\mathrm{u})\frac{dx(\mathrm{u})}{d\mathrm{u}}d\mathrm{u}$, while $\omega_{02}$, that is the torus Bergman kernel, is given by \eqref{eq:berg_ker_torus}. Now, in order to use \eqref{eq:w03_general}, the only thing left to specify are the $\mathrm{u}$-coordinates of the branch points $(x_1,x_2,x_3,x_4)$. As already discussed, these are the solutions of the equation $x'(\mathrm{u})=0$, that is:
\begin{equation}
    x'(\mathrm{u})=\frac{2(x_2-x_1)(x_3-x_1)(x_3-x_2)\mathrm{cn}(\mathrm{u},k)\mathrm{dn}(\mathrm{u},k)\mathrm{sn}(\mathrm{u},k)}{(x_1-x_3+(x_3-x_2)\mathrm{sn}(\mathrm{u},k)^2)^2}=0.
\end{equation}
If we denote by $\Tilde{\mathrm{u}}_i$ the coordinate corresponding to the position of the branch point on the real axis $x_i$, we obtain the following:
\begin{equation}
    \begin{aligned}
        &(x_1);\quad\mathrm{sn(\Tilde{u}_1)}=\infty\implies \Tilde{\mathrm{u}}_1=-iK(k')\approx_{k\approx 1} -\frac{i\pi}{2}-\frac{i\pi}{8}\frac{d_kd_{k-1}}{\delta_k^2},\\& (x_2);\quad\mathrm{sn(\Tilde{u}_2)}=0\implies \Tilde{\mathrm{u}}_2=0,\quad\qquad(x_3);\quad \mathrm{sn}(\Tilde{\mathrm{u}}_3)=1\implies \mathrm{\Tilde{u}}_3=K(k),\\&(x_4);\quad \mathrm{sn}(\Tilde{\mathrm{u}}_4)=\sqrt{\frac{(x_3-x_1)(x_4-x_2)}{(x_3-x_2)(x_4-x_1)}}\implies \mathrm{dn}(\Tilde{\mathrm{u}}_4)=0\implies \mathrm{\Tilde{u}}_4=K(k)-iK(k').
    \end{aligned}
\end{equation}
For convenience, let us define a $\Tilde{B}(\mathrm{u},\mathrm{u}')$ such that the Bergman kernel is expressed as $B(\mathrm{u},\mathrm{u}')=\Tilde{B}(\mathrm{u},\mathrm{u}') d\mathrm{u} d\mathrm{u}'$. Then \eqref{eq:w03_general} prescribes that we should compute the residues in $\mathrm{u}=\Tilde{\mathrm{u}}_i$ of the following expression:
\begin{equation}
    \frac{\Tilde{B}(\mathrm{u},\mathrm{u}_1)\Tilde{B}(\mathrm{u},\mathrm{u}_2)\Tilde{B}(\mathrm{u},\mathrm{u}_3)}{x'(\mathrm{u})y_j'(\mathrm{u})}\,d\mathrm{u}\,d\mathrm{u}_1\,d\mathrm{u}_2\,d\mathrm{u}_3.
\end{equation}
In particular, we find that the residues in $\Tilde{\mathrm{u}}_2$ and $\Tilde{\mathrm{u}}_4$ are zero, as the poles in $y_j'(\mathrm{u})$ simplify the zeroes of $x'(\mathrm{u})$. In order to compute the $\mathrm{\Tilde{u}_1}$ residue, it is convenient again to leverage the pinching approximation, in which we have:
\begin{equation}
     \begin{aligned}
   x'(\mathrm{u})\approx_{\mathrm{u}\sim \Tilde{\mathrm{u}}_1} -\frac{2k(x_2-x_1)(x_3-x_1)}{(x_3-x_2)\mathrm{sn}(\mathrm{u},k)},\quad \frac{y'(\mathrm{u})}{i\pi C_j}\approx\frac{k^2-1-k^2\,\mathrm{cn}(\mathrm{u},k)^4}{\mathrm{dn}(\mathrm{u},k)^2\mathrm{sn}(\mathrm{u},k)^2}\approx_{\mathrm{u}\sim \Tilde{\mathrm{u}}_1}1
    \end{aligned}
\end{equation}
At this point, if we expand the Jacobi sine around $\mathrm{\Tilde{u}_1}$, we obtain $\mathrm{sn}(\mathrm{\Tilde{u}},k)\approx \frac{1}{k(\mathrm{u}-\mathrm{\Tilde{u}_1})}+\mathcal{O}(\mathrm{u}-\mathrm{\Tilde{u}_1}))$, and the residue we are searching for becomes\footnote{useful information on the poles and residues of Jacobi elliptic functions can be found in \url{https://dlmf.nist.gov/22} and in \url{https://functions.wolfram.com}.}
\begin{equation}
    \mathrm{Res}_{\mathrm{u}\to \mathrm{\Tilde{u}_1}}\frac{\Tilde{B}_{0,2}(\mathrm{u},\mathrm{u}_1)\Tilde{B}_{0,2}(\mathrm{u},\mathrm{u}_2)\Tilde{B}_{0,2}(\mathrm{u},\mathrm{u}_3)}{x'(\mathrm{u}) y'(\mathrm{u})}du\approx-\frac{\Tilde{B}_{0,2}(\Tilde{\mathrm{u}}_1,\mathrm{u}_1)\Tilde{B}_{0,2}(\Tilde{\mathrm{u}}_1,\mathrm{u}_2)\Tilde{B}_{0,2}(\Tilde{\mathrm{u}}_1,\mathrm{u}_3)}{2k^2(x_2-x_1)(i\pi C_j)}.
\end{equation}
Finally, in an analogous manner we can determine the residue in $\mathrm{\Tilde{u}}_3$:
\begin{equation}
    \mathrm{Res}_{\mathrm{u}\to \mathrm{\Tilde{u}}_3}\frac{\Tilde{B}_{0,2}(\mathrm{u},\mathrm{u}_1)\Tilde{B}_{0,2}(\mathrm{u},\mathrm{u}_2)\Tilde{B}_{0,2}(\mathrm{u},\mathrm{u}_3)}{x'(\mathrm{u}) y'(\mathrm{u})}du=\frac{(x_2-x_1)\Tilde{B}_{0,2}(\mathrm{\Tilde{u}}_3,\mathrm{u}_1)\Tilde{B}_{0,2}(\mathrm{\Tilde{u}}_3,\mathrm{u}_2)\Tilde{B}_{0,2}(\mathrm{\Tilde{u}}_3,\mathrm{u}_3)}{2(1-k^2)(x_3-x_2)(x_3-x_1)(i\pi C_j)}.
\end{equation}
Now, if we sum the residues, we obtain from \eqref{eq:w03_general} the $\omega_{0,3}$ we were searching for:
\begin{equation}\label{eq:elliptic_3pt}
\begin{aligned}
    &\frac{w_{0,3}(u_1,u_2,u_3)}{du_1du_2du_3}\approx\\&\approx\frac{\Tilde{B}_{0,2}(\mathrm{\Tilde{u}}_3,\mathrm{u}_1)\Tilde{B}_{0,2}(\mathrm{\Tilde{u}}_3,\mathrm{u}_2)\Tilde{B}_{0,2}(\mathrm{\Tilde{u}}_3,\mathrm{u}_3)}{2d_k(i\pi C_j)}-\frac{\Tilde{B}_{0,2}(\Tilde{\mathrm{u}}_1,\mathrm{u}_1)\Tilde{B}_{0,2}(\Tilde{\mathrm{u}}_1,\mathrm{u}_2)\Tilde{B}_{0,2}(\Tilde{\mathrm{u}}_1,\mathrm{u}_3))}{2d_{k-1}(i\pi C_j)}.
\end{aligned}
\end{equation}
 We remind the reader that, due to the assumptions subtending the spectral curve equation \eqref{eq:spectral_curve_ansatz}, the above expression rigorously holds only for correlations inside the same bin, contained in some high-energy window. In principle, all other topological invariants $\omega_{gn}$ can be computed in an analogous way by running the recursion machinery. 
 Notice that \eqref{eq:elliptic_3pt} is
computed on the bin-localized genus one curve, before any averaging over the bins' positions has
been performed. Under the prescription of \cref{sec:mesoscopic_inbin} that averaging acts on the
mesoscopic density and collapses the curve to the genus zero case, so what we have computed here
is precisely the regularization dependence that the \textit{a priori} averaging removes.

Several limitations should be kept in mind before pushing the topological recursion machinery further. Removing the information inserted upon bin-regularization \textit{a posteriori} requires averaging the
$\omega_{gn}$ themselves over the bins' centres, which is very hard to do analytically.
Single-bin correlators are in any case likely to be too restrictive to capture interesting global-structure
effects, and for higher-point functions the neglected non-vacuum kernel contributions could
become important as well. More importantly, the whole construction rests on the eigenvalues
moving in a constant potential inside each bin, which holds because the Cardy potential depends
on them only through the occupation numbers. Once the remaining bootstrap constraints are
imposed through their own constraint$^2$ potentials, as discussed in \cref{sec:discussion}, this
will no longer be the case: they will survive within each bin as a residual potential, deforming
the mesoscopic density and with it the whole higher-genus structure computed above. For these
reasons, we will mostly focus on the random matrix features that can be extracted directly from
the Bergman kernel, namely the late-time ramp of the spectral form factor, which sees the
spectral curve only through its cut structure and is therefore the most robust of these
observables.

\subsubsection{Details on the spectral curve degeneration}\label{app:torus_deg}
In this section, we give some details on how the elliptic functions, used to describe the genus 1 spectral curve in the previous section, degenerate to the genus 0 case in certain limits.

We can rewrite the Abel map \eqref{eq:uniform_abel_map} in terms of simpler standard special functions provided that the spectral edges occupy special positions. We will choose the symmetric convention where such positions in the complex plane are $(-1/\kappa,-1,1,1/\kappa)$. In this case, the uniformization map can be written in terms of Jacobi elliptic functions as \cite{eynard2018randommatrices}:
\begin{equation}\label{eq:jacobi_unif}
    t(u)=\mathrm{sn}(u,\kappa),
\end{equation}
which, as expected for a torus coordinate, is doubly periodic with periods $2\omega_1=4K(\kappa)$ and $2\omega_2=2i K(\kappa')$. Notice that we can always bring the original branch points $(x_1,x_2,x_3,x_4)$ to $(-1/\kappa,-1,1,1/\kappa)$ using a Möbius transformation of the form:
\begin{equation}\label{eq:mobius_to_symm}
    x(u)=\frac{at(u)+b}{ct(u)+d}=A+D\left(\frac{t(u)+\sigma}{1+\sigma t(u)}\right)\quad\mathrm{with}\quad ad-bc\neq0,
\end{equation}
where we re-parametrized the complex numbers $d=1$, $c=\sigma$, $b=A+D\sigma$ and $a=D+A\sigma$. Before determining these parameters as a function of the original branch point positions $(x_1,x_2,x_3,x_4)$ we discuss the relation between the uniformization \eqref{eq:jacobi_unif} and \eqref{eq:uniform_abel_map}. Indeed, if we consider the Abel coordinates of \eqref{eq:uniform_abel_map} with branch point positions $(-1/\kappa,-1,1,1/\kappa)$, we obtain:\\
\begin{equation}\label{eq:abel_symm}
    z(t)=\int_{x_2=-1}^{t}\frac{\kappa ds}{\sqrt{(1-s^2)(1-\kappa^2s^2)}}=\frac{2\kappa}{\kappa+1} F\biggr(\arcsin{\sqrt{\frac{(\kappa+1)(t+1)}{2(\kappa t+1)}}},\frac{2\sqrt{\kappa}}{\kappa+1}\biggr).
\end{equation}
In particular, notice that $k=2\sqrt{\kappa}/(\kappa+1)$ so that in the pinching limit $k\to1\implies \kappa\to 1$.
When the branch point are in the special position $(-1/\kappa,-1,1,1/\kappa)$, a cleaner choice, which amounts only to a scaling plus shifting procedure, is the following:
\begin{equation}\label{eq:jacobi_def_int}
    u(t)=\int_{0}^{t}\frac{ ds}{\sqrt{(1-s^2)(1-\kappa^2s^2)}}=F(\arcsin{t},\kappa)=\mathrm{sn}^{-1}(t,\kappa),
\end{equation}
which upon inversion gives back \eqref{eq:jacobi_unif}.
Notice, with respect to \eqref{eq:abel_symm}, that we not only reabsorbed the constant $\kappa$, but also shifted the coordinates by a quarter-period to move the lower integration extremum, so they are connected by the following relation:
\begin{equation}\label{eq:Abel_to_Jacobi}
    z=\kappa(u+K(\kappa)).
\end{equation}
We can define the half-periods in the coordinates \eqref{eq:jacobi_def_int} analogously to \eqref{eq:ks_torus}. In particular, for torus parameters related by 
\begin{equation}\label{eq:landen_k}
    k=\frac{2\sqrt{\kappa}}{1+\kappa},\quad k'\equiv \sqrt{1-k^2}=\frac{1-\kappa}{1+\kappa},
\end{equation}
the following (Landen) identity holds for the elliptic integrals
\begin{equation}\label{eq:landen_int}
    K(k)=(1+\kappa)K(\kappa),\quad K(k')=\frac{1+\kappa}{2}K(\kappa')
\end{equation}
where we defined $\kappa'^2=1-\kappa^2$. We report that \eqref{eq:landen_k} is called an ascending Landen's transformation \cite{lawden1989elliptic}, which, as we can consistently check using \eqref{eq:landen_int}, halves the period of the torus, as anticipated after \eqref{eq:jacobi_unif}.

Now we determine the parameters appearing in \eqref{eq:mobius_to_symm} as a function of the original branch point positions $(x_1,x_2,x_3,x_4)$ and find:
\begin{equation}\label{eq:AD_param}
    A=\frac{x_2+x_3}{2},\quad D=\frac{x_3-x_2}{2}.
\end{equation}
It is slightly more involved to obtain the expression of $\sigma$. We define $\rho\equiv\frac{1-\kappa}{1+\kappa}$, which is a small parameter in the degeneration limit when $k\to1$, so that:
\begin{equation}\label{eq:krho_deg}
    \kappa=\frac{1-\rho}{1+\rho}\approx 1-2\rho+\mathcal{O}(\rho^2)
\end{equation}
Now, using \eqref{eq:mobius_to_symm}\eqref{eq:AD_param}, we can write, in the degeneration limit \eqref{eq:krho_deg}, the following expressions:
\begin{equation}
\begin{aligned}
    &x_2-x_1=D\frac{(1-\kappa)(1+\sigma)}{k-\sigma}\approx2D\rho\frac{1+\sigma}{1-\sigma},\\&x_4-x_3=D\frac{(1-\kappa)(1-\sigma)}{\kappa+\sigma}\approx2D\rho\frac{1-\sigma}{1+\sigma}.
\end{aligned}
\end{equation}
At this point, we obtain:
\begin{equation}\label{eq:sigma_value}
    \frac{x_2-x_1}{x_4-x_3}=\frac{d_{k-1}}{d_k}\approx\left(\frac{1+\sigma}{1-\sigma}\right)^2\implies \sigma=\frac{\sqrt{d_{k-1}}-\sqrt{d_k}}{\sqrt{d_{k-1}}+\sqrt{d_k}}.
\end{equation}
 Now, we re-parametrize $\sigma\equiv\tanh{\eta}$, so that \eqref{eq:mobius_to_symm} becomes:
\begin{equation}
    x(u)=A+D\tanh({u+\eta}),
\end{equation}
where we used that, in a compact region, $\mathrm{sn}(u,k)\to_{k\sim1}\tanh{u}$ and the hyperbolic tangent addition rule. Now define:
\begin{equation}
\begin{aligned}
     \Xi\equiv\frac{1+\sigma}{1-\sigma}\frac{1+t}{1-t}=&e^{2(u+\eta)}\equiv w^2\implies \frac{t+\sigma}{1+\sigma t}=\frac{\Xi-1}{\Xi+1}, \quad\mathrm{and}\quad Z\equiv\frac{w-i}{w+i}
\end{aligned}
\end{equation}
If we substitute these coordinate changes in \eqref{eq:mobius_to_symm}, we have:
\begin{equation}
    x(Z)=A+D\frac{\Xi-1}{\Xi+1}=A+D\frac{w^2-1}{w^2+1}=A+\frac{D}{2}\left(Z+\frac{1}{Z}\right)
\end{equation}
In particular, we recognize the above expression as the genus 0 Joukowski uniformization associated to the cut $(x_2,x_3)$ \cite{eynard2018randommatrices}:
\begin{equation}\label{eq:jouk_map_deg}
    x(Z)=\frac{x_3+x_2}{2}+\frac{x_3-x_2}{4}\left(Z+\frac{1}{Z}\right)
\end{equation}
However, at this point, if we go back through the coordinate changes above, we can obtain:
\begin{equation}\label{eq:Z_two_valued}
    Z(u)=\frac{\mathrm{sn}(u,\kappa)+\sigma-i\sqrt{1-\sigma^2}\mathrm{cn}(u,\kappa)}{1+\sigma\mathrm{sn}(u,\kappa)}.
\end{equation}
In particular, notice that when $u\to u+2iK(\kappa')$ we have $\mathrm{cn}(u,\kappa)\to-\mathrm{cn}(u,\kappa)$, so that $Z(u)$, as defined on the torus with periods $2\omega_1=4K(\kappa)$ and $2\omega_2=2i K(\kappa')$, is a two-valued function, but it is single-valued on the double-period cover given by the torus $\mathbb{C}/(4K(\kappa)\mathbb{Z}+4iK(\kappa')\mathbb{Z})$. This indicates that the pinched torus is a connected two-sheeted cover of the sphere, and the sphere is obtained by quotienting the sheet exchange involution $w\to-w$.

At this point, we wish to show that, in the coordinate $Z$ appearing in \eqref{eq:jouk_map_deg}, the genus 1 Bergman kernel reduces to the universal genus 0 expression.
The Bergman kernel on the torus, in the coordinates \eqref{eq:uniform_abel_map}, is given by \cite{eynard2018randommatrices}:
\begin{equation}\label{eq:torus_bg}
    B(z,z')=\biggr(\wp(z-z';\omega_1,\omega_2) +2\frac{\eta_1}{\omega_1}\biggr)dzdz',
\end{equation}
By taking the limit $\omega_1\to\infty$ in the Weierstrass functions appearing in \eqref{eq:berg_ker_torus} become:
\begin{equation}\label{eq:wp_pinched}
\begin{aligned}
    &\wp(z;\omega_1,\omega_2)\approx\left(\frac{\pi}{\omega_2}\right)^2\left(\frac{1}{\sin^2{\frac{\pi z}{\omega_2}}}-\frac{1}{3}\right)\\&\zeta\left(z;\omega_1,\omega_2\right)\approx_{\omega_1\to\infty}\frac{1}{3}\left(\frac{\pi}{\omega_2}\right)^2z+\frac{\pi}{\omega_2}\cot{\frac{\pi z}{\omega_2}}\implies\zeta\left(\frac{\omega_1}{2};\omega_1,\omega_2\right)\approx \frac{\pi^2}{6\omega_2^2}\omega_1
\end{aligned}
\end{equation}
If we insert them in \eqref{eq:berg_ker_torus} we obtain the following expression for the degenerate Bergman kernel:
\begin{equation}\label{eq:berg_torus_deg_sinh}
B(z,z')\to_{\omega_{1}\to\infty}\frac{dz dz'}{\sinh^2({z-z'})}.
\end{equation}
Now, let us consider the Bergman kernel for the double-cover torus, with coordinate $v=u/2$, that we introduced after \eqref{eq:Z_two_valued}. We can obtain, by either pulling back \cref{eq:torus_bg,eq:wp_pinched,eq:berg_torus_deg_sinh} under $u=2v$, or by keeping the coordinate $v$ and replacing $\omega_2\to\omega_2/2$, the Bergman kernel on the double cover torus as:
\begin{equation}\label{eq:berg0_deg_change_coords}
    \frac{du du'}{4\sinh^2{\left(\frac{u-u'}{2}\right)}}=\frac{dwdw'}{(w-w')^2}=\frac{dZdZ'}{(Z-Z')^2}=B(Z,Z').
\end{equation}
 So, if we go through all the appropriate coordinate changes described in this section, we obtain precisely the universal form of the Bergman kernel on a sphere in the $Z$-coordinates of \eqref{eq:jouk_map_deg}.

\subsection{The computation of the spectral form factor}\label{app:sff_details}

In this section, we explain in detail how to compute the spectral form factor in a random matrix theory, following for example \cite{Br_zin_1997, Saad:2019lba, okuyama2020multiboundarycorrelatorsjtgravity, okuyama2023discreteanalogueweilpeterssonvolume}, in the genus $0$ and genus $1$ cases described above. First, let us review in the general case how to obtain $Z_{02}(\beta_1,\beta_2)$, and hence the spectral form factor, from the Bergman kernel.

We can define the $2$-point connected resolvent correlator by subtracting the universal singular behavior from the Bergman kernel \eqref{eq:n_wgn_def}:
\begin{equation}\label{eq:w02_gen}
    W_{02}(x,x')dxdx'=B(z_1,z_2)-\frac{dxdx'}{(x-x')^2}.
\end{equation}
We can write the following genus expansion for the connected component of the $n$-partition function correlator
\begin{equation}
    \biggr\langle\prod^n Z(\beta_i)\biggr\rangle_{conn}=\sum_{g=0}^\infty N^{2-2g-n}Z_{g,n}(\beta_1,\dots,\beta_n),
\end{equation}
and, given $\Gamma$ a contour encircling the eigenvalue support, the $Z_{gn}$ are connected to $W_{gn}$ via:
\begin{equation}
    Z_{gn}(\beta_1,\dots,\beta_n)=\oint_\Gamma\dots\oint_\Gamma W_{gn}(z_1,\dots, z_n)\frac{\prod e^{-\beta_ix(z_i)}dx(z_i)}{(2\pi i)}
\end{equation}
In particular, we will focus on the intra-bin spectral form factor, that is when both $x,\;x'\in[x_2,x_3]$, and we have that:
\begin{equation}\label{eq:Z02_integral1_gen}
    Z_{02}(\beta_1,\beta_2)=\oint_\Gamma \frac{dx}{2\pi i}\oint_\Gamma \frac{dx'}{(2\pi i)} W_{02}(x,x')e^{-\beta_1 x}e^{-\beta_2 x'},
\end{equation}
 where the integration happens on nested copies of the closed loop contour $\Gamma$  encircling the $(x_2,x_3)$ cut.

\subsubsection{The genus 0 SFF}\label{eq:sff_genus_0}
In the genus 0 case with support $[x_2,x_3]$, reviewed in \cref{app:top_rec_recap}, from \eqref{eq:w02_gen}, as in \eqref{eq:W02_jouk}, we obtain
\begin{equation}
    W_{02}(x,x')dxdx'=B(Z_1,Z_2)-\frac{dxdx'}{(x-x')^2}=-B\left(Z_1,\frac{1}{Z_2}\right)=\frac{dZ_1dZ_2}{(1-Z_1Z_2)^2},
\end{equation}
so that from \eqref{eq:Z02_integral1_gen}
\begin{equation}\label{eq:Z02_integral1}
    Z_{02}(\beta_1,\beta_2)=\oint\oint \frac{dZdZ'}{(2\pi i)^2}\frac{e^{-\beta_1 x(Z)}e^{-\beta_2 x(Z')}}{(1-ZZ')^2},
\end{equation}
where in the $Z$ coordinates the integration contours surround $[-1,1]$.
Now we use the following mathematical identities respectively on \eqref{eq:jouk_def} and the denominator appearing in \eqref{eq:Z02_integral1}
\begin{equation}
    e^{-\frac{b}{2}(Z+1/Z)}=\sum_{m=-\infty}^\infty (-1)^mI_m(b)Z^m,\quad\frac{1}{(1-ZZ')^2}=\sum_{n=1}^\infty n(ZZ')^{n-1},
\end{equation}
so that we obtain:
\begin{equation}
    Z_{02}(\beta_1,\beta_2)=e^{-(\beta_1+\beta_2)A}\sum_{m,k,n}\oint\oint (-1)^{m+k}n\, I_m(\beta_1 D)I_k(\beta_2 D) Z^{m+n-1}Z'^{k+n-1}\frac{dZdZ'}{(2\pi i)^2}
\end{equation}
The result of the complex integral above, by the residue theorem, is the coefficient of $(ZZ')^{-1}$, that is the terms appearing in the summation with $k=m=-n$:
\begin{equation}\label{eq:Z02_besselsum}
    Z_{02}(\beta_1,\beta_2)=e^{-(\beta_1+\beta_2)A}\sum_{n\geq1}n\,I_n(\beta_1D)I_n(\beta_2D),
\end{equation}
where we used that a property of these Bessel functions is that $I_n(x)=I_{-n}(x)$. Another useful property is that $I_{n-1}(z)-I_{n+1}(z)=\frac{2n}{z}I_n(z)$ so that we obtain:
\begin{equation}
    \begin{aligned}
        &\frac{2n}{\beta_1 D}I_n(\beta_1D)I_n(\beta_2D)=(I_{n-1}(\beta_1 D)-I_{n+1}(\beta_1D))I_n(\beta_2D),\\&\frac{2n}{\beta_2 D}I_n(\beta_1D)I_n(\beta_2D)=(I_{n-1}(\beta_2 D)-I_{n+1}(\beta_2D))I_n(\beta_1D).
    \end{aligned}
\end{equation}
We define $A_n\equiv I_{n-1}(\beta_1D)I_n(\beta_2D)+I_n(\beta_1D)I_{n-1}(\beta_2D)$, so that if we add the two expressions above we obtain
\begin{equation}
    \frac{2(\beta_1+\beta_2)}{D\beta_1\beta_2}nI_{n}(\beta_1D)I_n(\beta_2D)=A_n-A_{n+1}.
\end{equation}
Using the identity above we can express \eqref{eq:Z02_besselsum} as a telescopic sum, so that:
\begin{equation}
    \sum_{n=1}^N nI_{n}(\beta_1D)I_n(\beta_2D)=\frac{D\beta_1\beta_2}{2(\beta_1+\beta_2)}(A_1-A_{N+1})\underset{N\to\infty}{\longrightarrow}\frac{D\beta_1\beta_2}{2(\beta_1+\beta_2)}A_1,
\end{equation}
where we used that, for fixed $z$, $I_N(z)\sim (z/2)^N/N!\to 0$, if we send $N\to\infty$. Then \eqref{eq:Z02_besselsum} becomes:
\begin{equation}\label{eq:Z02_fullbess}
     Z_{02}(\beta_1,\beta_2)=e^{-(\beta_1+\beta_2)A}\frac{D\beta_1\beta_2}{2(\beta_1+\beta_2)}(I_0(\beta_1D)I_1(\beta_2D)+I_1(\beta_1D)I_0(\beta_2D)).
\end{equation}
The expression above greatly simplifies in the limit where $\beta_{1,2}D\gg 1$, when we can use the asymptotic behavior of the Bessel functions $I_n(z)\approx_{z\gg1} e^z/\sqrt{2\pi z}$ to obtain:
\begin{equation}\label{eq:Z02_finitebetas}
    Z_{02}(\beta_1, \beta_2)\approx e^{-(\beta_1+\beta_2)(A-D)}\frac{\sqrt{\beta_1\beta_2}}{2\pi(\beta_1+\beta_2)}
\end{equation}
At this point, if we analytically continue $\beta_1\to\beta+it$ and $\beta_2\to\beta-it$ in \eqref{eq:Z02_finitebetas}, we obtain the late-time $t\gg1/D$ expression of the spectral form factor $SFF(\beta,t)$ at finite temperatures:
\begin{equation}\label{eq:SFF_genus0}
    SFF(\beta,t)\equiv Z_{02}(\beta+it, \beta-it)=e^{-(\beta_1+\beta_2)x_2}\frac{\sqrt{\beta^2+t^2}}{4\pi \beta}.
\end{equation}
Notice that this is the linear ramp of the connected spectral form factor when $t\gg\beta$, so, in particular, this designates $1/D$ as the Thouless time. For completeness, we also extract from \eqref{eq:Z02_fullbess} the infinite temperature limit $\beta\to0$:
\begin{equation}
\begin{aligned}
    &SFF(0,t)= \frac{Dt}{2}(Dt J_0(Dt)^2-J_0(Dt)J_1(Dt)+Dt J_1(Dt)^2)=\\&\approx\frac{Dt}{2}\biggr(\frac{2}{\pi}(\cos^2(Dt-\pi/4)+\sin^2(Dt-\pi/4))-\frac{2}{\pi Dt}\cos(Dt-\pi/4)\sin(Dt-\pi/4))\biggr),
\end{aligned}
\end{equation}
where we used that $I_n(it)=i^nJ_n(t)$ together with the asymptotic expansions $J_0(x)\sim \sqrt{\frac{2}{\pi x}}\cos(x-\pi/4)$ and $J_1(x)\sim \sqrt{\frac{2}{\pi x}}\sin(x-\pi/4)$. So, in the late time limit $Dt\gg 1$, we obtain:
\begin{equation}\label{eq:sff_inf_T}
    SFF(0,t)\approx \frac{Dt}{\pi}.
\end{equation}

\subsubsection{The elliptic SFF and its pinching limit}
\label{app:elliptic-sff}

We now compare the genus-zero result above with the spectral form factor
computed directly on the genus-one curve at finite pinching parameter. We use
the Jacobi parametrization of \cref{app:torus_deg}, which we report again here for convenience:
\begin{equation}
 x(u)
 =
 A+D\frac{\operatorname{sn}(u,\kappa)+\sigma}
 {1+\sigma\operatorname{sn}(u,\kappa)},
 \qquad
 A=\frac{x_2+x_3}{2},
 \qquad
 D=\frac{x_3-x_2}{2}.
 \label{eq:elliptic-x-map}
\end{equation}
We re-label the periods of the torus as
\begin{equation}
 u\sim u+L,
 \qquad
 u\sim u+iH,
 \qquad
 L=4K(\kappa),
 \qquad
 H=2K(\kappa'),
 \label{eq:elliptic-periods}
\end{equation}
and we introduce
\begin{equation}
 \tau=i\tau_2,
 \qquad
 \tau_2\equiv\frac{H}{L}
 =\frac{K(\kappa')}{2K(\kappa)}
 =\frac{K(k')}{K(k)},
 \qquad
 Q_\tau\equiv e^{2i\pi \tau}= e^{-2\pi\tau_2}.
 \label{eq:elliptic-tau}
\end{equation}
At finite $\kappa$, the $A$-cycle is the closed curve encircling the
physical cut $[x_2,x_3]$. On the torus it can be represented by
$u=r+ic$, $-K\leq r\leq 3K$, with the endpoints identified and with $c$
arbitrarily small. Its $x$-projection is the usual dogbone contour around
the physical $[x_2,x_3]$ cut. In the two-point function we use two nearby, non-intersecting
such contours $\Gamma_1,\Gamma_2$, chosen such that their $x$-projections
are nested. The universal Cauchy term in \eqref{eq:w02_gen} then integrates to zero:
\begin{equation}
 \oint_{\Gamma_2}\frac{dx'}{2\pi i}\oint_{\Gamma_1}\frac{dx}{2\pi i}
 \frac{e^{-s_1x-s_2x'}}{(x-x')^2}= \oint_{\Gamma_2}\frac{dx'}{2\pi i}
\left(-s_1e^{-(s_1+s_2) x'}\right)=0,
 \label{eq:cauchy-term-zero}
\end{equation}
and the integrated two-point function can be computed directly from the
torus Bergman kernel. At this point, we define the normalized coordinate on the torus and its periods as
\begin{equation}
 \xi=\frac{u-K}{L},
 \qquad
 \xi\sim\xi+1\sim\xi+\tau.
\end{equation}
If we use that $\pi^2\mathrm{csc}^2(\pi z)=-4\pi^2\sum_{n=1}^\infty n e^{2\pi i n z}$, together with the Fourier series expansion of the Weierstrass $\wp$-function\footnote{please refer to \url{https://dlmf.nist.gov/23}, but notice that in our conventions $\omega_{1,2}$ are full periods. The series used converge for $0<\operatorname{Im}(\xi-\xi')<\tau_2$, a condition which we can guarantee by ordering the small imaginary offsets of the nested integration contours.}, we can obtain the following expression for the Bergman kernel \eqref{eq:berg_ker_torus}:
\begin{equation}
 B(\xi,\xi')
 =
 -4\pi^2\sum_{n=1}^{\infty}
 \frac{n}{1-Q_\tau^n}
 \left[
 e^{2\pi in(\xi-\xi')}
 +Q_\tau^n e^{-2\pi in(\xi-\xi')}
 \right]d\xi\,d\xi'.
 \label{eq:bergman-strip-expansion}
\end{equation}
We introduce the Fourier--Laplace coefficients
\begin{equation}
 F_n(s;\kappa)
 =
 \int_{-1/2}^{1/2}d\xi\,
 e^{-s x_\kappa(K+L\xi)}
 e^{-2\pi in\xi}.
 \label{eq:elliptic-Fn}
\end{equation}
 From the periodicity of the Jacobi-sine function, we have that the $x$-map \eqref{eq:elliptic-x-map} is symmetric under $\xi\to-\xi$, which corresponds to $u\to 2K-u$. This implies, from the definition \eqref{eq:elliptic-Fn}, that we have the following relation:
\begin{equation}
 F_{-n}(s;\kappa)=F_n(s;\kappa).
 \label{eq:Fn-symmetry}
\end{equation}
Inserting \eqref{eq:bergman-strip-expansion} into the double contour
integral \eqref{eq:Z02_integral1_gen} gives the exact elliptic result
\begin{equation}
 Z^{\rm ell}_{0,2}(s_1,s_2)
 =
 \sum_{n=1}^{\infty}
 n\coth(\pi n\tau_2)
 F_n(s_1;\kappa)F_n(s_2;\kappa).
 \label{eq:exact-elliptic-Z02}
\end{equation}
In particular, since
$F_n(\bar s;\kappa)=\overline{F_n(s;\kappa)}$ for the chosen contour, if we insert $s_1=\Bar{s}_2=\beta+it$, we can obtain the genus 1 elliptic spectral form factor:
\begin{equation}
 \mathrm{SFF}^{\rm ell}(\beta,t)
 =
 \sum_{n=1}^{\infty}
 n\coth(\pi n\tau_2)
 \left|F_n(\beta+it;\kappa)\right|^2.
 \label{eq:exact-elliptic-sff}
\end{equation}

\subsubsection*{Oscillatory neck contribution and decay at long times}
To build intuition for the general result for the elliptic spectral form factor \eqref{eq:exact-elliptic-sff}, we examine its pinching limit. As $\kappa\to1$, the torus may be viewed intuitively as separating into a sphere together with an increasingly long, thin neck joining the two points that become identified; correspondingly, these two sectors are probed by different Fourier. We begin by studying the contribution of the low-momentum modes supported along the long neck.

Let us consider the pinching limit, where we recall that
\begin{equation}
 d_{k-1}=x_2-x_1,
 \qquad
 d_k=x_4-x_3,
 \qquad
 \delta_k=x_3-x_2=2D.
\end{equation}
From \eqref{eq:K_Kp_pinched} we have:
\begin{equation}
 \tau_2
 \simeq
 \frac{\pi}{
 \log\!\left(16\delta_k^2/(d_{k-1}d_k)\right)}.
 \label{eq:tau-pinching}
\end{equation}
At this point, we will make the assumption that the limit is a \textit{balanced pinching}, by which we mean, recalling \eqref{eq:sigma_value}
\begin{equation}
 \frac{|\eta|}{K(\kappa)}\longrightarrow0,
 \qquad
 \sigma=\tanh\eta,
 \qquad
 \eta\approx\frac14\log\frac{d_{k-1}}{d_k}.
 \label{eq:balanced-pinching}
\end{equation}
In particular, this includes the case in which $d_{k-1}/d_k$ remains
finite\footnote{it is also possible to discuss an \textit{asymmetric pinching}, where $d_k,d_{k-1}\to 0$ at parametrically different rates. This asymmetry translates in the possibility of the two plateaux in
\eqref{eq:step-profile} not occupying equal fractions of the cycle. In particular, if
$x_3$ occupies a fraction $\alpha$, the coefficient $1/8$ in
\eqref{eq:neck-sff} will be replaced by $\alpha(1-\alpha)/2$. However, here now we only stress that both balanced and asymmetric pinching are assumptions: in principle the $d_k$'s should be obtained as solutions of \eqref{eq:epsk_integralCk}, and, even though we expect the existence of regions controlled by $E$ where they are parametrically small, they are not directly tunable parameters.}. On the normalized $A$-cycle one then has, almost everywhere,
\begin{equation}
 x(K+L\xi)
 \longrightarrow
 \begin{cases}
 x_3, & |\xi|<\frac14,\\[1mm]
 x_2, & \frac14<|\xi|<\frac12.
 \end{cases}
 \label{eq:step-profile}
\end{equation}
Consequently, for every fixed non-zero $n$, from \eqref{eq:elliptic-Fn} we obtain:
\begin{equation}
 F_n(s;\kappa)
 \longrightarrow
 \left(e^{-s x_3}-e^{-s x_2}\right)
 \frac{\sin(\pi n/2)}{\pi n}.
 \label{eq:Fn-step-limit}
\end{equation}
At the same time, when $\tau_2\to0$ we have
\begin{equation}
 n\coth(\pi n\tau_2)
 =
 \frac{1}{\pi\tau_2}
 +O(n^2\tau_2),
\end{equation}
so that using $\sum_{\substack{n\geq1\\ n\ {\rm odd}}}\frac{1}{n^2}
 =\frac{\pi^2}{8}$ in \eqref{eq:exact-elliptic-sff} we obtain:
\begin{equation}
\begin{aligned}
    \mathrm{SFF}_{\rm neck}(\beta,t)
 &=
 \frac{1}{8\pi\tau_2}
 \left|
 e^{-(\beta+it)x_3}
 -
 e^{-(\beta+it)x_2}
 \right|^2
 +O(1)=\\&=\frac{1}{8\pi\tau_2}
 \Big[
 e^{-2\beta x_2}+e^{-2\beta x_3}-2e^{-2\beta A}\cos(2Dt)
 \Big]
 +O(1).
\end{aligned}
 \label{eq:neck-sff}
\end{equation}
Therefore, the spectral form factor contains a term that oscillates with frequency $ \omega_{\rm neck}=x_3-x_2=2D$ and amplitude $e^{-2\beta A}/(4\pi\tau_2)$, exponentially suppressed by $A\approx E$ in the region where the cut-pinching approximation holds. Similarly to the genus 0 case of \eqref{eq:sff_genus_0}, if we consider the infinite temperature case, this suppression vanishes, and only the logarithmic enhancement in the pinching parameter remains:
\begin{equation}
 \mathrm{SFF}_{\rm neck}(0,t)
 =
 \frac{\sin^2(Dt)}{2\pi\tau_2}
 \simeq
 \frac{1}{2\pi^2}
 \log\!\left(\frac{16\delta_k^2}{d_{k-1}d_k}\right)
 \sin^2(Dt).
 \label{eq:neck-beta-zero}
\end{equation}
At this point, we clarify which modes \eqref{eq:neck-sff} is capturing. The procedure defined here is equivalent to considering a finite number of modes in \eqref{eq:exact-elliptic-sff} before taking the pinching limit $\kappa \to 1$, and then remove the mode cutoff. This captures all finite $n$ modes, with momentum $p_n=2\pi n/L\to 0$ and wavelengths of the same order of the parametrically long A-cycle. Conversely, the modes with fixed momentum as $L\to\infty$ contribute to the $\mathcal{O}(1)$ term in \eqref{eq:neck-sff}, which contains the universal sphere SFF linear contribution, as we will momentarily discuss.

Before moving on, we discuss the effect of retaining the first correction in the pinching parameters in \eqref{eq:step-profile}, which becomes important at late times. We provide the expansion in a local coordinate $v$, near the two
stationary points of $x_\kappa(u)$:
\begin{equation}\label{eq:endpoint-expansions}
    \begin{aligned}
        x(K+v)
 &=
 x_3-\nu_+v^2+O(v^4),
 \qquad \nu_+
 =
 \frac{D\kappa'^2}{2}\frac{1-\sigma}{1+\sigma}
 \approx d_k,\\
 x(3K+v)
 &=
 x_2+\nu_-v^2+O(v^4),
 \qquad\nu_-
 =
 \frac{D\kappa'^2}{2}\frac{1+\sigma}{1-\sigma}
 \approx d_{k-1}.
    \end{aligned}
\end{equation}
For the fixed $n$ neck sector, the pinching-parameter corrections to \eqref{eq:elliptic-x-map} presented above, become important at times such that $t\nu_\pm\gg 1$. In this regime, we can compute \eqref{eq:elliptic-Fn} using the stationary phase approximation around $x_2$ and $x_3$:
\begin{equation}
    F_n(\beta+it;\kappa)
 \sim
 \frac{\sqrt{\pi}}{L\sqrt{t}}
 \Bigg(
 \frac{e^{-(\beta+it)x_3+i\pi/4}}
 {\sqrt{\nu_+}}
 e^{-ip_n^2/(4t\nu_+)}+
 (-1)^n
 \frac{e^{-(\beta+it)x_2-i\pi/4}}
 {\sqrt{\nu_-}}
 e^{ip_n^2/(4t\nu_-)}
 \Bigg).
 \label{eq:Fn-stationary-phase-detailed}
\end{equation}
If we insert the result above in \eqref{eq:exact-elliptic-sff}, it follows that at sufficiently late times we have the following single-mode contribution to the SFF:
\begin{equation}\label{eq:neck-mode-tminusone}
    \begin{aligned}
        \operatorname{SFF}^{(n)}_{\rm neck}(\beta,t)
 \approx
 \frac{1}{HLt}
 \Bigg(
 \frac{e^{-2\beta x_3}}{d_k}
 +
 \frac{e^{-2\beta x_2}}{d_{k-1}}
 +
 \frac{2(-1)^ne^{-2\beta A}}
 {\sqrt{d_k d_{k-1}}}
 \sin\!\left(
 2D t
 \right)
 \Bigg).
    \end{aligned}
\end{equation}
Thus, once $t d_{k-1},t d_k\gg1$, every fixed neck mode, and more
generally every fixed finite band of such modes, develops a $t^{-1}$
envelope, together with the overall $L^{-1}$ suppression displayed in
\eqref{eq:neck-mode-tminusone}. This statement applies to the
low-momentum neck sector, $n=o(L)$. Since this sector contains only
$o(L)$ modes, summing \eqref{eq:neck-mode-tminusone} over all such modes
does not compensate the $L^{-1}$ prefactor: its total contribution
remains parametrically suppressed relative to the $O(L)$ band of modes
forming the sphere sector. Finite-pinching corrections to the step
profile \eqref{eq:step-profile} therefore turn on a time decay for the amplitude of the neck oscillations, whose late-time envelope is proportional
to $t^{-1}$ once $t d_{k-1},t d_k\gg1$.
\subsubsection*{Sphere contribution and the linear ramp}
From the Fourier coefficients expression \eqref{eq:elliptic-Fn}, we understand that the fixed-$n$ limit isolates modes whose wavelength is of order the full
$A$-cycle and hence probes the long neck. The local sphere sector instead
comes from modes whose momenta is such that:
\begin{equation}
 p_n=\frac{2\pi n}{L}=O(1)\implies
 n=O(L).
 \label{eq:sphere-mode-scaling}
\end{equation}
Indeed, given that there are $\mathcal{O}(L)$ such modes, we have
\begin{equation}
 n\coth(\pi n\tau_2)
 =
 \frac{Lp_n}{2\pi}
 \coth\left(\frac{Hp_n}{2}\right),
 \qquad
 F_n=O(L^{-1}),
 \label{eq:sphere-mode-weight}
\end{equation}
so they can give a finite contribution to \eqref{eq:exact-elliptic-sff}. Equivalently,
for $u-u'$ in a fixed compact region, the expansion of \eqref{eq:torus_bg}
\begin{equation}
 B(u,u')
 =
 \left[
 \left(\frac{\pi}{H}\right)^2
 \frac{1}{\sinh^2\!\left(\pi(u-u')/H\right)}
 +
 \frac{2\pi}{LH}
 +
 O\!\left(e^{-2\pi L/H}\right)
 \right]du\,du' .
 \label{eq:local-cylinder-kernel}
\end{equation}
The first term is the genus $0$ Bergman kernel of the original cylinder of compact
period $iH$, which we already obtained in \cref{app:torus_deg}. The remaining terms, although subleading with respect to \eqref{eq:berg_torus_deg_sinh}, can conspire to give $\mathcal{O}(L)$ contributions to the spectral form factor, if the two integration contours have length $\mathcal{O}(L)$. This is indeed what happens in the pinching limit and the corrections to \eqref{eq:berg_torus_deg_sinh} generate the long-neck contribution.

Let
$\iota:\Sigma\to\Sigma$ be the elliptic Galois
involution on the torus with parameter $\kappa$. In the Jacobi coordinate, this is defined by the conditions
\begin{equation}
 \iota(u)=2K(\kappa)-u,
 \qquad
 x_\kappa(\iota u)=x(u).
\end{equation}
Now we define
\begin{equation}
 \Omega(u,u')
 \equiv
 W_{0,2}(x(u),x(u'))\,dx(u)\,dx(u'),
\end{equation}
so that, given the above definition and \eqref{eq:w02_gen}, we have the following identity:
\begin{equation}
 \frac{dx(u)\,dx(u')}{(x(u)-x(u'))^2}
=B(u,u')+B(u,\iota u'),
 \qquad
 \Omega(u,u')=-B(u,\iota u'),
 \label{eq:elliptic-image-identity}
\end{equation}
Parametrize the two lips of the cut from $x_2$ to $x_3$ by
$\gamma_+$ and $\gamma_-=\iota(\gamma_+)$, so that the oriented
$A$-cycle is
\begin{equation}
 \Gamma=\gamma_+-\gamma_- .
\end{equation}
By separating in such way the contour $\Gamma$ appearing in \eqref{eq:Z02_integral1_gen}, we can rewrite the SFF as a standard integral on a real interval as
\begin{equation}
    SFF_{g=0}(\beta, t)=\int_{x_2}^{x_3}dx \int_{x_2}^{x_3}dx' e^{-\beta(x+x')}e^{-it (x-x')}\rho_{0,2}^c(x,x').
\end{equation}
We now proceed to determine $\rho_{0,2}^c(x,x')$ by summing over all the lip contributions. The Bergman kernel on the cylinder:
\begin{equation}
 B_{\rm cyl}(u,u')
 \equiv
 \left(\frac{\pi}{H}\right)^2
 \frac{du\,du'}
 {\sinh^2\!\left(\pi(u-u')/H\right)}
 =
 \frac{dx(u)\,dx(u')}
 {(x(u)-x(u'))^2}.
 \label{eq:two-sheet-cylinder-kernel}
\end{equation}
Consequently, from \eqref{eq:elliptic-image-identity} and \eqref{eq:local-cylinder-kernel}, when $u$ and $u'$ live in the compact region away from the endpoints that we are restricting to here, we have:
\begin{equation}
\begin{aligned}
 &\Omega_{\pm\mp}=-\left(\frac{\pi}{H}\right)^2\frac{dudu'}{\sinh^2{(\pi(u-u')/H})}+\mathcal{O}(L^{-1})\longrightarrow_{K\to\infty}-\frac{dxdx'}{(x-x'\pm i0)^2},\\&
\Omega_{\pm\pm}=-\left(\frac{\pi}{H}\right)^2\frac{dudu'}{\sinh^2{(\pi(2K(\kappa)-u-u')/H})}+\mathcal{O}(L^{-1})\longrightarrow_{K\to\infty}0,
\end{aligned}
 \label{eq:two-sheet-cylinder-sectors}
\end{equation}
where the $\pm\pm$ at the pedices correspondingly label the lip containing respectively the first and second argument of $\Omega$. Notice that the $\pm i\epsilon$ prescription used depends on the nesting chosen for the integration contours. However this convention does not change the expression for $\rho_{0,2}^c$, determined by the sum of the two cross sectors, and therefore, in the Hadamard finite-part sense, given by:
\begin{equation}
\begin{aligned}
    \rho_{0,2}^{\,c}(x,x')dxdx'
 &=\frac{1}{(2\pi i)^2}\left(\Omega_{++}-\Omega_{-+}-\Omega_{+-}+\Omega_{--}\right)\\&\implies
 \rho_{0,2}^{\,c}(x,x')=-\frac{1}{2\pi^2}
 \operatorname{Pf}\frac{1}{(x-x')^2}+\mathrm{reg.\,in\,} x=x',
 \label{eq:bulk-cylinder-density}
\end{aligned}
\end{equation}
where the regular part is $\mathcal{O}(L^{-1})$, and can be neglected when studying the compact region contribution.
At this point, if we use $\int_{\mathbb R}d\omega\,
 e^{-it\omega}
 \operatorname{Pf}\frac{1}{\omega^2}
 =
 -\pi |t|$, we obtain that the singular part produces the universal ramp
\begin{equation}
 \mathrm{SFF}_{\rm sphere}(\beta,t)
 \approx
 \frac{|t|}{4\pi\beta}
 \left(e^{-2\beta x_2}-e^{-2\beta x_3}\right)
 +O(1),
 \qquad
 D|t|\gg1.
 \label{eq:sphere-ramp-general}
\end{equation}
In particular, if we consider the limit $\beta D\gg1$, as we did in \cref{eq:sff_genus_0}, the upper endpoint contribution is suppressed and
\eqref{eq:sphere-ramp-general} reduces to \eqref{eq:SFF_genus0}. Similarly, we can obtain \eqref{eq:sff_inf_T} for the infinite temperature case
\begin{equation}\label{eq:sff_sphere_app}
 \mathrm{SFF}_{\rm sphere}(0,t)
 \approx
 \frac{D}{\pi}|t|.
\end{equation}
The fact that we are able to recover the genus 0 sphere results is an indication that
a convenient global interpretation of the sheet bookkeeping in
\eqref{eq:two-sheet-cylinder-sectors} is obtained by passing to the degree-two cover selected
by the surviving cut, which we already considered in \cref{app:torus_deg} to reconstruct the sphere Bergman kernel.
Each component is the genus-zero double cover associated with the
surviving cut $[x_2,x_3]$, and consequently, either component is uniformized
by the same coordinate $Z$ appearing in \eqref{eq:jouk_map_deg}, and carries the
genus-zero Bergman kernel already obtained in \eqref{eq:berg0_deg_change_coords}. We stress that the role of the cover is therefore only to repackage the
two lips and their sheet structure into single global data. Notice that 
choosing one lift of the physical contour localizes us to the sphere contribution
\eqref{eq:sphere-ramp-general}, while the full elliptic answer is the sum of \eqref{eq:sff_sphere_app} and \eqref{eq:neck-beta-zero}, so it additionally contains the long-neck
sector. This fact suggests a possible prescription to get rid of the oscillatory terms appearing in \cref{app:elliptic-sff}: if we keep the lifted contour fixed before taking
$\kappa\to1$ we isolate the sphere term, whereas collapsing the original
$A$-cycle first also retains the fixed-mode contribution \eqref{eq:neck-sff}.

\section{Numerical implementation and run parameters}
\label{app:numerics}
 
\paragraph{Sampled measure.}
All Monte Carlo data sample the eigenvalue measure of the Cardy matrix
model \eqref{eq:part_funct_cardy}: the Vandermonde $\Delta(h)^2$ times the constraint$^2$ weight
$\exp[-(1/4a\epsilon^2)\sum_P (N_{P,\epsilon}-\epsilon\sum_i
\mathbb{S}_{PP_i}(\mathds{1})-s_P)^2]$, where we separated the source term $s_P$ from the dynamical heavy-states sum. In the code the constraint is
written in the equivalent form with deviation
$(N_{P,\epsilon}-\epsilon\sum_i \mathbb{S}_{PP_i}-s_P)/\epsilon$ and weight
$e^{-(\cdot)^2/4a}$, so the reported coupling $a$ is exactly that of
\eqref{eq:part_funct_cardy}. We work at $c=25$ ($b=1$) unless noted, with
$\mathbb{S}_{\mathds{1}P}(\mathds{1})=4\sqrt2\sinh^2(2\pi P)$ (2.30) and
$\mathbb{S}_{P'P}(\mathds{1})=2\sqrt2\cos(4\pi P'P)$; the identity sits at
fixed imaginary momentum and enters as the external source $s_P$. Two
boundary conditions at the cutoff are sampled. In the \emph{clipped} runs the source is the restriction of the Cardy density to the
band, $s_P=\epsilon\,\mathbb{S}_{P\mathds{1}}(\mathds{1})$. In the
\emph{completed} runs of \cref{sec:numerics} the identity drags its
modular tail above the cutoff with it, and the source is its finite-window
modular defect,
\begin{equation}
 s_P\;=\;\epsilon\Bigl[\mathbb{S}_{P\mathds{1}}(\mathds{1})
 -\!\int_{\rm band}\!dP'\,\mathbb{S}_{PP'}(\mathds{1})\,
 \mathbb{S}_{P'\mathds{1}}(\mathds{1})\Bigr].
 \label{eq:completed-source}
\end{equation}
Both boundary conditions use the same midpoint discretization of the
kernel, $\epsilon\,\mathbb{S}_{P_bP_i}$ at the bin centre; the exact
bin-integrated kernel differs from it by the multiplicative factor
$\mathrm{sinc}(2\pi\epsilon P_i)$, i.e.\ by at most $0.4\%$ per element at
our coarsest $\epsilon$. The discretization fixes the count bookkeeping:
the preferred Cardy count of a run is the bin sum
$\epsilon\sum_b\mathbb{S}_{P_b\mathds{1}}(\mathds{1})$ and
source and $N$ are always chosen in the same convention.
 
\paragraph{Proposal kernel and annealing.}
We use Metropolis--Hastings on the $N$ momenta, with the same procedure for
both boundary conditions. Each update perturbs one eigenvalue with a
three-scale mixture proposal: with probabilities $0.60/0.25/0.15$ a local
Gaussian step of width $\sigma$, a bin-scale Gaussian step of width
$\epsilon$, or a uniform step over the band --- mixing eigenvalues within a
bin, across neighboring bins, and between distant bins respectively. The
local width $\sigma$ is adapted to hold the acceptance in $[0.18,0.42]$,
and frozen before production sampling. Because the
constraint raises high barriers between occupation sectors, we anneal $a$
from $a_{\rm hi}=1$ to its target along a logarithmic schedule of
$n_{\rm anneal}$ stages (the numerical analogue of $a\to0$) before
collecting production samples at fixed $a$, thinned by $N$ sweeps.
High-energy window runs are initialized from the Cardy density
$\mathbb{S}_{P\mathds{1}}(\mathds{1})$; full-band and vacuum-off runs from
a uniform distribution. The same conventions apply verbatim to the
completed runs.
 
\paragraph{Level statistics.}
The spectral form factor and level spacing (fig.~5) probe the spectrum at
the mean level spacing, $\sim3\times10^{-5}$, far below the proposal scales
above. For these we re-equilibrate the saved configurations with an
additional fine local move of width $\sigma_f\simeq5\times10^{-5}$ (mixed
$0.70/0.20/0.08/0.02$ with the bin-, $\epsilon$- and band-scale moves),
which thermalizes the within-bin structure that the coarse moves leave
unresolved; levels are then unfolded by the ensemble-mean cumulative count
before computing $P(s)$ and $K_c(\tau)$. All level-statistics data refer to
the clipped runs.
 
\begin{table}[t]
\centering
\footnotesize
\setlength{\tabcolsep}{3.5pt}
\begin{tabular}{llcccll}
\toprule
run & source & $\epsilon$ & band/window & $N$ & $a$ &
\shortstack[l]{steps/chain (anneal/equil/prod),\\ chains} \\
\midrule
\cref{fig:mcmc_occupations}(a)      & clipped   & 0.027 & $[0,0.918]$  & 11458  & 0.004 & $80\times12$k$/60$k$/400$k, 8 \\
\cref{fig:mcmc_crossover}(a) (vacuum on)        & clipped   & 0.005 & $[0.35,0.9]$ & 9172   & 0.04  & $100\times10$k$/60$k$/400$k, 6 \\
\cref{fig:mcmc_crossover}(a) (vacuum off)       & clipped   & 0.005 & $[0.35,0.9]$ & 9172   & 0.04  & $60\times8$k$/60$k$/250$k, 6 \\
\cref{fig:mcmc_sff} (SFF, $P(s)$)      & clipped   & 0.005 & $[0.35,0.9]$ & 9172   & 0.04  & --/80k/700k (+fine), 8 \\
\addlinespace
\cref{fig:mcmc_occupations}(b)   & completed & 0.027 & $[0,0.918]$  & 11458  & 0.004 & $80\times12$k$/60$k$/400$k, 8 \\
\cref{fig:mcmc_crossover}(b) (vacuum on)     & completed & 0.005 & $[0.35,0.9]$ & 9172   & 0.04  & $100\times10$k$/60$k$/400$k, 6 \\
\cref{fig:mcmc_crossover}(b) (vacuum off)    & completed & 0.005 & $[0.35,0.9]$ & 9172   & 0.04  & $60\times8$k$/60$k$/250$k, 6 \\
\bottomrule
\end{tabular}
\caption{Run parameters at $c=25$. Robustness runs repeat the crossover
setup at $c=26$ and $c=38.5$.
In every case $N$ is set to the preferred Cardy count $M_{\mathds{1}}$ of the
run in the discretization used.}
\label{tab:runs}
\end{table}
 
\paragraph{Reproducibility.}
Each run records its seed and parameters; clipped chains are seeded
independently by a \texttt{SeedSequence} fan-out and combined. For the
completed ensembles we perform an independent verification on
every such run: the positivity-constrained minimizer of the sampled
functional (constraint$^2$ plus the Vandermonde energy on a sub-bin grid)
is computed as a convex quadratic program, and the equilibrated chains are
required to reproduce it --- the mean occupations agree with the minimizer
bin by bin, including the count-neutral deformation along the
near-invariant prolate modes discussed in section~\ref{sec:vac_occ_gap},
with the sampled potential sitting at the integer-occupation floor above
the continuum minimum. The universal outputs --- the Cardy slope $2\pi Q$,
the gap at $(c-1)/24$, the agreement of the completed high-momentum
density with $\epsilon\,\mathbb{S}_{P\mathds{1}}(\mathds{1})$ at the
$10^{-3}$ level, and the GUE level statistics --- are independent of the
regulator and sampling choices above, as we checked by varying $c$,
$\epsilon$, and the chain length.

\bibliography{ref}
\end{document}